\documentclass{article}

\usepackage[margin=1.5cm, includefoot, footskip=30pt]{geometry}
\usepackage{hyperref}
\usepackage{multicol}
\usepackage{booktabs}
\usepackage{xstring}
\usepackage{subcaption}
\usepackage{graphicx}
\usepackage{amsmath}
\usepackage{standalone}
\usepackage{fancyvrb}
\usepackage{algorithm,algorithmic}
\usepackage{authblk}

\usepackage{tikz}
\usetikzlibrary{calc, shapes, patterns}

\title{Evolution Reinforces Cooperation with the Emergence of Self-Recognition
       Mechanisms: an empirical study of the Moran process for the iterated
       Prisoner's dilemma}
\author[1]{Vincent Knight}
\author[2]{Marc Harper}
\author[1]{Nikoleta E. Glynatsi}
\author[3]{Owen Campbell}

\affil[1]{Cardiff University, School of Mathematics, UK}
\affil[2]{Google Inc., Mountain View, CA, USA}
\affil[3]{Not affiliated}

\date{}

\begin{document}

\maketitle

\begin{abstract}
    We present insights and empirical results from an extensive numerical
    study of the evolutionary dynamics of the iterated prisoner's dilemma.
    Fixation probabilities for Moran processes are obtained
    for all pairs of 164 different strategies including classics such as TitForTat, zero
    determinant strategies, and many more sophisticated strategies.
    Players with long memories and sophisticated behaviours outperform
    many strategies that perform well in a two player setting. Moreover we
    introduce several strategies trained with evolutionary algorithms to
    excel at the Moran process. These strategies are excellent invaders
    and resistors of invasion and in some cases naturally evolve handshaking
    mechanisms to resist invasion. The best invaders were those trained
    to maximize total payoff while the best resistors invoke handshake mechanisms.
    This suggests that while maximizing individual payoff
    can lead to the evolution of cooperation through invasion, the relatively
    weak invasion resistance of payoff maximizing strategies are not as
    evolutionarily stable as strategies employing handshake mechanisms.
\end{abstract}

\section{Introduction}\label{sec:introduction}

The Prisoner's Dilemma (PD) \cite{Flood1958} is a fundamental two player game
used to model a variety of strategic interactions. Each player chooses
simultaneously and independently
between cooperation (C) or defection (D). The payoffs of
the game are defined by the matrix $\begin{pmatrix} R & S \\ T & P
\end{pmatrix}$, where $T > R > P > S$ and $2R > T + S$. The PD is a one
round game, but is commonly studied in a manner where the prior outcomes
matter. This repeated form is called the Iterated Prisoner's
Dilemma (IPD). As described in \cite{Axelrod1980a, Knight2016, Press2012} a
number of strategies have been developed to take advantage of the history of
play. Recently, some strategies referred to as zero determinant (ZD) strategies
\cite{Press2012} can manipulate some players through extortionate mechanisms.

The Moran Process \cite{Moran1957} is a model of
evolutionary population dynamics that has been used to gain insights about the
evolutionary stability in a number of settings (more details given in
Section~\ref{sec:the_moran_process}).
Several earlier
works have studied iterated games in the context of the prisoner's dilemma
\cite{Nowak, stewart2013extortion}, however these often make simplifying assumptions
or are limited to classes of strategies
such as memory-one strategies that only use the previous round of play.

This manuscript provides a detailed numerical analysis of agent-based simulations
of \textbf{164}complex and adaptive strategies for the
IPD\@. This is made possible by the Axelrod library \cite{axelrodproject}, an
effort to provide software for reproducible research for the IPD\@. The library
now contains over 186parameterized
strategies including classics like TitForTat and WinStayLoseShift, as well as
recent variants such as OmegaTFT, zero determinant and other memory one
strategies, strategies based on finite state machines, lookup tables, neural
networks, and other machine learning based strategies, and a collection of novel
strategies. Not all strategies have been considered for this study: excluded
are those that make use of knowledge of the number of turns in a match
and others that have a high
computational run time. The large number of strategies are available thanks to
the open source nature of the project with over 50 contributors from around the
world, made by programmers and researchers \cite{Knight2016}. Three of the considered
strategies are finite state machines trained specifically for Moran processes
(described further in Section~\ref{sec:strategies}).

In addition to providing a large collection of strategies, the Axelrod library
can conduct matches, tournaments and population
dynamics with variations including noise and spatial structure.
The strategies and simulation frameworks are
automatically tested to an extraordinarily high degree of coverage in accordance
with best research software practices.

Using the Axelrod library and the many strategies it contains, we obtain
fixation probabilities for all pairs of strategies, identifying
those that are effective invaders and those resistant to invasion, for
population sizes $N=2$ to $N=14$. Moreover we present a number of strategies
that were created via reinforcement algorithms (evolutionary and particle
swarm algorithms) that are among the best invaders and resistors of invasion
known to date, and show that handshaking mechanisms naturally arise from these
processes as an invasion-resistance mechanism.

Recent work has argued that agent-based
simulations can provide insights in evolutionary game theory not available
via direct mathematical analysis \cite{adami2016evolutionary}. The results
and insights contained in this paper would be difficult to derive analytically.

In particular the following questions are addressed:
\begin{enumerate}
    \item What strategies are good invaders?
    \item What strategies are good at resisting invasion?
    \item How does the population size affect these findings?
\end{enumerate}

While the results agree with some of the published literature, it is found that:

\begin{enumerate}
 \item Zero determinant strategies are not particularly effective for $N > 2$
 \item Complex strategies can be effective, and in fact can naturally evolve
     through evolutionary processes to outperform designed strategies.
 \item The strongest resistors specifically evolve or have a handshake mechanism.
 \item Strong invaders are generally cooperative strategies that do not defect
 first but retaliate to varying degrees of intensity against strategies that defect.
 \item Strategies evolved to maximize their total payoff can be strong invaders
 and achieve mutual cooperation with many other strategies.
\end{enumerate}

\subsection{The Moran Process}\label{sec:the_moran_process}

\begin{figure}[!hbtp]
    \centering
    \begin{tikzpicture}[scale=.9]
	\node (A1) at (-1, -1) [circle, pattern=north west lines, pattern
        color=blue!70, draw=blue] {};
	\node (A2) at (-1, 1) [circle, pattern=north west lines, pattern
        color=blue!70, draw=blue] {};
	\node (A3) at (0, .2) [circle, pattern=north west lines, pattern
        color=blue!70, draw=blue] {};
	\node (A4) at (.2, -.6) [circle, pattern=north west lines, pattern
        color=blue!70, draw=blue] {};
	\node (A5) at (.5, -0.1) [circle, pattern=north west lines, pattern
        color=blue!70, draw=blue] {};
	\node (B1) at (-1, -.2) [circle, pattern=north east lines, pattern
        color=red!70, draw=red] {};
	\node (B2) at (1, -.5) [circle, pattern=north east lines, pattern
        color=red!70, draw=red] {};
	\node (B3) at (.5, .8) [circle, pattern=north east lines, pattern
        color=red!70, draw=red] {};
	\node (B4) at (-.6, .2) [circle, pattern=north east lines, pattern
        color=red!70, draw=red] {};
	\node (B5) at (.6, -.8) [circle, pattern=north east lines, pattern
        color=red!70, draw=red] {};

	\draw [->] (1, 0) -- (3, 0) node [above, pos=0.5] {Selection};

	\node (A1) at ($(A1) + (4.5, 0)$) [circle, pattern=north west lines,
        pattern color=blue!70, draw=blue] {};
	\node (A2) at ($(A2) + (4.5, 0)$) [circle, pattern=north west lines,
        pattern color=blue!70, draw=blue] {};
	\node (A3) at ($(A3) + (4.5, 0)$) [circle, pattern=north west lines,
        pattern color=blue!70, draw=blue] {};
	\node (A4) at ($(A4) + (4.5, 0)$) [circle, pattern=north west lines,
        pattern color=blue!70, draw=blue] {};
    \node (A5) at ($(A5) + (4.5, 0)$) [circle, pattern=north west lines,
        pattern color=blue!70, draw=blue] {};
	\node (B1) at ($(B1) + (4.5, 0)$) [circle, pattern=north east lines,
        pattern color=red!70, draw=red] {};
	\node (B2) at ($(B2) + (4.5, 0)$) [circle, pattern=north east lines,
        pattern color=red!70, draw=red] {};
	\node (B3) at ($(B3) + (4.5, 0)$) [circle, pattern=north east lines,
        pattern color=red!70, draw=red] {};
	\node (B4) at ($(B4) + (4.5, 0)$) [circle, pattern=north east lines,
        pattern color=red!70, draw=red] {};
	\node (B5) at ($(B5) + (4.5, 0)$) [circle, pattern=north east lines,
        pattern color=red!70, draw=red] {};

	\draw [dotted, very thick] (B4) circle (.25cm);

	\draw [->] (5.5, 0) -- (7.5, 0) node [above, pos=0.5] {Birth};

	\node (A1) at ($(A1) + (4.5, 0)$) [circle, pattern=north west lines,
        pattern color=blue!70, draw=blue] {};
	\node (A2) at ($(A2) + (4.5, 0)$) [circle, pattern=north west lines,
        pattern color=blue!70, draw=blue] {};
	\node (A3) at ($(A3) + (4.5, 0)$) [circle, pattern=north west lines,
        pattern color=blue!70, draw=blue] {};
	\node (A4) at ($(A4) + (4.5, 0)$) [circle, pattern=north west lines,
        pattern color=blue!70, draw=blue] {};
    \node (A5) at ($(A5) + (4.5, 0)$) [circle, pattern=north west lines,
        pattern color=blue!70, draw=blue] {};
	\node (B1) at ($(B1) + (4.5, 0)$) [circle, pattern=north east lines,
        pattern color=red!70, draw=red] {};
	\node (B2) at ($(B2) + (4.5, 0)$) [circle, pattern=north east lines,
        pattern color=red!70, draw=red] {};
	\node (B3) at ($(B3) + (4.5, 0)$) [circle, pattern=north east lines,
        pattern color=red!70, draw=red] {};
	\node (B4) at ($(B4) + (4.5, 0)$) [circle, pattern=north east lines,
        pattern color=red!70, draw=red] {};
	\node (B5) at ($(B5) + (4.5, 0)$) [circle, pattern=north east lines,
        pattern color=red!70, draw=red] {};

	\draw [dotted, very thick] (B4) circle (.25cm);
	\node (B6) at ($(B4) + (0.5, 0.5)$) [circle, pattern=north east lines,
        pattern color=red!70, draw=red] {};
	\draw [->, dotted, very thick] (B4) [out=90, in=180] to (B6);

	\draw [->] (10, 0) -- (12, 0) node [above, pos=0.5] {Selection};

	\node (A1) at ($(A1) + (4.5, 0)$) [circle, pattern=north west lines,
        pattern color=blue!70, draw=blue] {};
	\node (A2) at ($(A2) + (4.5, 0)$) [circle, pattern=north west lines,
        pattern color=blue!70, draw=blue] {};
	\node (A3) at ($(A3) + (4.5, 0)$) [circle, pattern=north west lines,
        pattern color=blue!70, draw=blue] {};
	\node (A4) at ($(A4) + (4.5, 0)$) [circle, pattern=north west lines,
        pattern color=blue!70, draw=blue] {};
    \node (A5) at ($(A5) + (4.5, 0)$) [circle, pattern=north west lines,
        pattern color=blue!70, draw=blue] {};
	\node (B1) at ($(B1) + (4.5, 0)$) [circle, pattern=north east lines,
        pattern color=red!70, draw=red] {};
	\node (B2) at ($(B2) + (4.5, 0)$) [circle, pattern=north east lines,
        pattern color=red!70, draw=red] {};
	\node (B3) at ($(B3) + (4.5, 0)$) [circle, pattern=north east lines,
        pattern color=red!70, draw=red] {};
	\node (B4) at ($(B4) + (4.5, 0)$) [circle, pattern=north east lines,
        pattern color=red!70, draw=red] {};
	\node (B5) at ($(B5) + (4.5, 0)$) [circle, pattern=north east lines,
        pattern color=red!70, draw=red] {};
	\node (B6) at ($(B6) + (4.5, 0)$) [circle, pattern=north east lines,
        pattern color=red!70, draw=red] {};

	\draw [dotted, very thick] (A3) circle (.25cm);

	\draw [->] (14.5, 0) -- (16.5, 0) node [above, pos=0.5] {Death};

	\node (A1) at ($(A1) + (4.5, 0)$) [circle, pattern=north west lines,
        pattern color=blue!70, draw=blue] {};
	\node (A2) at ($(A2) + (4.5, 0)$) [circle, pattern=north west lines,
        pattern color=blue!70, draw=blue] {};
	\node (A3) at ($(A3) + (4.5, 0)$) {};
	\node (A4) at ($(A4) + (4.5, 0)$) [circle, pattern=north west lines,
        pattern color=blue!70, draw=blue] {};
    \node (A5) at ($(A5) + (4.5, 0)$) [circle, pattern=north west lines,
        pattern color=blue!70, draw=blue] {};
	\node (B1) at ($(B1) + (4.5, 0)$) [circle, pattern=north east lines,
        pattern color=red!70, draw=red] {};
	\node (B2) at ($(B2) + (4.5, 0)$) [circle, pattern=north east lines,
        pattern color=red!70, draw=red] {};
	\node (B3) at ($(B3) + (4.5, 0)$) [circle, pattern=north east lines,
        pattern color=red!70, draw=red] {};
	\node (B4) at ($(B4) + (4.5, 0)$) [circle, pattern=north east lines,
        pattern color=red!70, draw=red] {};
	\node (B5) at ($(B5) + (4.5, 0)$) [circle, pattern=north east lines,
        pattern color=red!70, draw=red] {};
	\node (B6) at ($(B6) + (4.5, 0)$) [circle, pattern=north east lines,
        pattern color=red!70, draw=red] {};

	\draw [dotted, very thick] (A3) circle (.25cm);
\end{tikzpicture}
    \caption{A diagrammatic representation of a Moran process.}
    \label{fig:moran_process}
\end{figure}
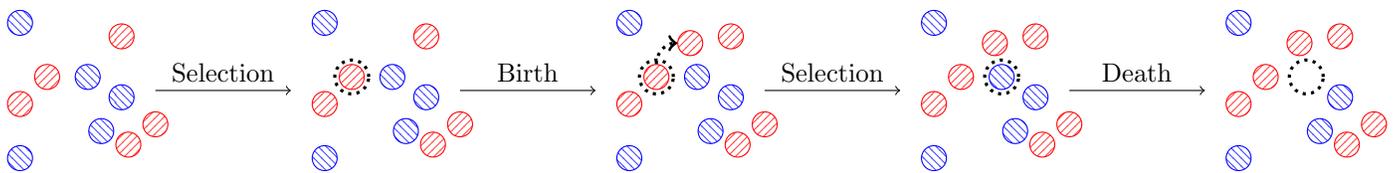

Figure~\ref{fig:moran_process} shows a diagrammatic representation of the Moran
process, a stochastic birth death process on a finite population in which the
population size stays constant over time. Individuals are \textbf{selected}
according to a given fitness landscape. Once selected, the individual is
reproduced and similarly another individual is chosen to be removed from the
population. In some settings mutation is also considered but without mutation
(the case considered in this work) this process will arrive at an absorbing
state where the population is entirely made up of players of one strategy. The
probability with which a given strategy takes over a population is called the
\textit{fixation probability}. A more detailed analytic description of this is
given in Section~\ref{sec:validation}. In our simulations offspring do not
inherit any knowledge or history from parent replicants.

The Moran process was initially introduced in \cite{Moran1957}. It has since
been used in a variety of settings including the understanding of the spread of
cooperative and non-cooperative behaviour such as cancer \cite{West2016} and the
emergence of cooperative behaviour in spatial topologies \cite{Nowak2017}.
However these works mainly consider relatively simple strategies. Some work has
looked at evolutionary stability of agent-based strategies within the Prisoner's Dilemma
\cite{Li2014} but this is not done in the more widely used setting of the Moran
process, rather in terms of infinite population stability. In \cite{Baek2016}
Moran processes are studied in a theoretical framework for a small subset of
strategies. The subset included memory one strategies: strategies that recall
the events of the previous round only.

Of particular interest are the zero determinant strategies introduced in
\cite{Press2012}. It was argued in \cite{stewart2013extortion} that generous
ZD strategies are robust against invading strategies. However, in \cite{Lee2015},
a strategy using machine learning techniques was capable of resisting invasion
and also able to invade any memory one strategy. Recent work \cite{Hilbe2017}
has investigated the effect of memory length on strategy performance and the
emergence of cooperation but this is not done in a Moran process context and only
considers specific cases of memory 2 strategies. In \cite{Adami2013} it was
recognised that many zero determinant strategies do not fare well against
themselves. This is a disadvantage for the Moran process where the best
strategies cooperate well with other players using the same strategy.

\subsection{Strategies considered}\label{sec:strategies}

To carry out this numerical experiment, 
strategies, listed (with their properties) in Appendix~\ref{app:list_of_players},
are used from the Axelrod library. There are
43stochastic and
121deterministic strategies. Their memory depth,
defined by the number of rounds of history used by the strategy each round, is
shown in Table~\ref{tbl:memory_depth_count}. The memory depth is infinite if the
strategy uses the entire history of play (whatever its length). For example, a
strategy that utilizes a handshaking mechanism where the opponent's actions on
the first few rounds of play determines the strategies subsequent behavior would
have infinite memory depth.

A number of these strategies have been trained with reinforcement learning
algorithms prior to this study and not specifically for the Moran process.

\begin{itemize}
    \item Evolved ANN: a neural network based strategy;
    \item Evolved LookerUp: a lookup table based strategy;
    \item PSO Gambler: a stochastic version of the lookup table based strategy;
    \item Evolved HMM: a hidden Markov model based strategy.
\end{itemize}

Apart from the PSO Gambler strategy, which was trained using a particle swarm
optimisation algorithm, these strategies are trained with an evolutionary
algorithm that perturbs strategy parameters and optimizes the mean total score
against all other opponents \cite{affenzeller2009genetic}. They were trained to
win IPD tournaments by maximizing their mean total payoffs against a variety
of opponents. Variation is
introduced via mutation and crossover of parameters, and the best performing
strategies are carried to the next generation along with new variants. Similar
methods appear in the literature \cite{Ashlock2006}.

More information about each player can be obtained in the documentation for
\cite{axelrodproject} and a detailed description of the performance
of these strategies in IPD tournaments is described in~\cite{Harper2017}.

All of the training code is archived at \cite{marc_harper_2017_824264}. This
software is (similarly to the Axelrod library) available on github
\url{https://github.com/Axelrod-Python/axelrod-dojo} with documentation to
train new strategies easily. Training
typically takes less than 100 generations and can be completed within several
hours on commodity hardware.

\begin{figure}[!hbtp]
    \centering
        \begin{tikzpicture}

    \tikzstyle{state}=[minimum width=0.5cm, font=\boldmath];

    \node[circle, draw=black, thick]  (0) at (0,0) [state] {$1$};
    \node[circle, draw=black, thick]  (1) at ($(0)+(0,2)$) [state] {$2$};
    \node[circle, draw=black, thick]  (2) at ($(1)+(2,1.5)$) [state] {$3$}; 
    \node[circle, draw=black, thick]  (3) at ($(2)+(3,0)$) [state] {$4$};
    \node[circle, draw=black, thick]  (4) at ($(1)+(8,0)$) [state] {$5$};  
    \node[circle, draw=black, thick]  (5) at ($(0)+(8,0)$) [state] {$6$};
    \node[circle, draw=black, thick]  (6) at ($(3)+(0,-4.5)$) [state] {$7$};
    \node[circle, draw=black, thick]  (7) at ($(2)+(0,-4.5)$) [state] {$8$}; 

    \coordinate[left of=0] (s);

    \draw (s) edge[out=0, in=180, ->, thick] node [above] {$C$} (0);

    \draw (0) edge[out=-45, in=-100, loop, thick] node [below] {$C/C$} (0);
    \draw (6) edge[out=-45, in=-100, loop, thick] node [below] {$C/D$} (6);
    \draw (6) edge[out=190, in=135, loop, thick] node [below, yshift=-0.5cm] {$D/D$} (6);
    \draw (7) edge[out=190, in=135, loop, thick] node [above, xshift=1cm] {$C/D$} (7);
    \draw (2) edge[out=190, in=135, loop, thick] node [left] {$D/D$} (2);

    \draw (2) edge[out=0,in=180,->,thick] node [above] {$C/C$} (3);
    \draw (3) edge[out=-35,in=170,->,thick] node [above right] {$C/D$} (4);
    \draw (5) edge[out=-135,in=0,->,thick] node [below] {$C/C$} (6);

    \draw (0) edge[out=45,in=-135,->,thick] node [above, rotate=45, xshift=2cm, yshift=-0.5cm] {$D/C$} (3);

    \draw (1) edge[out=-45,in=180,->,thick] node [below right, xshift=2cm] {$C/D$} (5);
    \draw (1) edge[out=-135,in=135,->,thick] node [left] {$D/C$} (0);
    \draw (3) edge[out=-90,in=90,->,thick] node [above, rotate=90] {$D/D$} (6);
    \draw (4) edge[out=90,in=0,->,thick] node [above] {$C/C$} (3);
    \draw (4) edge[out=180,in=0,->,thick] node [above left, xshift=-2.5cm] {$D/D$} (1);
    \draw (5) edge[out=135,in=-45,->,thick] node [below, rotate=-45] {$D/D$} (3);
    \draw (7) edge[out=-90,in=-90,->,thick] node [below] {$D/C$} (5);
    \end{tikzpicture}
    \caption{TF3: an 8 state finite state machine.}
    \label{fig:tf3}
\end{figure}
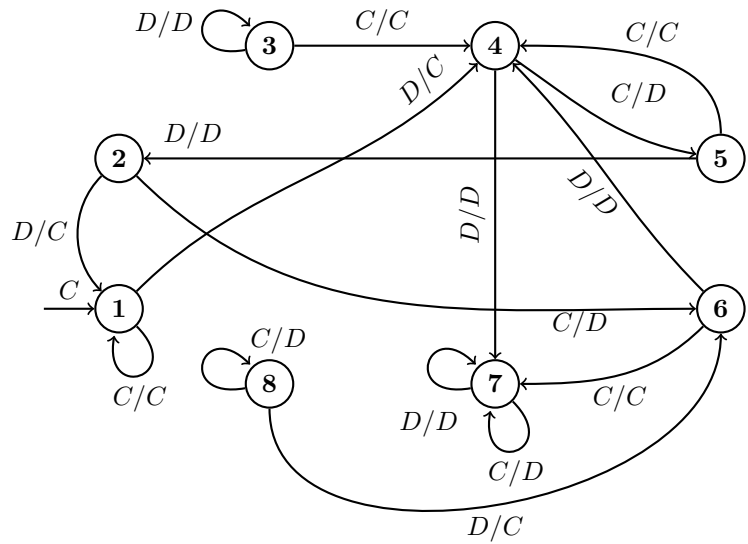

There are three further strategies trained specifically for this study; Trained
FSM 1, 2, and 3 (TF1 - TF3). These are based on finite state machines of 16, 16,
and 8 states respectively (see Figures~\ref{fig:tf1},~\ref{fig:tf2}
and~\ref{fig:tf3}).

\begin{figure}[!hbtp]
    \centering
        \begin{tikzpicture}

    \tikzstyle{state}=[minimum width=0.8cm, font=\boldmath];
    \node[circle, draw=black, thick] (5) at (0, 0) [state] {$6$};
	\node[circle, draw=black, thick] (2) at (4, 0) [state] {$3$};
	\node[circle, draw=black, thick] (15) at (-2, -2) [state] {$16$};
	\node[circle, draw=black, thick] (8) at ($(2)+(2,-2)$) [state] {$9$};
	\node[circle, draw=black, thick] (10) at ($(15)+(4, 0)$) [state] {$11$};
	
	\node[circle, draw=black, thick] (7) at ($(5)+(-1,-4)$) [state] {$8$};
	\node[circle, draw=black, thick] (11) at ($(7)+(2, 0)$) [state] {$12$};

	\node[circle, draw=black, thick] (9) at ($(10)+(0,-4)$) [state] {$10$};
	\node[circle, draw=black, thick] (0) at ($(9)+(-5,0)$) [state] {$1$};
	\node[circle, draw=black, thick] (14) at ($(9)+(5,0)$) [state] {$15$};

	\node[circle, draw=black, thick] (1) at ($(0)+(0,-2)$) [state] {$2$};
	\node[circle, draw=black, thick] (4) at ($(1)+(2,0)$) [state] {$5$};
	\node[circle, draw=black, thick] (6) at ($(4)+(4,0)$) [state] {$7$};
	\node[circle, draw=black, thick] (13) at ($(6)+(2,0)$) [state] {$14$};

	\node[circle, draw=black, thick] (3) at ($(4)+(0,-2)$) [state] {$4$};
	\node[circle, draw=black, thick] (12) at ($(6)+(0,-2)$) [state] {$13$};

    \coordinate[left of=0] (s);

    \draw (s) edge[out=0, in=180, ->, thick] node [above] {$C$} (0);
    \draw (0) edge[out=90, in=135, ->, thick] node [above left] {$C/C$} (7);
    \draw (0) edge[out=-90, in=90, ->, thick] node [left] {$D/C$} (1);

    \draw (1) edge[out=45, in=-135, ->, thick] node [left] {$C/D$} (11);
    \draw (1) edge[out=45, in=-135, ->, thick] node [right] {$D/D$} (11);
    
    \draw (2) edge[out=-45, in=135, ->, thick] node [above right] {$C/D$} (8);
    \draw (2) edge[out=-45, in=135, ->, thick] node [below left] {$D/C$} (8);

    \draw (3) edge[out=180, in=135, ->, thick, loop] node [left] {$C/C$} (3);
    \draw (3) edge[out=0, in=180, ->, thick] node [below] {$D/D$} (12);

    \draw (4) edge[out=0, in=180, ->, thick] node [above] {$C/C$} (6);
    \draw (4) edge[out=-90, in=90, ->, thick] node [left] {$D/C$} (3);

    \draw (5) edge[out=0, in=180, ->, thick] node [below, yshift=-1cm, xshift=2cm] {$D/D$} (8);
    \draw (5) edge[out=-45, in=90, ->, thick] node [left, yshift=-0.8cm, xshift=-0.3cm, rotate=90] {$C/C$} (11);

    \draw (6) edge[out=0, in=180, ->, thick] node [below] {$C/D$} (13);
    \draw (6) edge[out=45, in=180, ->, thick] node [above] {$D/C$} (14);

    \draw (7) edge[out=-135, in=135, ->, thick] node [yshift=1.8cm] {$C/D$} (4);
    \draw (7) edge[out=45, in=180, ->, thick] node [above, yshift=1.2cm, xshift=1.8cm] {$D/D$} (2);

    \draw (8) edge[out=0, in=45, ->, thick, loop] node [above] {$D/D$} (8);
    \draw (8) edge[out=-90, in=130, ->, thick] node [right] {$C/D$} (14);

    \draw (9) edge[out=-135, in=-45, ->, thick] node [above] {$C/C$} (0);
    \draw (9) edge[out=100, in=-100, ->, thick] node [above left, yshift=1.5cm, rotate=90] {$D/D$} (10);

    \draw (10) edge[out=0, in=180, ->, thick] node [below left, xshift=-0.5cm] {$C/C$} (8);
    \draw (10) edge[out=180, in=0, ->, thick] node [above left, xshift=-0.5cm] {$D/C$} (15);

    \draw (11) edge[out=0, in=70, ->, thick] node [above right] {$C/D$} (6);
    \draw (11) edge[out=155, in=-90, ->, thick] node [right, yshift=1cm] {$D/D$} (5);

    \draw (12) edge[out=90, in=-90, ->, thick] node [above right] {$C/D$} (6);
    \draw (12) edge[out=155, in=-90, ->, thick] node [left, yshift=-1cm] {$D/D$} (9);

    \draw (13) edge[out=135, in=0, ->, thick] node [below left, xshift=-0.4cm, yshift=0.4cm] {$C/D$} (9);
    \draw (13) edge[out=70, in=-135, ->, thick] node [right] {$D/D$} (8);

    \draw (14) edge[out=70, in=-45, ->, thick] node [right] {$C/D$} (8);
    \draw (14) edge[out=-90, in=0, ->, thick] node [right] {$D/D$} (13);

    \draw (15) edge[out=-45, in=30, ->, thick] node [left, xshift=-1.8cm, yshift=2.5cm] {$C/C$} (4);
    \draw (15) edge[out=90, in=180, ->, thick] node [above left] {$D/C$} (5);

    \end{tikzpicture}
    \caption{TF1: a 16 state finite state machine with a handshake leading to
    mutual cooperation at state 4.}
    \label{fig:tf1}
\end{figure}
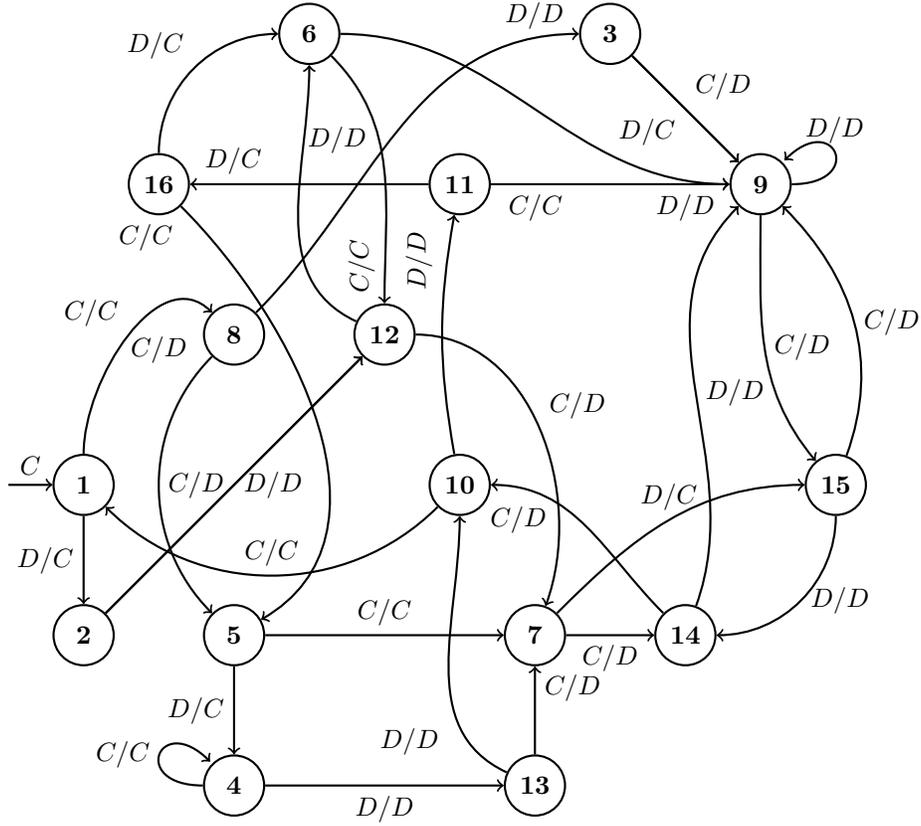

\begin{figure}[!hbtp]
    \centering
        \begin{tikzpicture}

    \tikzstyle{state}=[minimum width=0.8cm, font=\boldmath];
    \node[circle, draw=black, thick] (1) at (0, 0) [state] {$2$};
	\node[circle, draw=black, thick] (4) at (4, 0) [state] {$5$};

	\node[circle, draw=black, thick] (5) at (-2, -1) [state] {$6$};
    \node[circle, draw=black, thick] (6) at (2, -1) [state] {$7$};
    \node[circle, draw=black, thick] (3) at ($(5)+(-1,-2)$) [state] {$4$};
    \node[circle, draw=black, thick] (12) at ($(5)+(0,-4)$) [state] {$13$};
    \node[circle, draw=black, thick] (8) at ($(12)+(0,-2)$) [state] {$9$};
    \node[circle, draw=black, thick] (9) at ($(8)+(-2,0)$) [state] {$10$};
    \node[circle, draw=black, thick] (0) at ($(8)+(0,-2)$) [state] {$1$};

    \node[circle, draw=black, thick] (2) at ($(6)+(0,-2)$) [state] {$3$};
    \node[circle, draw=black, thick] (14) at ($(2)+(2,0)$) [state] {$15$};
    \node[circle, draw=black, thick] (7) at ($(2)+(0,-2)$) [state] {$8$};
    \node[circle, draw=black, thick] (10) at ($(14)+(0,-2)$) [state] {$11$};
    \node[circle, draw=black, thick] (13) at ($(7)+(0,-2)$) [state] {$14$};
    \node[circle, draw=black, thick] (11) at ($(10)+(0,-2)$) [state] {$12$};
    \node[circle, draw=black, thick] (15) at ($(11)+(0,-2)$) [state] {$16$};

    \coordinate[below of=0] (s);

    \draw (s) edge[out=90, in=-90, ->, thick] node [left] {$C$} (0);

    \draw (0) edge[out=0, in=-90, ->, thick] node [below right] {$C/D$} (13);
    \draw (0) edge[out=45, in=0, ->, thick] node [above, rotate=90] {$D/D$} (12);

    \draw (1) edge[out=-90, in=0, ->, thick] node [above, yshift=1.3cm, xshift=0.3cm] {$C/D$} (3);
    \draw (1) edge[out=0, in=180, ->, thick] node [above] {$D/D$} (4);

    \draw (2) edge[out=-45, in=-135, ->, thick] node [above] {$C/D$} (14);
    \draw (2) edge[out=180, in=90, ->, thick] node [above, yshift=0.8cm, xshift=3cm] {$D/D$} (9);

    \draw (3) edge[out=180, in=-135, ->, thick] node [left, yshift=2cm, xshift=-0.1cm] {$C/C$} (0);
    \draw (3) edge[out=135, in=180, ->, thick] node [above, yshift=0.2cm] {$D/D$} (1);

    \draw (4) edge[out=90, in=90, ->, thick] node [above] {$C/D$} (1);
    \draw (4) edge[out=-90, in=0, ->, thick] node [right] {$D/D$} (2);

    \draw (5) edge[out=-90, in=90, ->, thick] node [right, yshift=1cm] {$C/C$} (12);
    \draw (5) edge[out=0, in=180, ->, thick] node [below right] {$D/C$} (6);

    \draw (6) edge[out=135, in=-45, ->, thick] node [right, xshift=1cm] {$C/C$} (1);
    \draw (6) edge[out=-45, in=135, ->, thick] node [above left, yshift=-0.5cm] {$D/D$} (14);

    \draw (7) edge[out=135, in=45, ->, thick] node [above] {$C/D$} (12);
    \draw (7) edge[out=90, in=-90, ->, thick] node [left] {$D/D$} (2);

    \draw (8) edge[out=90, in=-135, ->, thick] node [above right] {$C/D$} (7);
    \draw (8) edge[out=180, in=0, ->, thick] node [above] {$D/D$} (9);

    \draw (9) edge[out=-45, in=-135, ->, thick] node [below] {$C/D$} (8);
    \draw (9) edge[out=-90, in=180, ->, thick] node [right] {$D/D$} (0);

    \draw (10) edge[out=180, in=-60, ->, thick] node [below, yshift=-0.5cm, xshift=0.4cm] {$C/C$} (2);
    \draw (10) edge[out=0, in=0, ->, thick] node [right] {$D/C$} (15);

    \draw (11) edge[out=90, in=-45, ->, thick] node [below right, xshift=0.7cm] {$C/D$} (7);
    \draw (11) edge[out=180, in=0, ->, thick] node [below] {$D/D$} (13);

    \draw (12) edge[out=135, in=-45, ->, thick] node [above left] {$C/C$} (3);
    \draw (12) edge[out=-90, in=135, ->, thick] node [left] {$D/D$} (8);

    \draw (13) edge[out=90, in=-90, ->, thick] node [left] {$C/C$} (7);
    \draw (13) edge[out=45, in=-135, ->, thick] node [below, yshift=-0.2cm, rotate=45] {$D/D$} (10);

    \draw (14) edge[out=-90, in=90, ->, thick] node [right] {$C/D$} (10);
    \draw (14) edge[out=-100, in=45, ->, thick] node [above] {$D/D$} (7);

    \draw (15) edge[out=-135, in=180, ->, thick, loop] node [left] {$C/C$} (15);
    \draw (15) edge[out=90, in=-90, ->, thick] node [left] {$D/D$} (11);

    \end{tikzpicture}
    \caption{TF2: a 16 state finite state machine with a handshake leading to
    mutual cooperation at state 16.}
    \label{fig:tf2}
\end{figure}

As opposed to the previously described strategies, these strategies were trained
with the objective function of \textbf{mean fixation probabilities for Moran
processes} starting at initial population states consisting of \(N/2\)
individuals of the training candidates and \(N/2\) individuals of an opponent
strategy, taken from a selection of 150 opponents from the Axelrod library:

\begin{itemize}
	\item TF1 \(N=12\), 0\% noise.
	\item TF2 \(N=10\), 0\% noise.
	\item TF3 \(N=8\), 1\% noise.
\end{itemize}

Each matchup of players was run to fixation to estimate the absorption
probabilities. The trained algorithms
were run for fewer than 50 generations. Training data for this is available at
\cite{data}.

TF3 cooperates and defects with
various cycles depending on the opponent's actions. TF3 will mutually
cooperate with any strategy and only tolerates a few defections before
defecting for the rest of match. It is similar to but not exactly the same as
Fool Me Once, a strategy that cooperates until the opponent has defected twice
(not necessarily consecutively), and defects indefinitely thereafter. Though a
product of training with a Moran objective, it differs from TF1 and TF2
in that it lacks a handshake mechanism. Figure~\ref{fig:tf3} shows
all 8 states of the strategy produced by the training process (states 3 and 8
are not reachable).

TF2 always starts with CD and will defect against opponents that start with
DD\@. It plays CDD against itself and then cooperates thereafter; Fortress3 and
Fortress4 also use a similar handshake and cooperate with TF2. Cooperation
can be rescued after a failed handshake by a complex sequence of plays
which sometimes results in mutual cooperation with Firm but Fair, Grofman, and
GTFT, and a few others with low probability. TF2 defects against all other
players in the study, barring
unusual cases arising from particular randomizations. Figure~\ref{fig:tf2} shows
all 16 states of the strategy (states 6 and 7 are not reachable).

TF1 has an initial handshake of CCD and cooperates if the opponent matches.
However if the opponent later defects, TF1 will respond in kind, so the
handshake is not permanent. Only one player (Prober 4 \cite{Prison1998}) manages to
achieve cooperation with TF1 after about 20 rounds of play. TF1 is functionally
very similar to a strategy known as ``Collective Strategy'', which has a
handshake of CD and cooperates with opponents that matched the handshake
until they defect, defecting thereafter if the opponent ever defects \cite{Li2009}.
This strategy was specifically designed for evolutionary processes.

For both TF1 and TF2 a handshake
mechanism naturally emerges from the structure of the underlying finite state
machine. This behavior is an outcome of the evolutionary process and is in no
way hard-coded or included via an additional mechanism.

\begin{table}[!hbtp]
    \centering
        \begin{tabular}{lrrrrrrrrrrrrrrrr}
\toprule
Memory Depth &   0   &   1   &   2   &   3   &   4   &   5   &   6   &   9   &   10  &   11  &   12  &   16  &   20  &   40  &   200 &  \(\infty\)   \\
\midrule
Count &     3 &    29 &    12 &     8 &     2 &     6 &     1 &     1 &     5 &     1 &     1 &     2 &     2 &     2 &     1 &    88 \\
\bottomrule
\end{tabular}

        \caption{Memory depth}
        \label{tbl:memory_depth_count}
\end{table}

\subsection{Data collection}\label{sec:data_collection}

Each strategy pair is run
with starting population distributions of $(1, N-1)$, $(N/2, N/2)$ and $(N-1 ,
1)$, for \(N\) from 2 through 14. The fixation probability is then empirically
computed for each combination of starting distribution and value of \(N\).  The
Axelrod library can carry out exact simulations of the Moran process. Since some
of the strategies have a high computational cost or are stochastic, samples are
taken from a large number of match outcomes for the pairs of players for use in
computing fitnesses in the Moran process. This approach was verified to agree
with unsampled calculations to a high degree of accuracy in specific cases.
This is described in Algorithms~\ref{alg:data_collection}
and~\ref{alg:moran_process}.

\begin{algorithm}[!hbtp]
        \caption{Data Collection}
        \label{alg:data_collection}
        
  \begin{algorithmic}[1]
    \FOR{player one \textbf{in} players list}
      \FOR{player two \textbf{in} (players list - player one)}
        \STATE pair $\gets$ (player one, player two)
        \FOR{starting population distributions \textbf{in} [$(1, N-1), (\frac{N}{2}, \frac{N}{2}), (N-1, 1)]$}
        \STATE \textbf{simulate} moran process*(pair, starting distribution)
        \RETURN fixation probabilities
        \ENDFOR
      \ENDFOR
    \ENDFOR
  \end{algorithmic}

\end{algorithm}

\begin{algorithm}[!hbtp]
        \caption{Moran process}
        \label{alg:moran_process}

  \begin{algorithmic}[1]
  \STATE initial population $\gets$ (pair, starting distribution) \
  \STATE population $\gets$ initial population

    \WHILE {population not uniform}
      \FOR{player in population}
      \FOR{opponent in (population - player)}
        \STATE match $\gets$ (player, opponent)
        \STATE results $\gets$ cache (match)
        \ENDFOR
      \ENDFOR
      \STATE population $\gets$ sorted(results)
      \STATE parent $\gets$ selected randomly in proportion to total match payoffs
      \STATE child $\gets$ parent
      \STATE kill off $\gets$ random player from population
      \STATE population $\gets$ child replaces kill off
    \ENDWHILE
  \end{algorithmic}

\end{algorithm}

Section~\ref{sec:validation} will further validate the methodology by comparing
simulated results to analytical results in some cases. The main results of this
manuscript are presented in Section~\ref{sec:empirical_results} which will
present a detailed analysis of all the data generated. Finally,
Section~\ref{sec:conclusion} will conclude and offer future avenues for the work
presented here.

\section{Validation}\label{sec:validation}

As described in \cite{Nowak} consider the payoff matrix:

\begin{equation}\label{equ:payoff_matrix}
    M = \begin{pmatrix}
        a, b\\
        c, d
        \end{pmatrix}
\end{equation}

The expected payoffs of \(i\) players of the first type in a population with \(N
- i\) players of the second type are given by:

\begin{equation}\label{equ:expected_payoff_one}
    f_i = \frac{a(i - 1) + b(N - i)}{N - 1}
\end{equation}

\begin{equation}\label{equ:expected_payoff_two}
    g_i = \frac{ci + d(N - i - 1)}{N - 1}
\end{equation}

The transitions within the birth death process that underpins the Moran process
are then given by:

\begin{align}
    p_{i, i+1}&= \frac{if_i}{if_i+(N-i)g_i}\frac{N-i}{N}\label{equ:p_up}\\
    p_{i, i-1}&= \frac{(N-i)g_i}{if_i+(N-i)g_i}\frac{i}{N}\label{equ:p_down}\\
    p_{ii} &= 1 - p_{i, i+1} - p_{i, i-1}\label{equ:p_stay}
\end{align}

Using this it is a known result \cite{Nowak2017} that the fixation probability
of the first strategy in a population of \(i\) individuals of the first type
and \(N-i\) individuals of the second:

\begin{equation}\label{equ:fixation_probability}
x_i = \frac{1 + \sum_{j=1}^{i-1}\prod_{k=1}^{j}\gamma_j}{1 + \sum_{j=1}^{N-1}
      \prod_{k=1}^{j}\gamma_j}
\end{equation}

where:

\[
\gamma_j = \frac{p_{j, j-1}}{p_{j, j+1}}
\]

A neutral strategy will have fixation probability $x_i = i/N$.

\begin{figure}[!hbtp]
    \centering
    \begin{subfigure}[t]{.3\textwidth}
        \centering
        \includegraphics[width=.8\textwidth]{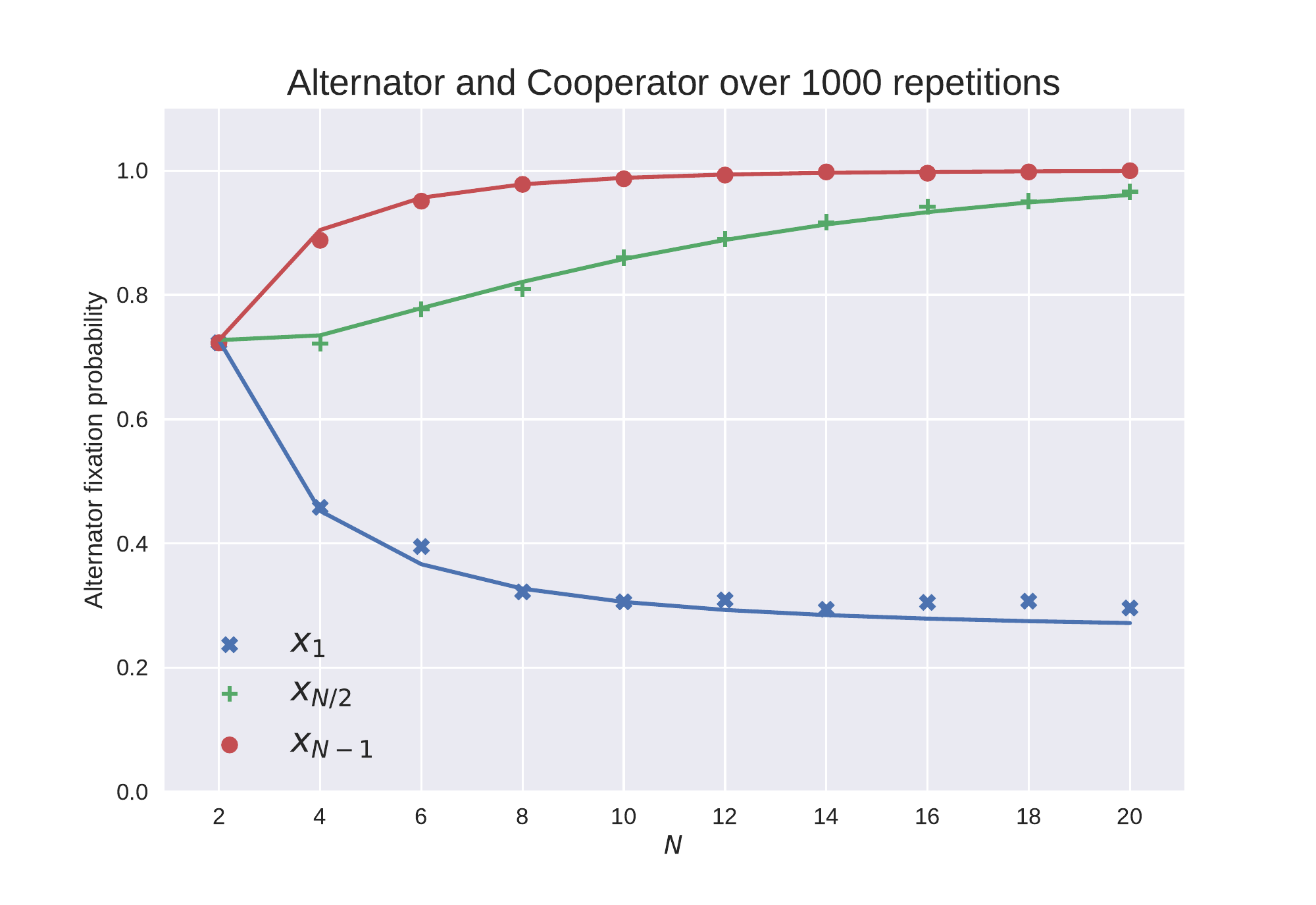}
        \caption{Alternator and Cooperator}
    \end{subfigure}%
    ~
    \begin{subfigure}[t]{.3\textwidth}
        \centering
        \includegraphics[width=.8\textwidth]{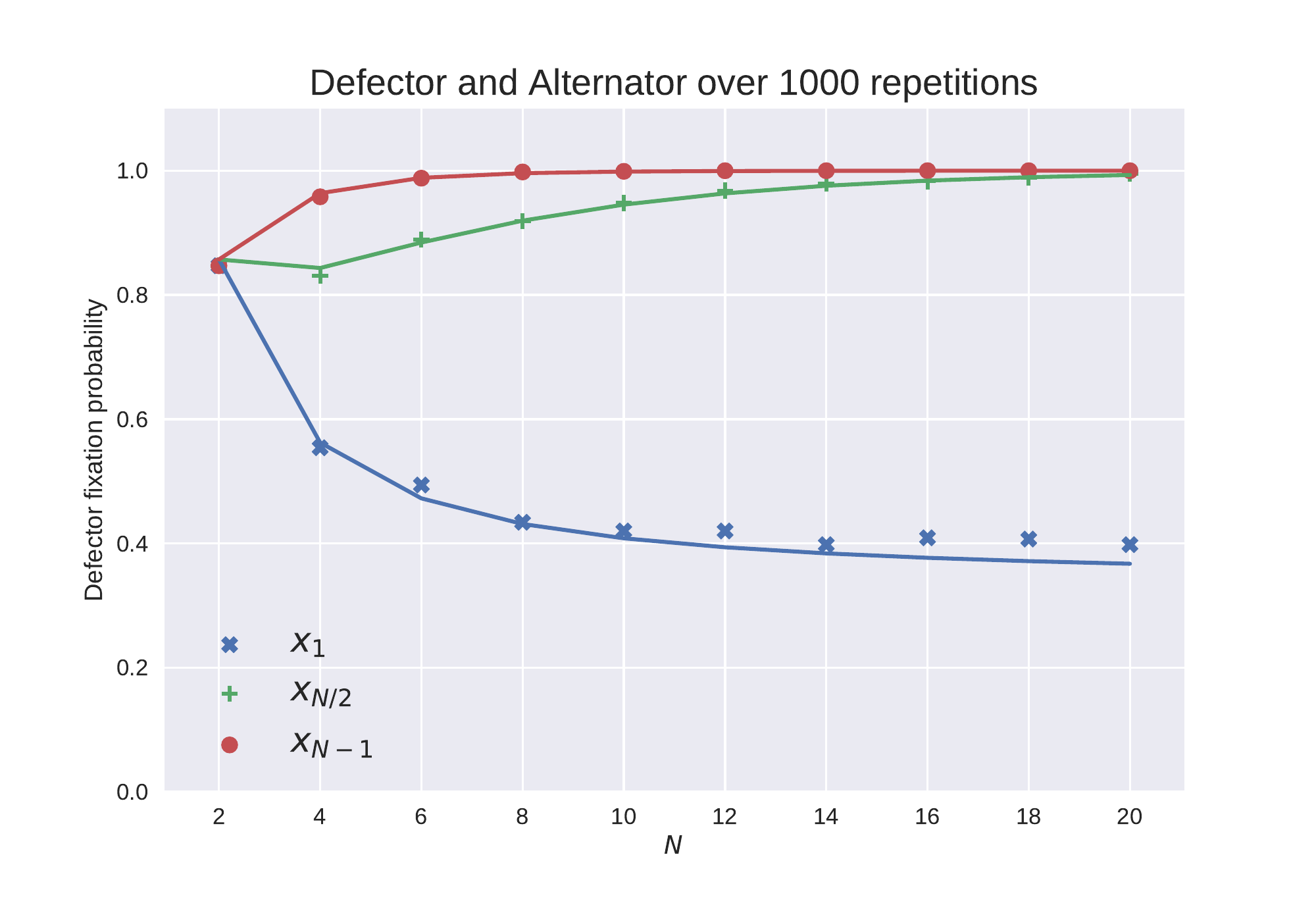}
        \caption{Defector and Alternator}
    \end{subfigure}%
    ~
    \begin{subfigure}[t]{.3\textwidth}
        \centering
        \includegraphics[width=.8\textwidth]{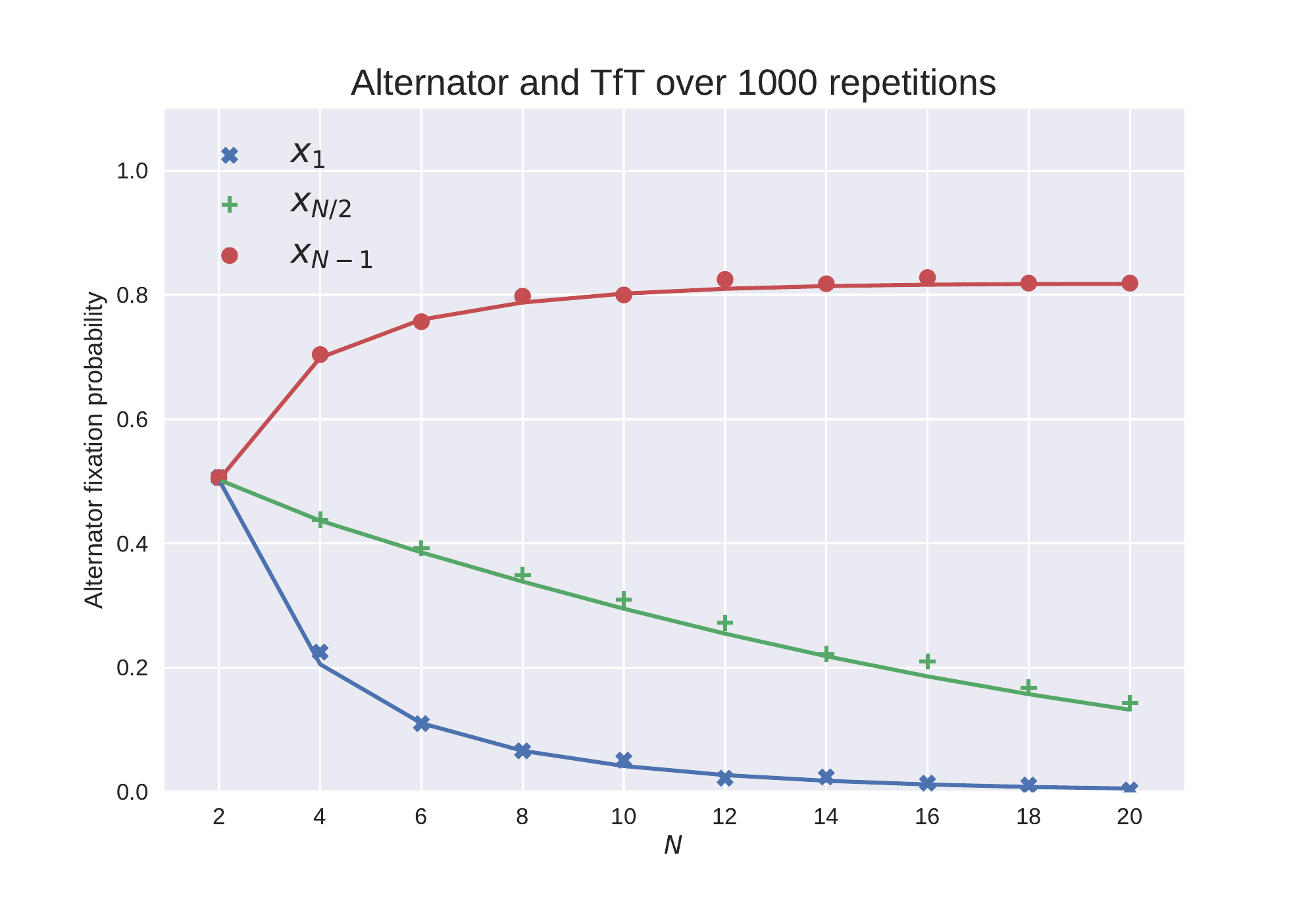}
        \caption{Alternator and Tit For Tat}
    \end{subfigure}%

    \begin{subfigure}[t]{.3\textwidth}
        \centering
        \includegraphics[width=.8\textwidth]{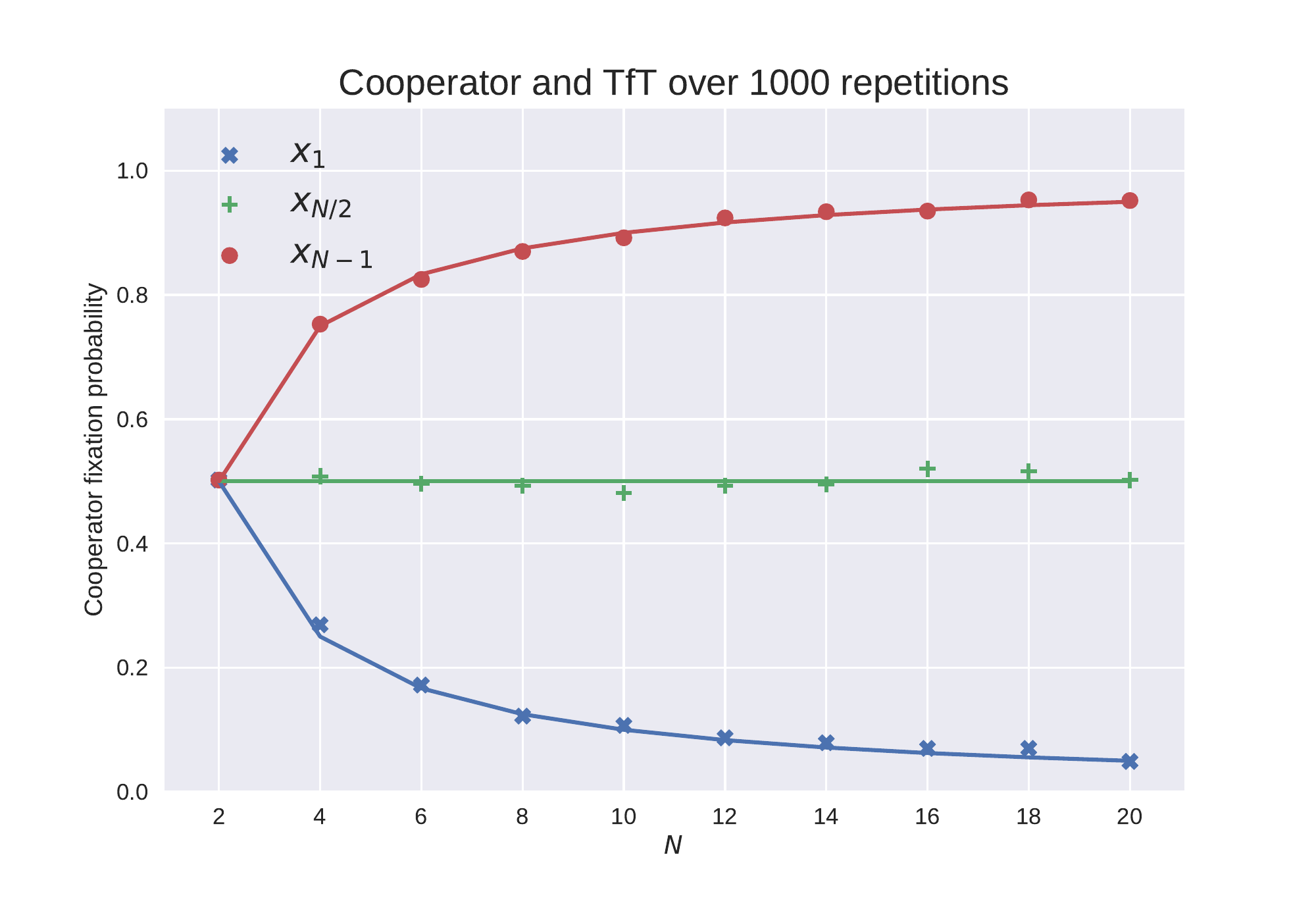}
        \caption{Cooperator and Tit For Tat}
    \end{subfigure}%
    ~
    \begin{subfigure}[t]{.3\textwidth}
        \centering
        \includegraphics[width=.8\textwidth]{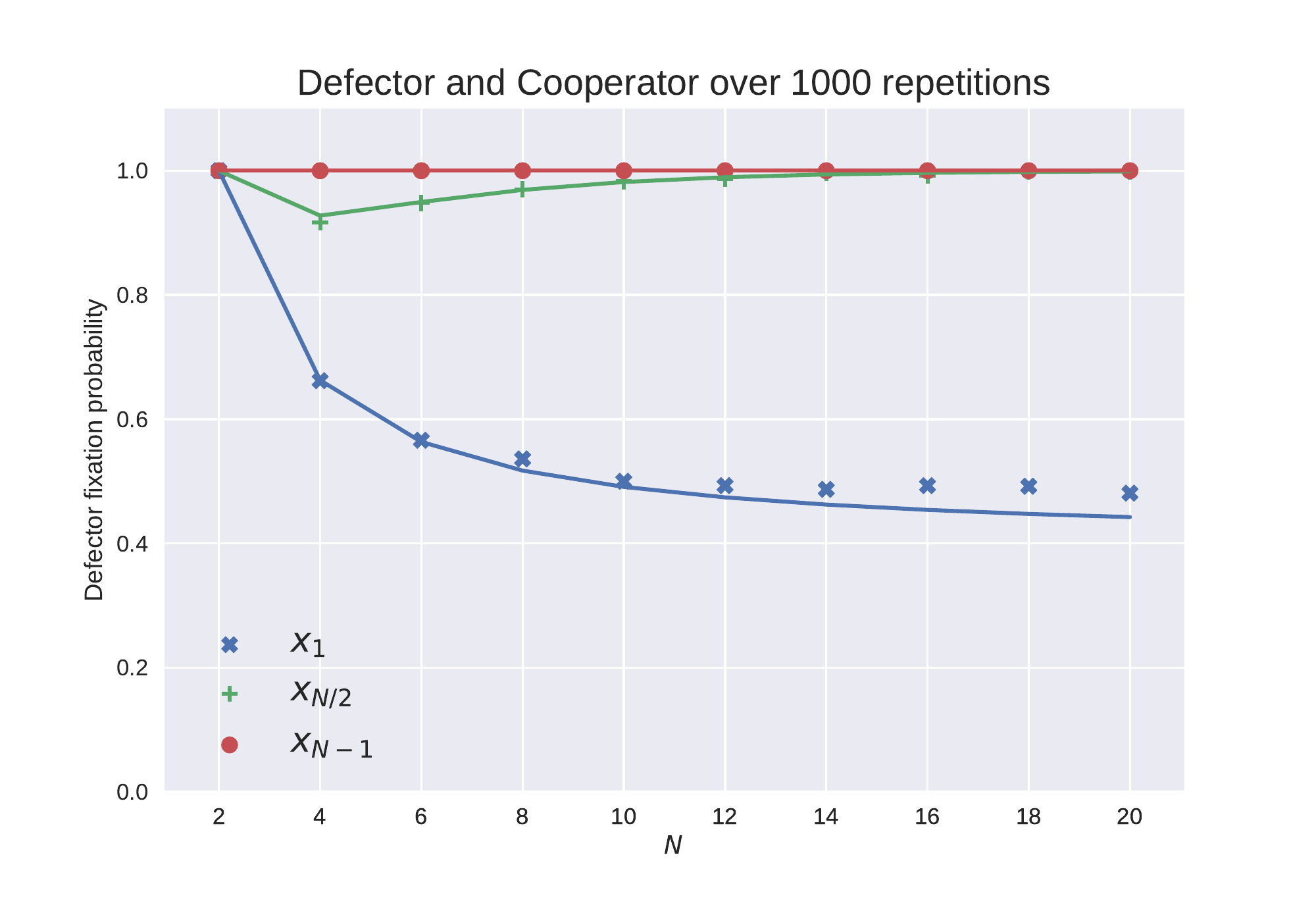}
        \caption{Defector and Cooperator}
    \end{subfigure}%
    ~
    \begin{subfigure}[t]{.3\textwidth}
        \centering
        \includegraphics[width=.8\textwidth]{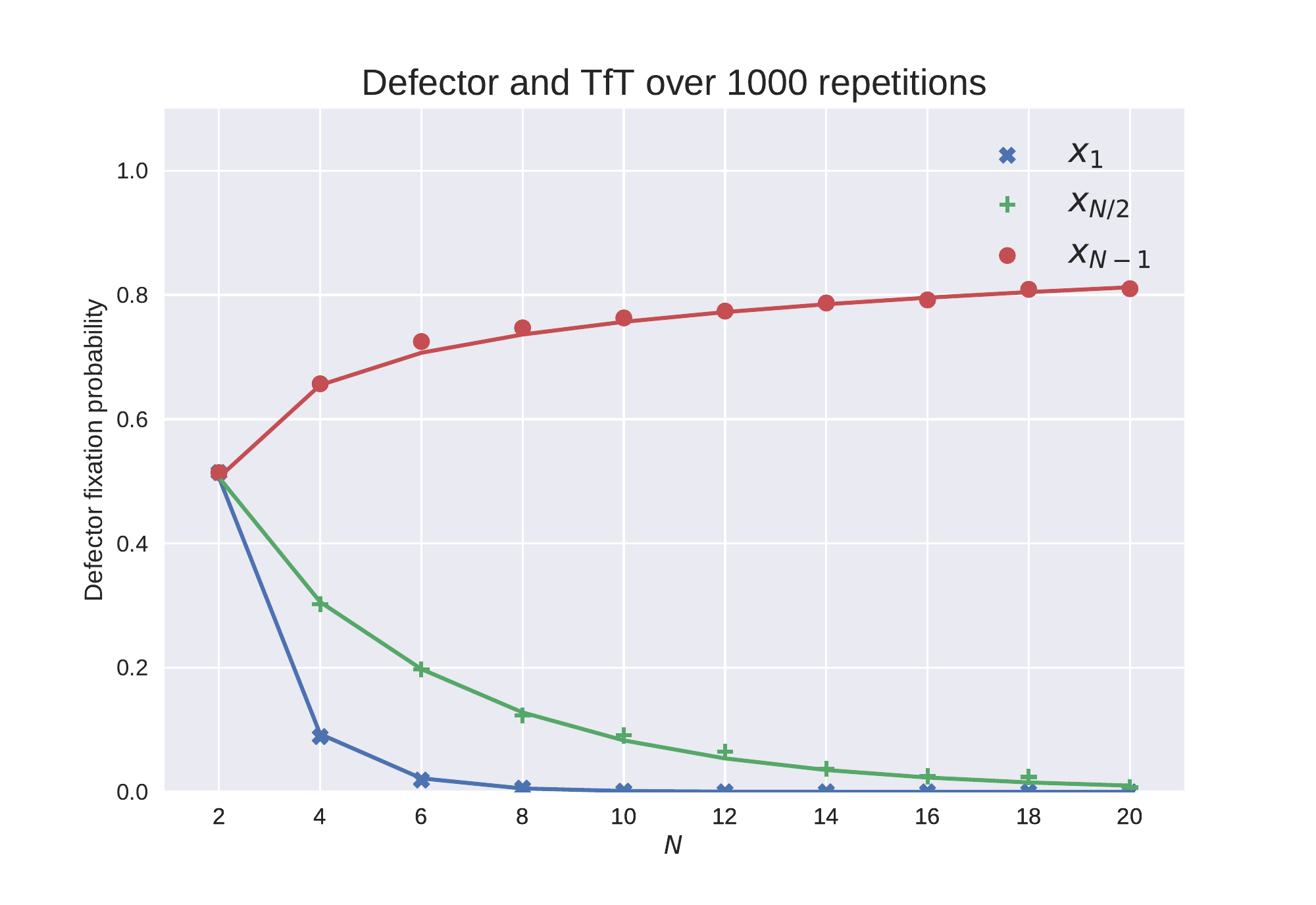}
        \caption{Defector and Tit For Tat}
    \end{subfigure}%

    \begin{subfigure}[t]{.3\textwidth}
        \centering
        \includegraphics[width=.8\textwidth]{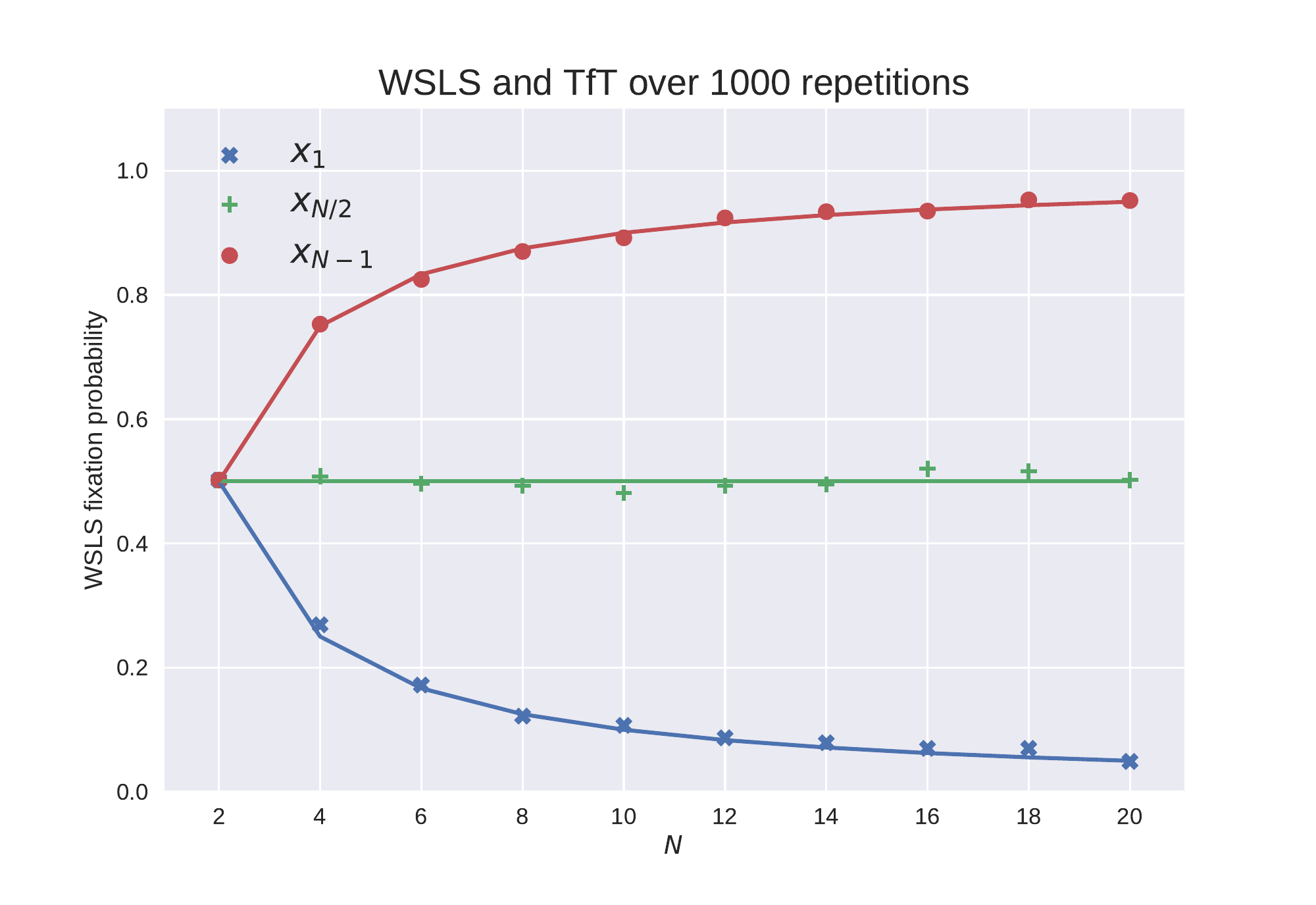}
        \caption{Win Stay Lose Shift and Tit For Tat}
    \end{subfigure}%
    ~
    \begin{subfigure}[t]{.3\textwidth}
        \centering
        \includegraphics[width=.8\textwidth]{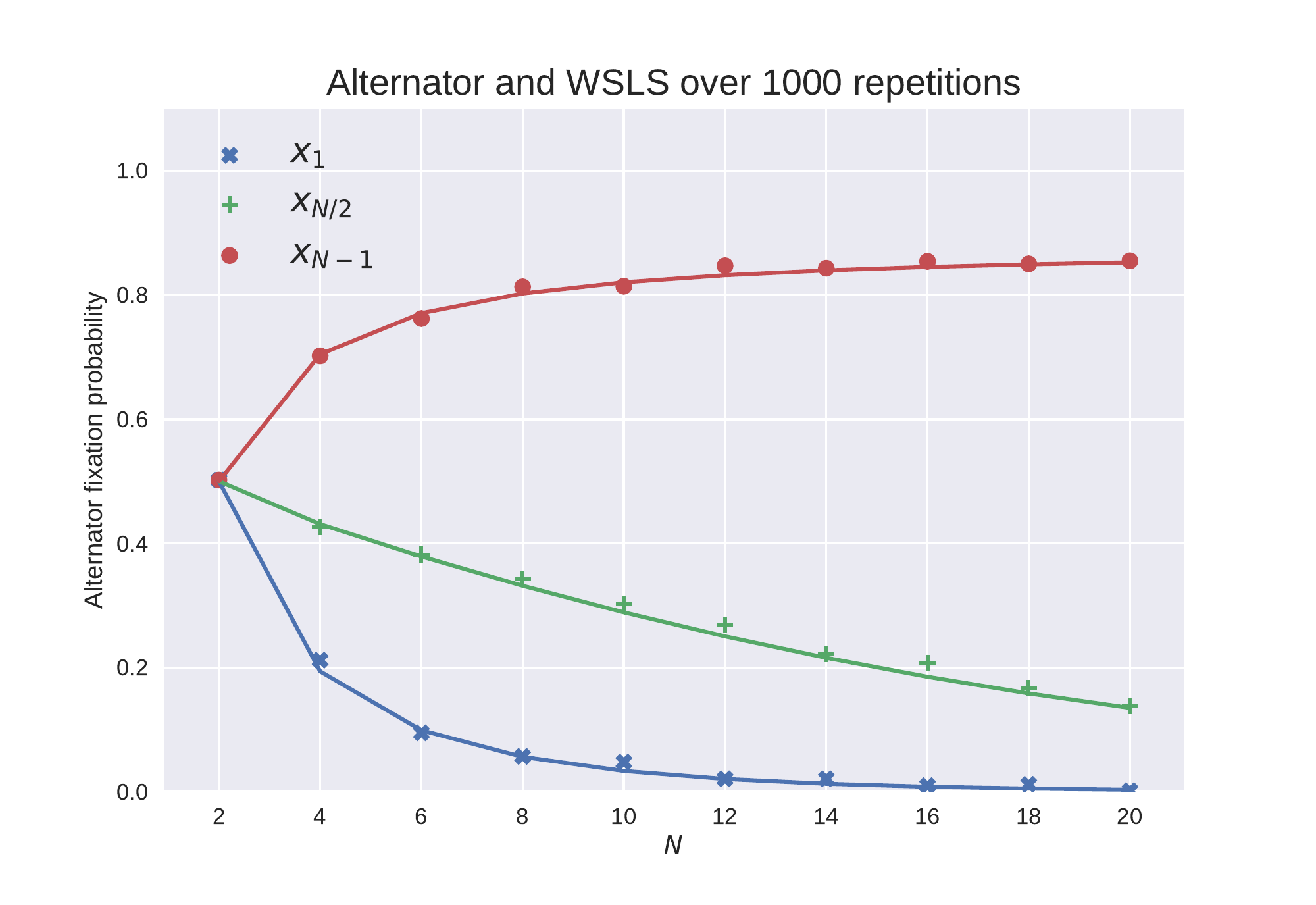}
        \caption{Alternator and Win Stay Lose Shift}
    \end{subfigure}%
    ~
    \begin{subfigure}[t]{.3\textwidth}
        \centering
        \includegraphics[width=.8\textwidth]{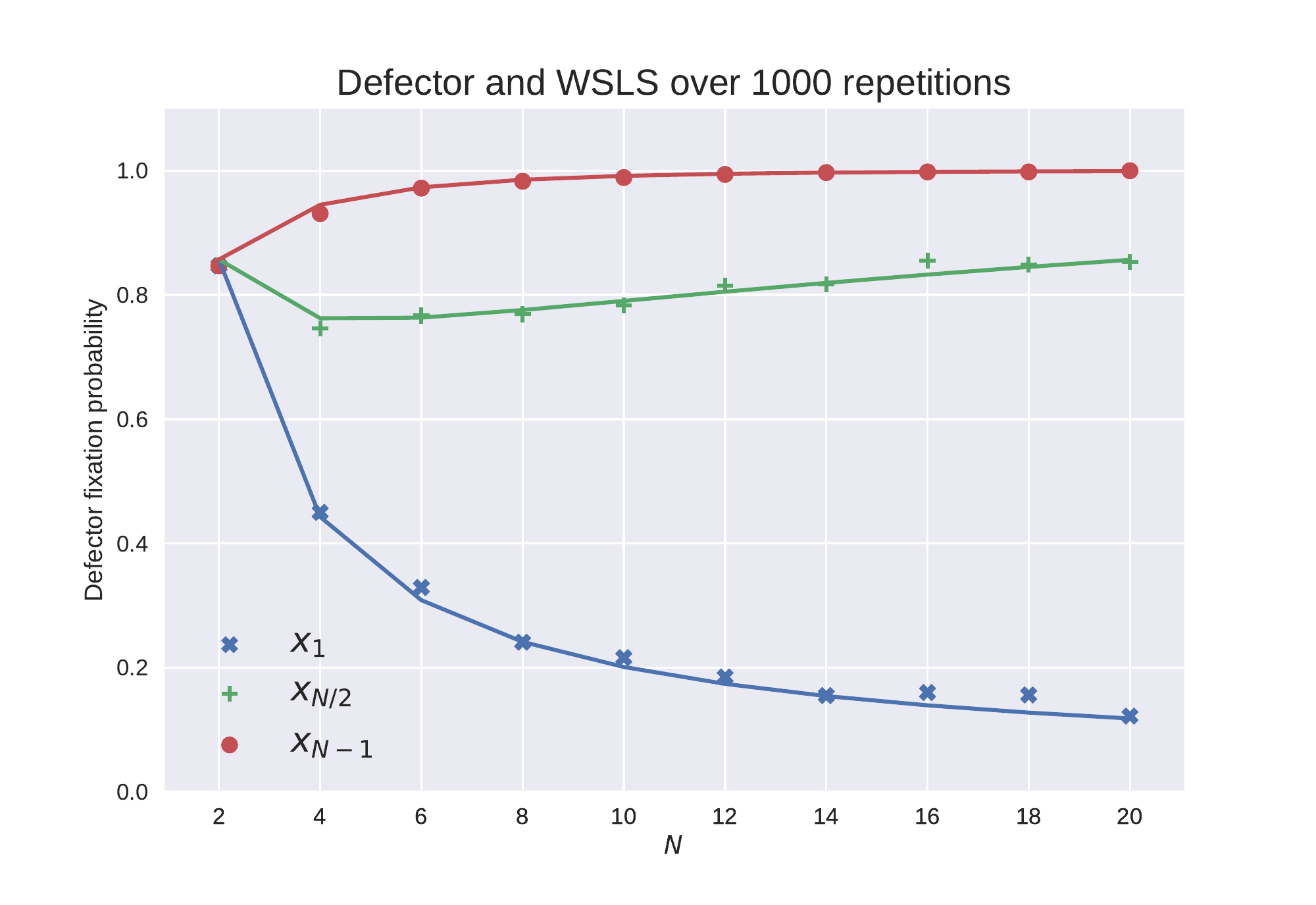}
        \caption{Defector and Win Stay Lose Shift}
    \end{subfigure}%
    \caption{Comparison of theoretic and actual Moran Process fixation
             probabilities for \textbf{deterministic} strategies.}
    \label{fig:comparison_deterministic}
\end{figure}

\begin{figure}[!hbtp]
    \centering
    \begin{subfigure}[t]{.3\textwidth}
        \centering
        \includegraphics[width=.8\textwidth]{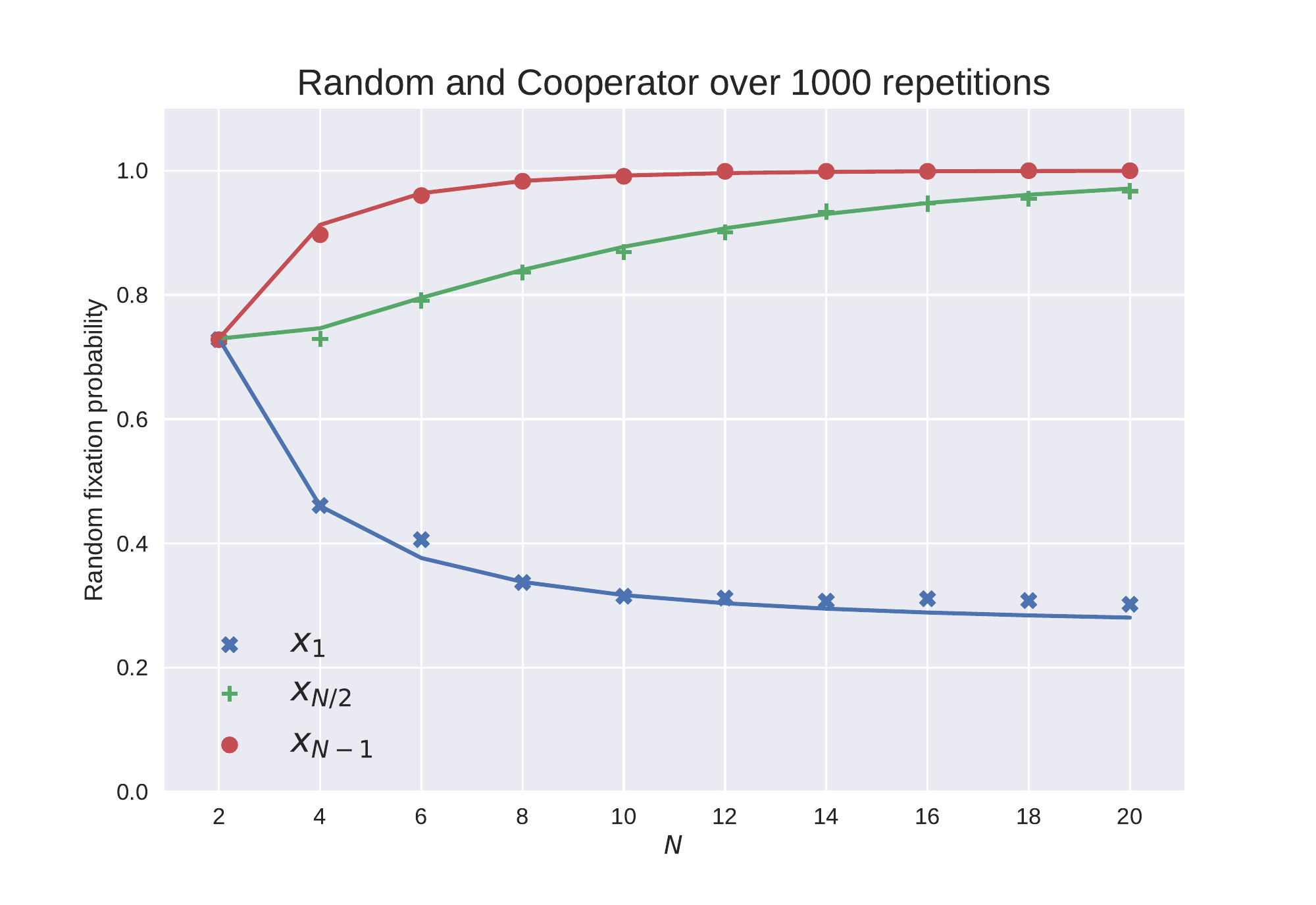}
        \caption{Random and Cooperator}
    \end{subfigure}%
    ~
    \begin{subfigure}[t]{.3\textwidth}
        \centering
        \includegraphics[width=.8\textwidth]{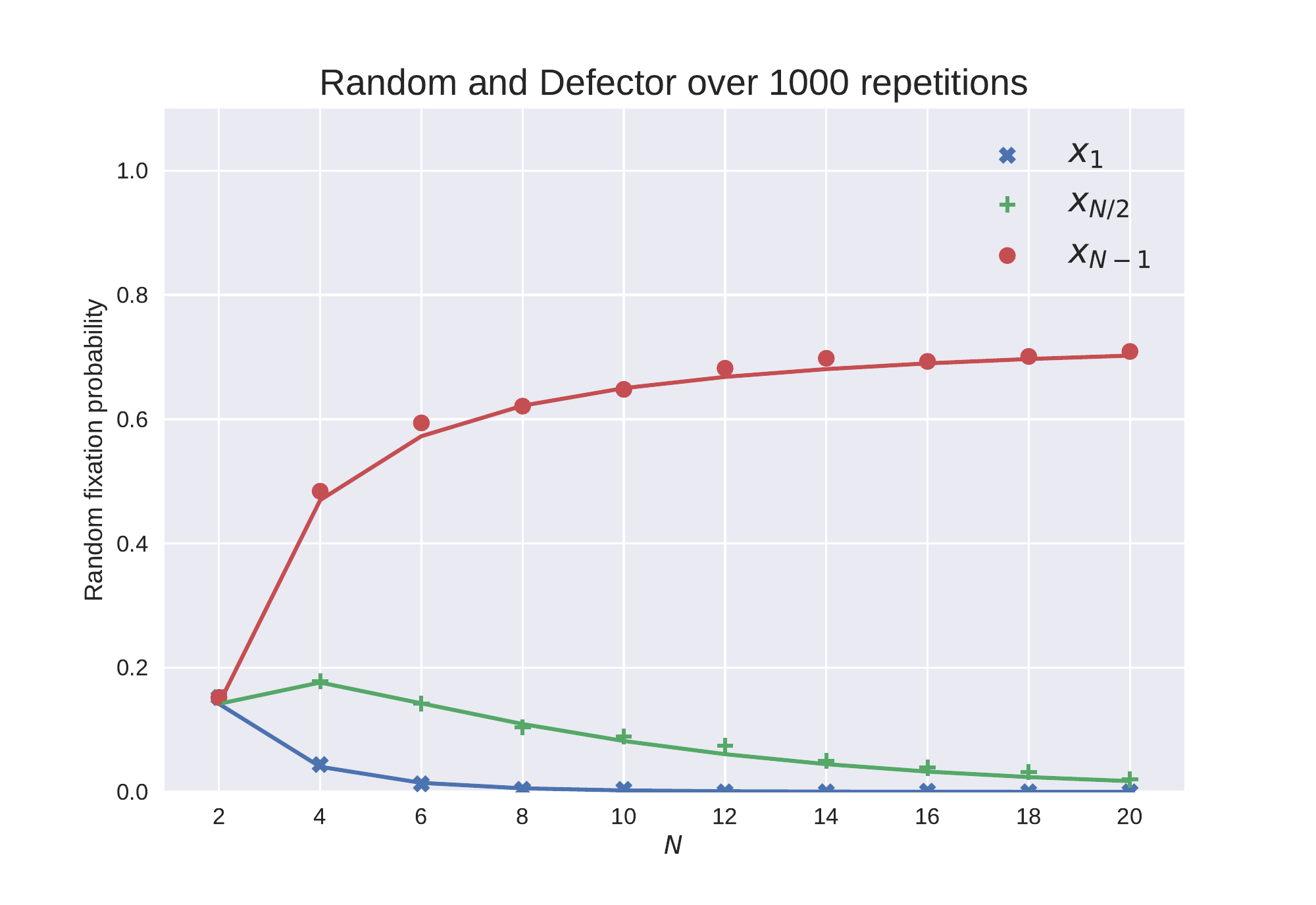}
        \caption{Random and Defector}
    \end{subfigure}%
    ~
    \begin{subfigure}[t]{.3\textwidth}
        \centering
        \includegraphics[width=.8\textwidth]{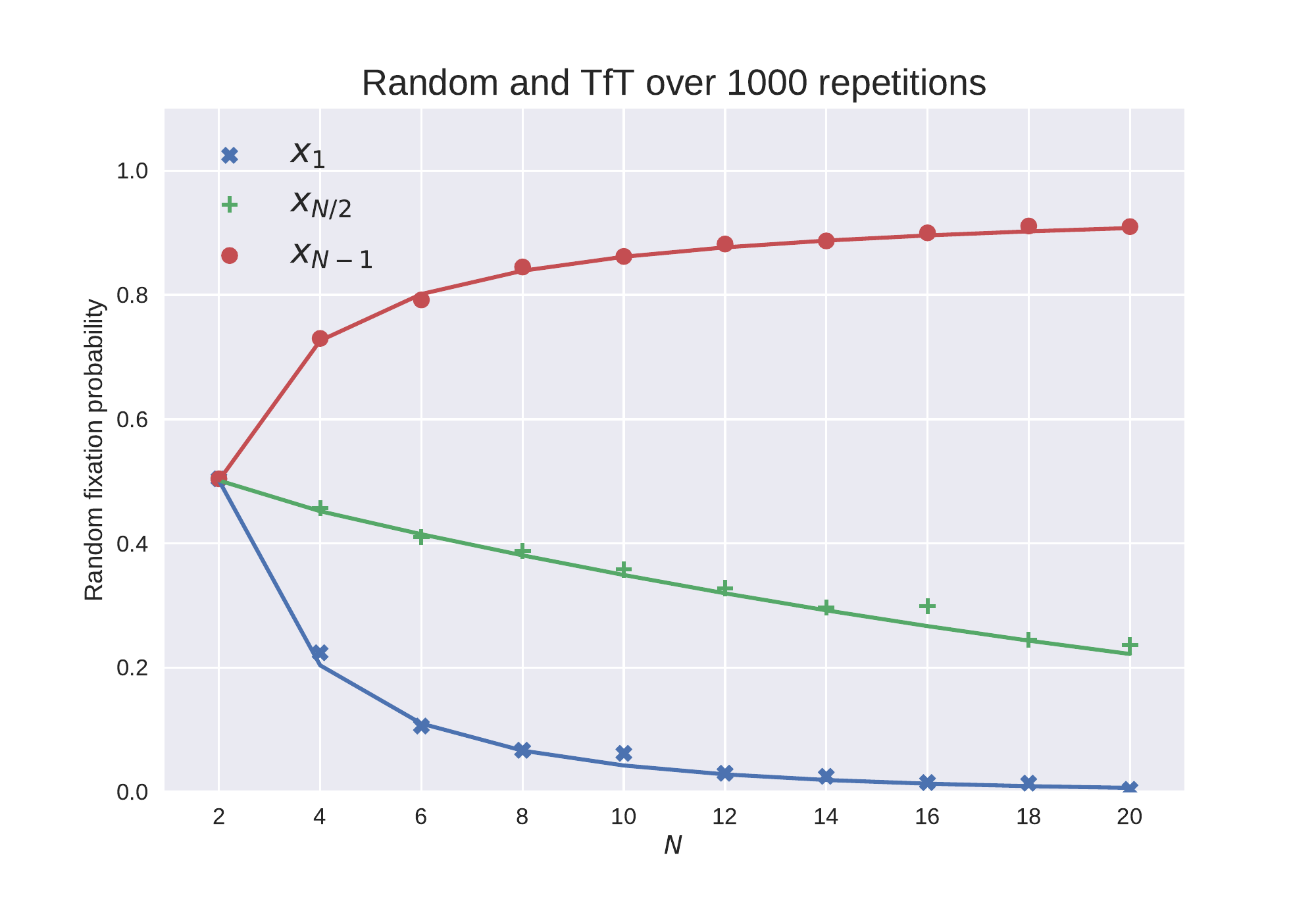}
        \caption{Random and Tit For Tat}
    \end{subfigure}%

    \begin{subfigure}[t]{.3\textwidth}
        \centering
        \includegraphics[width=.8\textwidth]{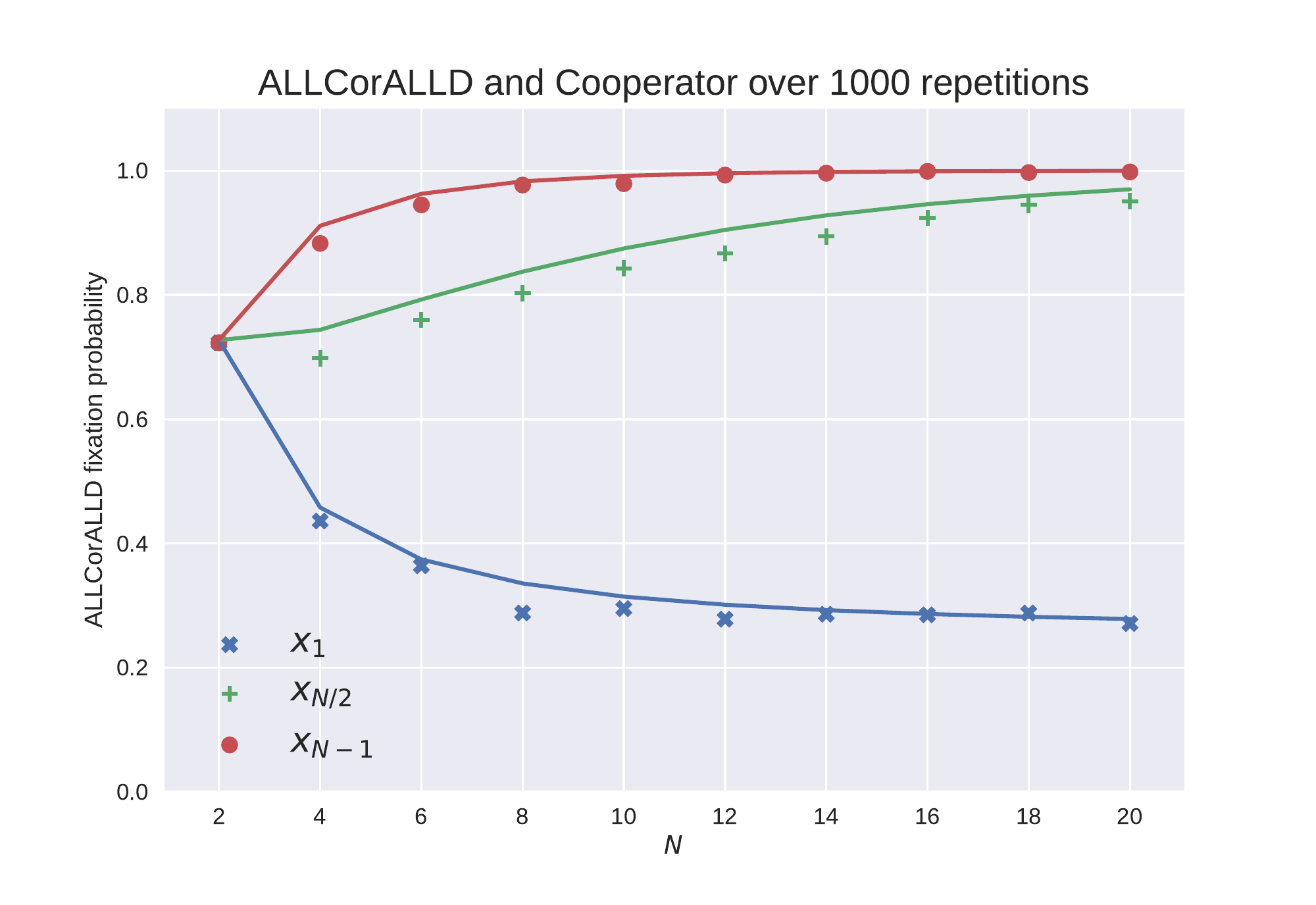}
        \caption{All C or all D and Cooperator}
    \end{subfigure}%
    ~
    \begin{subfigure}[t]{.3\textwidth}
        \centering
        \includegraphics[width=.8\textwidth]{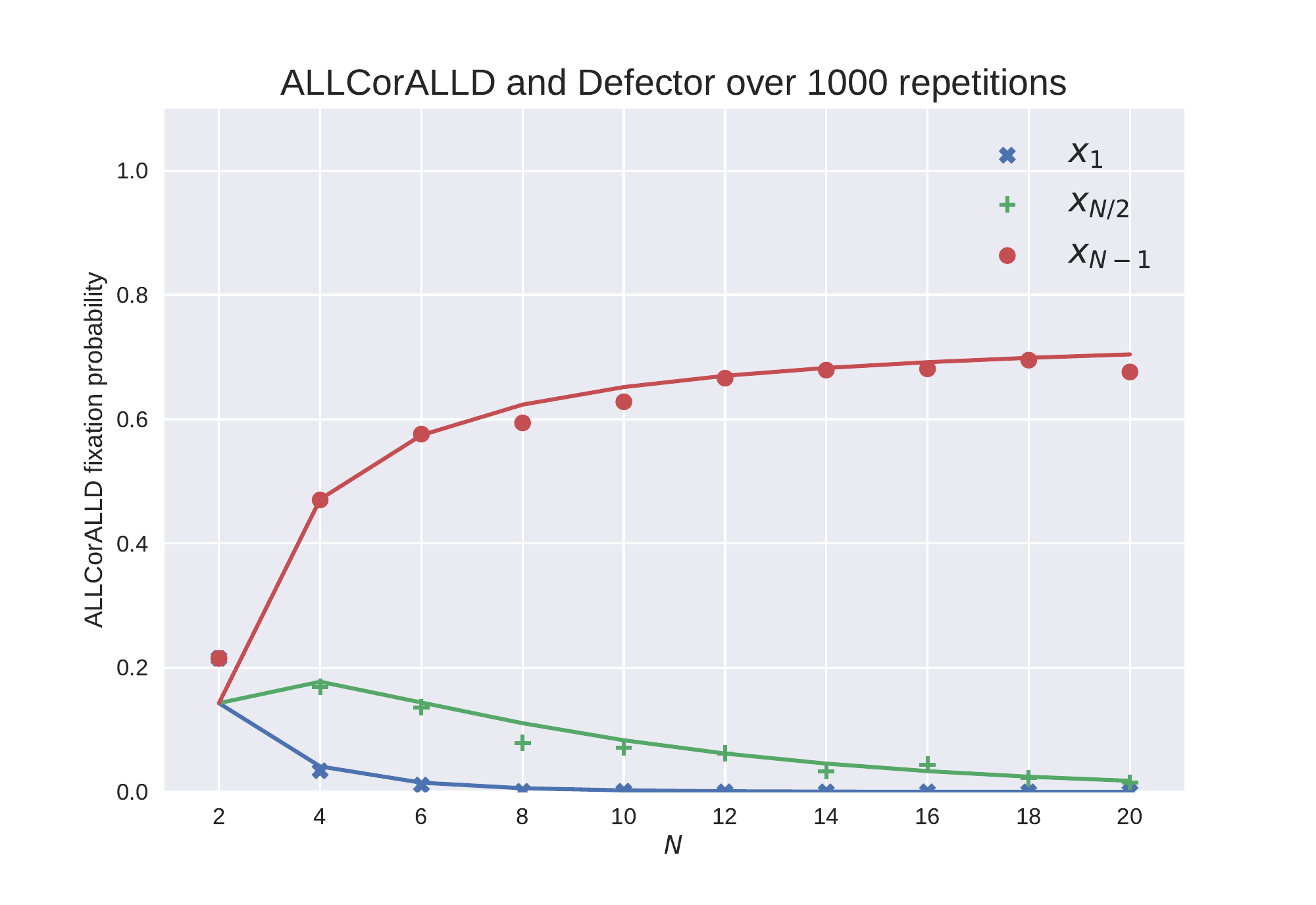}
        \caption{All C or all D and Defector}
    \end{subfigure}%
    ~
    \begin{subfigure}[t]{.3\textwidth}
        \centering
        \includegraphics[width=.8\textwidth]{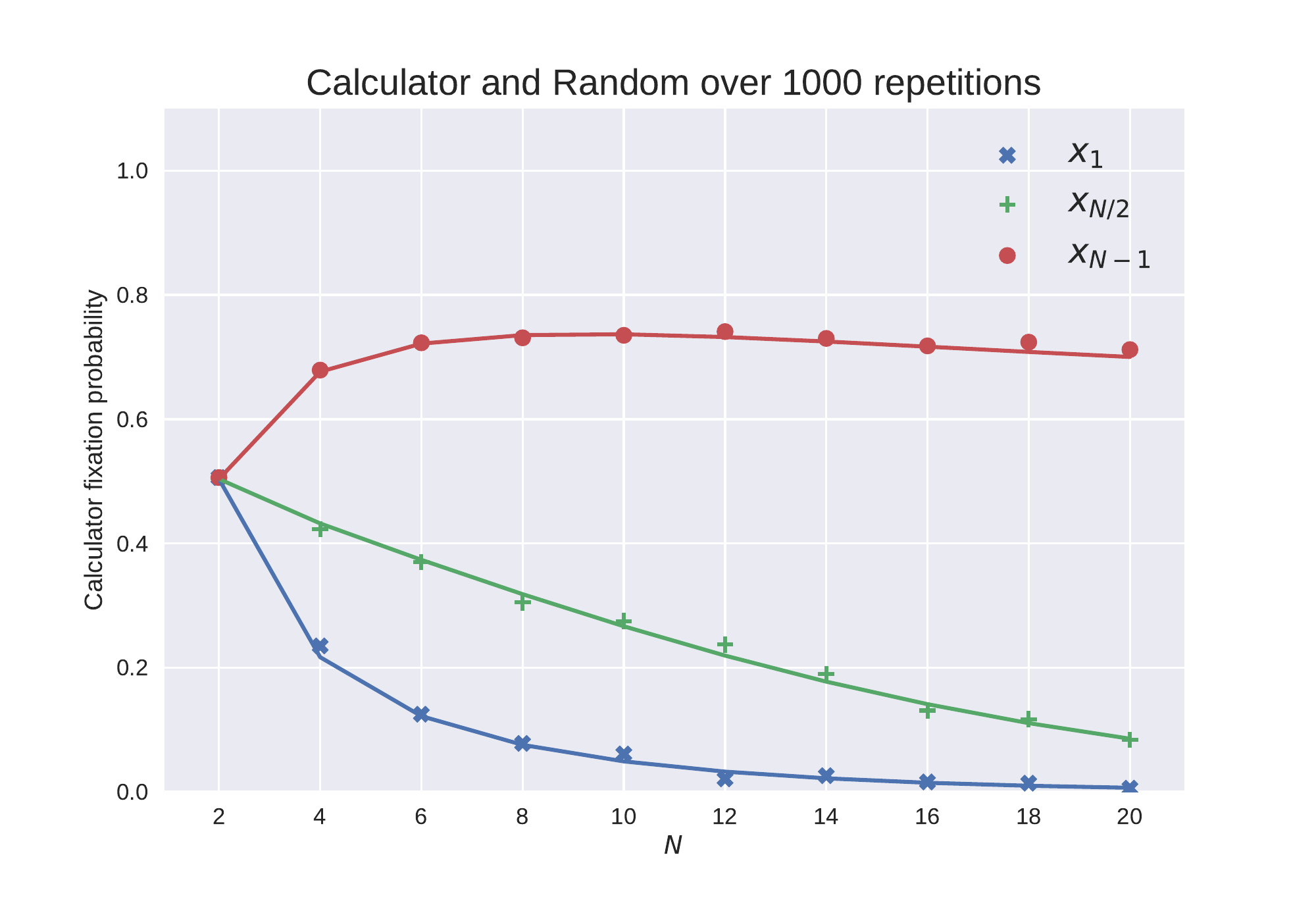}
        \caption{Calculator and Random}
    \end{subfigure}%

    \begin{subfigure}[t]{.3\textwidth}
        \centering
        \includegraphics[width=.8\textwidth]{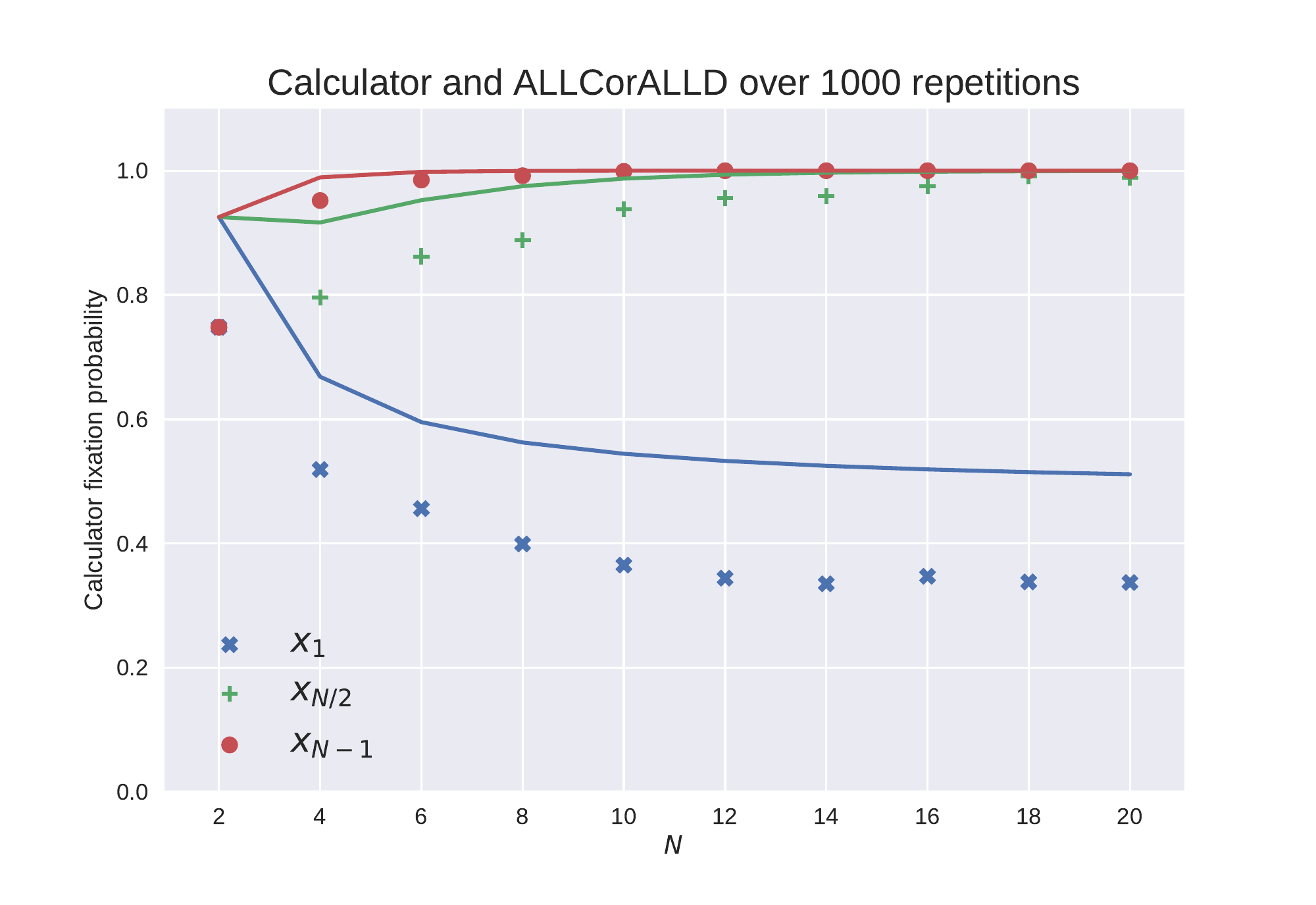}
        \caption{Calculator and All C or all D}
    \end{subfigure}%
    ~
    \begin{subfigure}[t]{.3\textwidth}
        \centering
        \includegraphics[width=.8\textwidth]{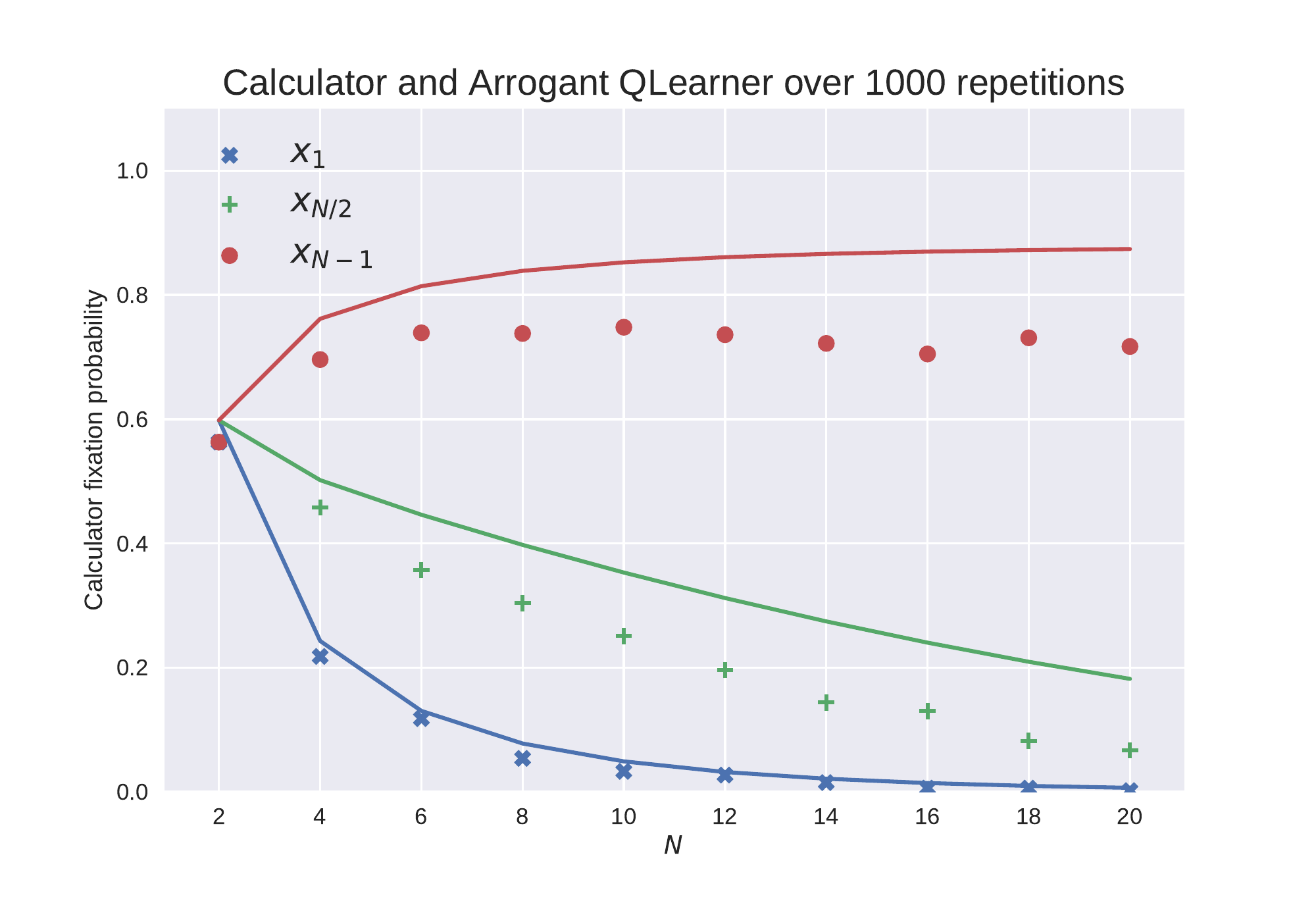}
        \caption{Calculator and Arrogant Q learner}
    \end{subfigure}%
    ~
    \begin{subfigure}[t]{.3\textwidth}
        \centering
        \includegraphics[width=.8\textwidth]{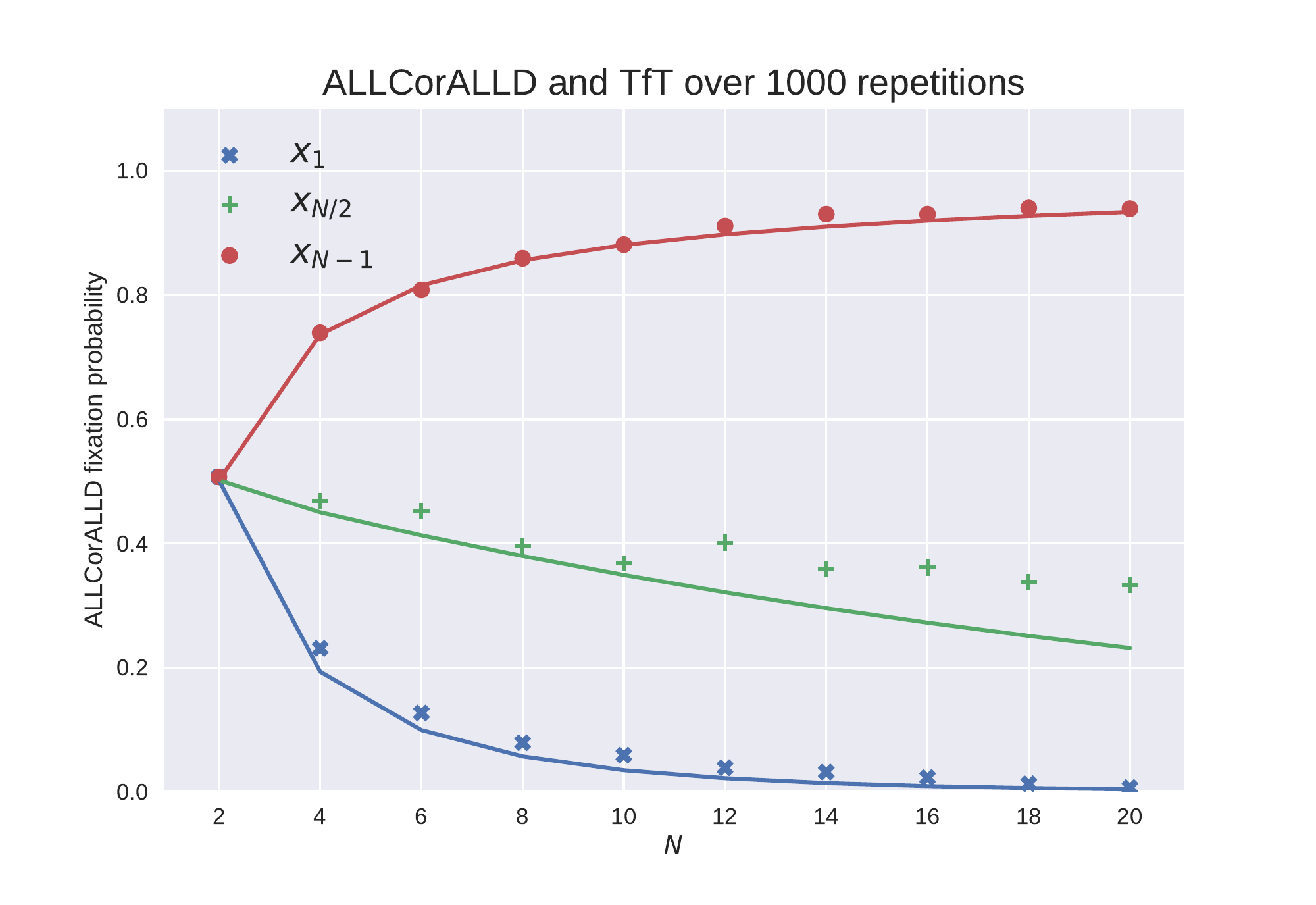}
        \caption{All C or all D and Tit For Tat}
    \end{subfigure}%
    \caption{Comparison of theoretic and actual Moran Process
             fixation probabilities for \textbf{stochastic} strategies.}
    \label{fig:comparison_stochastic}
\end{figure}

Comparisons of \(x_1, x_{N/2}, x_{N-1}\) are shown in
Figure~\ref{fig:comparison_deterministic}. The points represent the simulated
values and the line shows the theoretical value. Note that these are all
deterministic strategies and show a perfect match between the expected value
of (\ref{equ:fixation_probability}) and the actual Moran process for all
strategy pairs. Figure~\ref{fig:comparison_stochastic} shows the fixation probabilities for
stochastic strategies. These are no longer a good match which highlights the
weakness of assuming a given interaction between two IPD strategies can be
summarised with a set of utilities as shown in
(\ref{equ:payoff_matrix}). For any given pair of strategies it is possible to
obtain \(p_{i,i-1}, p_{i,i+1}, p_{ii}\) exactly (as opposed to the
approximations offered by (\ref{equ:p_up}), (\ref{equ:p_down}) and
(\ref{equ:p_stay})). Obtaining these requires particular analysis for a given
pair and can be quite a complex endeavour for stochastic strategies with long
memory: this is not necessary for the purposes of this work.
All data generated for this validation exercise can be found
at \cite{data}.

\section{Empirical results}\label{sec:empirical_results}

This section outlines the data analysis carried out, all data for this study is
available at~\cite{data}:

\begin{itemize}
    \item Section~\ref{sec:two_individuals} considers the specific case of
        \(N=2\).
    \item Section~\ref{sec:strong_invaders} investigates the effect of
        population size on the ability of a strategy to invade another
        population. This will highlight how complex strategies with long
        memories outperform simpler strategies.
    \item Section~\ref{sec:strong_resistors} similarly investigates the
        ability to defend against an invasion.
    \item Section~\ref{sec:population_size} investigates the relationship
        between performance for differing population sizes as well as
        taking a close look at zero determinant strategies \cite{Press2012}.
\end{itemize}

\subsection{The special case of \(N=2\)}\label{sec:two_individuals}

\begin{figure}[!hbtp]
    \begin{subfigure}{.5\textwidth}
        \centering
        \includegraphics[height=.7\textheight]{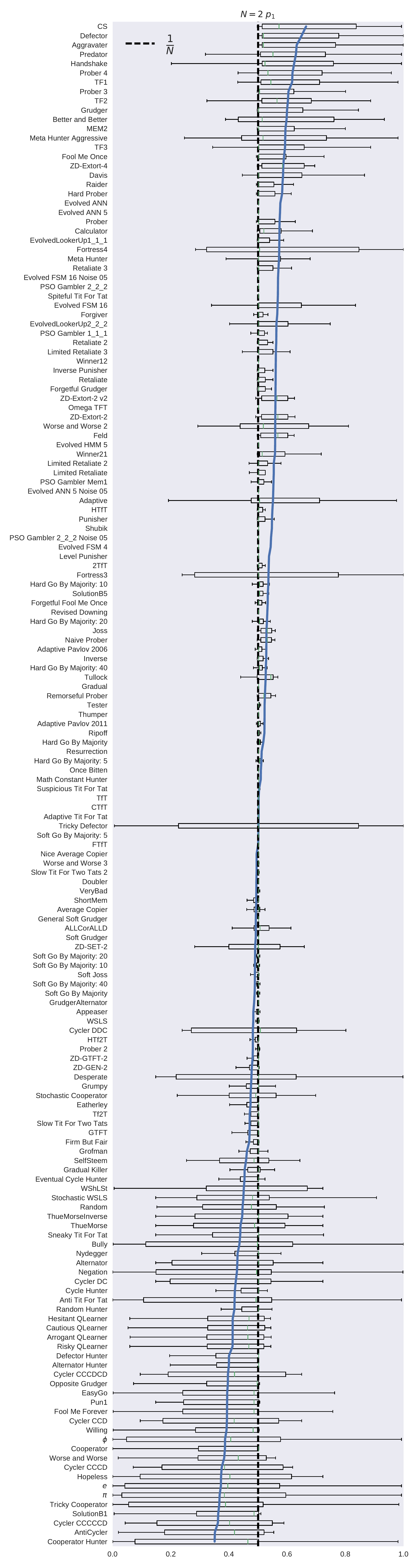}
        \caption{The fixation probabilities for \(N=2\)}
        \label{fig:boxplot_2}
    \end{subfigure}%
    ~
    \begin{subfigure}{.5\textwidth}
        \scriptsize
        \centering
        \begin{tabular}{llr}
\toprule
{} &                  Player &  Mean $p_1$ \\
\midrule
1  &                      CS &      0.6651 \\
2  &                Defector &      0.6496 \\
3  &              Aggravater &      0.6328 \\
4  &                Predator &      0.6301 \\
5  &               Handshake &      0.6240 \\
6  &                Prober 4 &      0.6183 \\
7  &                     \textbf{TF1} &      0.6171 \\
8  &                Prober 3 &      0.6044 \\
9  &                     \textbf{TF2} &      0.6026 \\
10 &                 Grudger &      0.5996 \\
11 &       Better and Better &      0.5980 \\
12 &                    MEM2 &      0.5942 \\
13 &  Meta Hunter Aggressive &      0.5933 \\
14 &                     \textbf{TF3} &      0.5927 \\
15 &            Fool Me Once &      0.5892 \\
16 &             ZD-Extort-4 &      0.5867 \\
\bottomrule
\end{tabular}

        \caption{Top strategies for \(N=2\) (neutral fixation is \(p=0.5\))}
        \label{tbl:summary_top_2}
    \end{subfigure}
    \caption{Performance of strategies for \(N=2\).}
\end{figure}

When $N=2$ the Moran process is effectively a measure of the distribution of relative
mean payoffs over all possible matches between two players. The strategy
that scores higher than the other more often will fixate more often. For \(N=2\)
the two cases of \(x_1\) and \(x_{N-1}\) coincide, but will be
considered separately for larger \(N\) in sections~\ref{sec:strong_invaders}
and~\ref{sec:strong_resistors}. Figure~\ref{fig:boxplot_2} shows all fixation
probabilities for the strategies considered. The top 16 (10\%) strategies are
shown in
Table~\ref{tbl:summary_top_2}. The top five ranking strategies are:

\begin{enumerate}
    \item The top strategy is the Collective Strategy (CS) which has a simple
        handshake mechanism described above.
    \item Defector: it always defects. Since it has no interactions with other
        defectors (recall that \(N=2\)), its aggressiveness is rewarded.
    \item Aggravater, which plays like Grudger (responding to any
        defections with unconditional defections throughout) however starts by
        playing 3 defections.
    \item Predator, a finite state machine described in \cite{Ashlock2006}.
    \item Handshake, a slightly less aggressive version of the Collective
        Strategy \cite{robson1989}. As long as the initial sequence is played
        then it cooperates. Thus it will do well in a population consisting of
        many members of itself just as the Collective Strategy does. The
        difference is that CS will defect after the handshake if the opponent
        defects while handshake will not.
\end{enumerate}

It is also noted that TF1, TF2 and TF3 all perform well. This is also the \(N\)
for which a zero determinant strategy does appear in the top 10\% ranking
strategies: ZD-extort-4. The performance of zero determinant strategies will be
examined more closely in Section~\ref{sec:population_size}.

As will be demonstrated in Section~\ref{sec:population_size} the results for
\(N=2\) differ from those of larger $N$. Hence these results do not concur with
the literature which suggests that zero determinant strategies should be
effective for larger population sizes, but these analyses consider stationary
behaviour, while this work runs for a fixed number of rounds. \cite{stewart2013extortion}
The stationarity assumption allows for a deterministic payoff matrix
leading to the conclusions about zero determinant strategies in the space
of memory-one strategies that do not generalize to this context.


\subsection{Strong Invaders}\label{sec:strong_invaders}

\begin{figure}[!hbtp]
    \centering
    \includegraphics[height=.9\textheight]{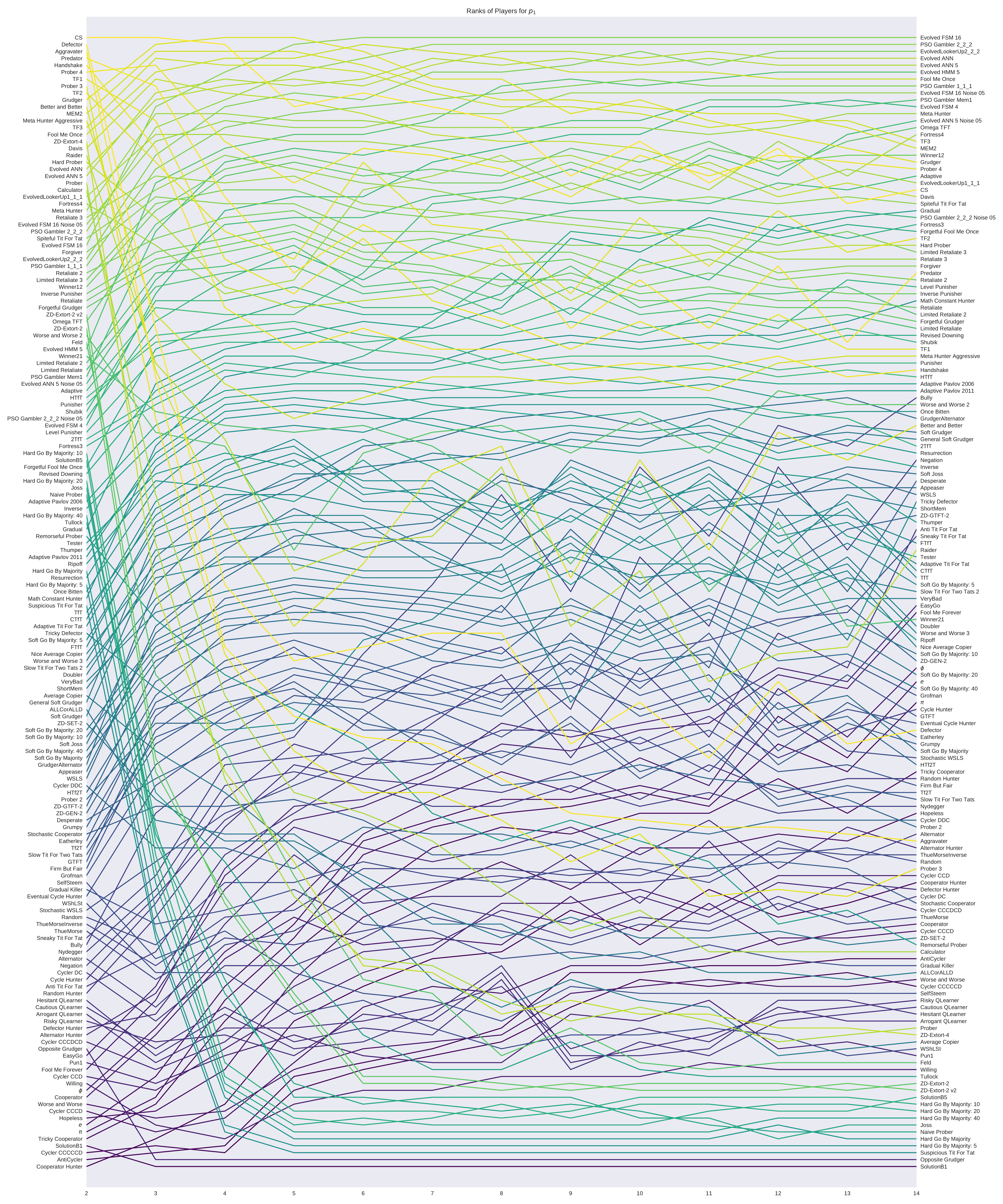}
    \caption{\textbf{Invasion}: Ranks of all strategies according to \(x_1\) for different
    population sizes.}
    \label{fig:ranks_v_size_invade}
\end{figure}

In this section the focus is on the ability of a mutant strategy to invade: the
probability of one individual of a given type successfully fixating in a
population of \(N - 1\) other individuals, denoted by \(x_1\).
The ranks of each strategy for all considered values of \(N\) according to mean
\(x_1\) are shown in Figure~\ref{fig:ranks_v_size_invade}.

\begin{figure}[!hbtp]
    \centering
    \begin{subfigure}{.3\textwidth}
        \centering
        \includegraphics[height=.9\textheight]{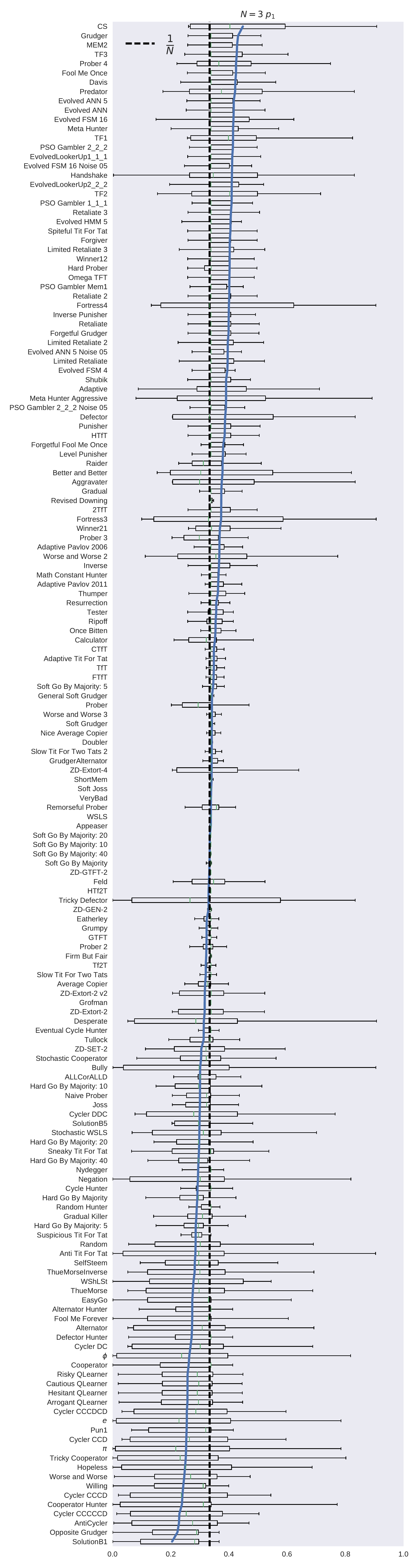}
        \caption{\(N=3\)}
        \label{fig:boxplot_3_invade}
    \end{subfigure}%
    ~
    \begin{subfigure}{.3\textwidth}
        \centering
        \includegraphics[height=.9\textheight]{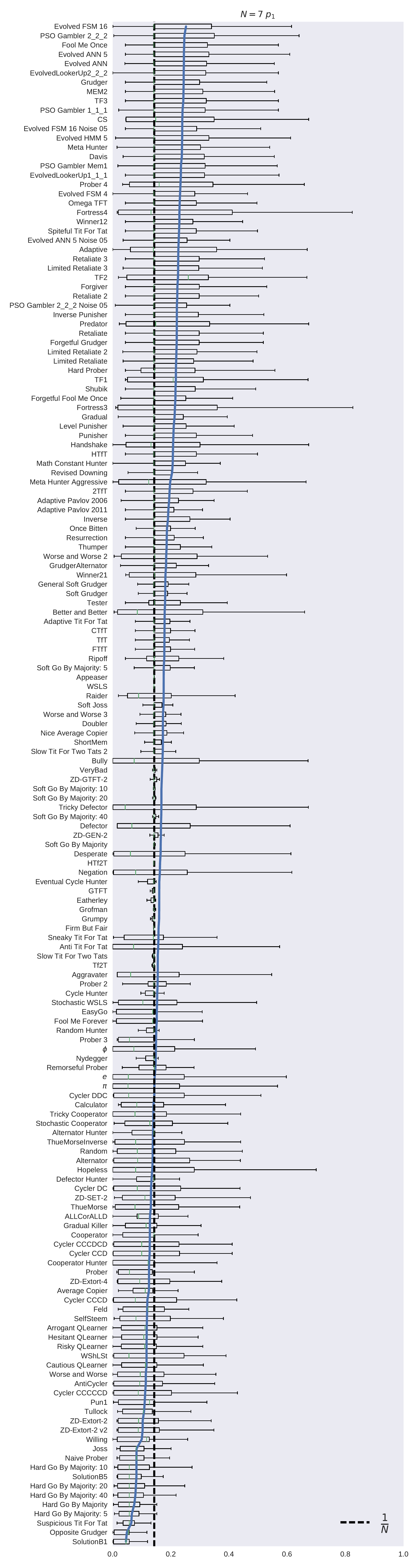}
        \caption{\(N=7\)}
        \label{fig:boxplot_7_invade}
    \end{subfigure}%
    ~
    \begin{subfigure}{.3\textwidth}
        \centering
        \includegraphics[height=.9\textheight]{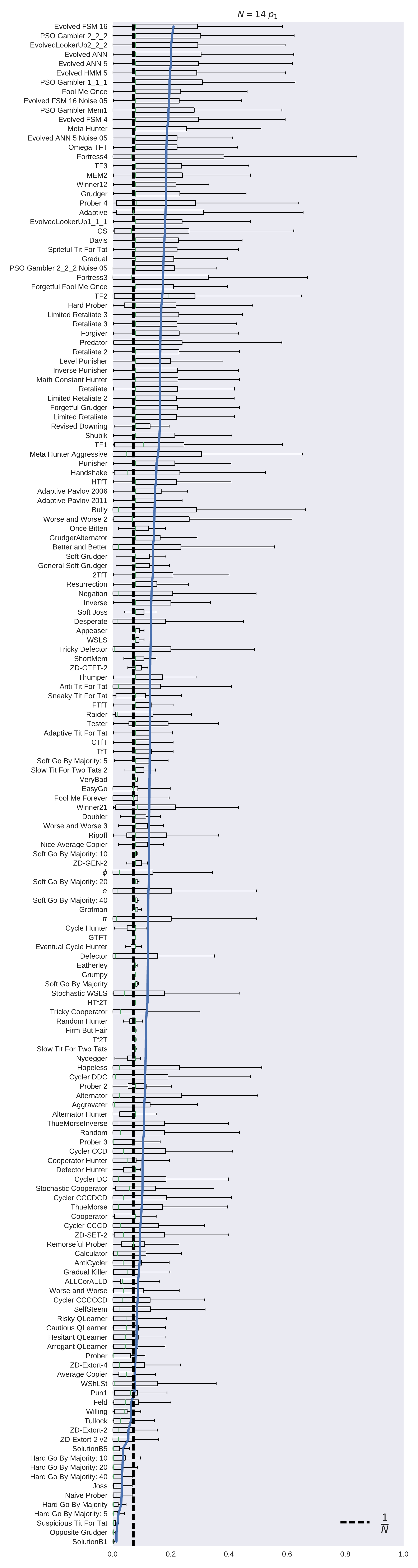}
        \caption{\(N=14\)}
        \label{fig:boxplot_14_invade}
    \end{subfigure}
    \caption{The fixation probabilities \(x_1\).}
\end{figure}

The fixation
probabilities are shown in
Figures~\ref{fig:boxplot_3_invade},~\ref{fig:boxplot_7_invade}
and~\ref{fig:boxplot_14_invade} for \(N\in\{3, 7, 14\}\) showing the mean
fixation as well as the neutral fixation for each given scenario.

The top 16 strategies are given in Tables~\ref{tbl:top_invade}.

\begin{table}[!hbtp]
    \centering
    \scriptsize
    \begin{subfigure}[t]{.5\textwidth}
        \centering
        \begin{tabular}{llr}
\toprule
{} &                   Player &  Mean $p_1$ \\
\midrule
1  &                       CS &      0.4478 \\
2  &                  Grudger &      0.4313 \\
3  &                     MEM2 &      0.4278 \\
4  &                      \textbf{TF3} &      0.4267 \\
5  &                 Prober 4 &      0.4242 \\
6  &             Fool Me Once &      0.4242 \\
7  &                    Davis &      0.4218 \\
8  &                 Predator &      0.4210 \\
9  &            Evolved ANN 5 &      0.4163 \\
10 &              Evolved ANN &      0.4163 \\
11 &           Evolved FSM 16 &      0.4154 \\
12 &              Meta Hunter &      0.4140 \\
13 &                      \textbf{TF1} &      0.4139 \\
14 &        PSO Gambler 2\_2\_2 &      0.4134 \\
15 &     EvolvedLookerUp1\_1\_1 &      0.4113 \\
16 &  Evolved FSM 16 Noise 05 &      0.4107 \\
\bottomrule
\end{tabular}

        \caption{\(N=3\)}
    \end{subfigure}%
    ~
    \begin{subfigure}[t]{.5\textwidth}
        \centering
        \begin{tabular}{llr}
\toprule
{} &                   Player &  Mean $p_1$ \\
\midrule
1  &           Evolved FSM 16 &      0.2523 \\
2  &        PSO Gambler 2\_2\_2 &      0.2467 \\
3  &             Fool Me Once &      0.2459 \\
4  &            Evolved ANN 5 &      0.2450 \\
5  &              Evolved ANN &      0.2449 \\
6  &     EvolvedLookerUp2\_2\_2 &      0.2443 \\
7  &                  Grudger &      0.2442 \\
8  &                     MEM2 &      0.2436 \\
9  &                      \textbf{TF3} &      0.2430 \\
10 &        PSO Gambler 1\_1\_1 &      0.2404 \\
11 &                       CS &      0.2395 \\
12 &  Evolved FSM 16 Noise 05 &      0.2394 \\
13 &            Evolved HMM 5 &      0.2390 \\
14 &              Meta Hunter &      0.2385 \\
15 &                    Davis &      0.2379 \\
16 &         PSO Gambler Mem1 &      0.2348 \\
\bottomrule
\end{tabular}

        \caption{\(N=7\)}
    \end{subfigure}

    \begin{subfigure}[t]{.3\textwidth}
        \centering
        \begin{tabular}{llr}
\toprule
{} &                   Player &  Mean $p_1$ \\
\midrule
1  &           Evolved FSM 16 &      0.2096 \\
2  &        PSO Gambler 2\_2\_2 &      0.2042 \\
3  &     EvolvedLookerUp2\_2\_2 &      0.2014 \\
4  &              Evolved ANN &      0.2014 \\
5  &            Evolved ANN 5 &      0.2004 \\
6  &            Evolved HMM 5 &      0.1972 \\
7  &        PSO Gambler 1\_1\_1 &      0.1955 \\
8  &             Fool Me Once &      0.1955 \\
9  &  Evolved FSM 16 Noise 05 &      0.1943 \\
10 &         PSO Gambler Mem1 &      0.1920 \\
11 &            Evolved FSM 4 &      0.1918 \\
12 &              Meta Hunter &      0.1869 \\
13 &   Evolved ANN 5 Noise 05 &      0.1858 \\
14 &                Omega TFT &      0.1849 \\
15 &                Fortress4 &      0.1848 \\
16 &                      \textbf{TF3} &      0.1846 \\
\bottomrule
\end{tabular}

        \caption{\(N=14\)}
    \end{subfigure}
    \caption{Top invaders for \(N\in\{3, 7, 14\}\)}
    \label{tbl:top_invade}
\end{table}

It can be seen that apart from CS, none of the strategies
of Table~\ref{tbl:summary_top_2} perform well for \(N\in\{3, 7, 14\}\). The new
top performing strategies are:

\begin{itemize}
    \item Grudger (which only performs well for \(N=3\)), starts by cooperating
        but will defect if at any point the opponent has defected.
    \item MEM2, an infinite memory strategy that switches between TFT, TF2T, and
        Defector \cite{Li2014}.
    \item TF3, the finite state machine trained specifically for Moran processes
        described in Section~\ref{sec:introduction}.
    \item Prober 4, a strategy which starts with a specific 20 move sequence of
        cooperations and defections \cite{Prison1998}. This initial sequence serves
        as approximate handshake.
    \item  PSO Gambler and Evolved Lookerup 2 2 2: are strategies that make use
        of a lookup table mapping the first 2 moves of the opponent as well as
        the last 2 moves of both players to an action. The PSO gambler is a
        stochastic version of the Lookerup which maps those states to probabilities of cooperating. The Lookerup was described in \cite{Knight2016}.
    \item The Evolved ANN strategies are neural networks that map a number of
	    attributes (first move, number of cooperations, last move, etc.) to
	    an action. Both of these have been trained using an evolutionary
	    algorithm.
    \item The Evolved FSM 16 is a 16 state finite state machine trained to
        perform well in tournaments.
\end{itemize}

Only one of the above strategies is stochastic although close inspection of the
source code of PSO Gambler shows that it makes stochastic decisions rarely, and
is functionally very similar to its deterministic cousin Evolved Looker Up.
The PSO Gambler Mem1 strategy is a memory one strategy that has been trained to
maximise its utility and does perform well.
Apart from TF3, the finite state machines trained specifically for
Moran processes do not appear in the top 5, while strategies trained for
tournaments do. This is due to the nature of invasion: most of the opponents
will initially be different strategies. The next section will consider the
converse situation.

\subsection{Strong resistors}\label{sec:strong_resistors}

\begin{figure}[!hbtp]
    \centering
    \includegraphics[height=.9\textheight]{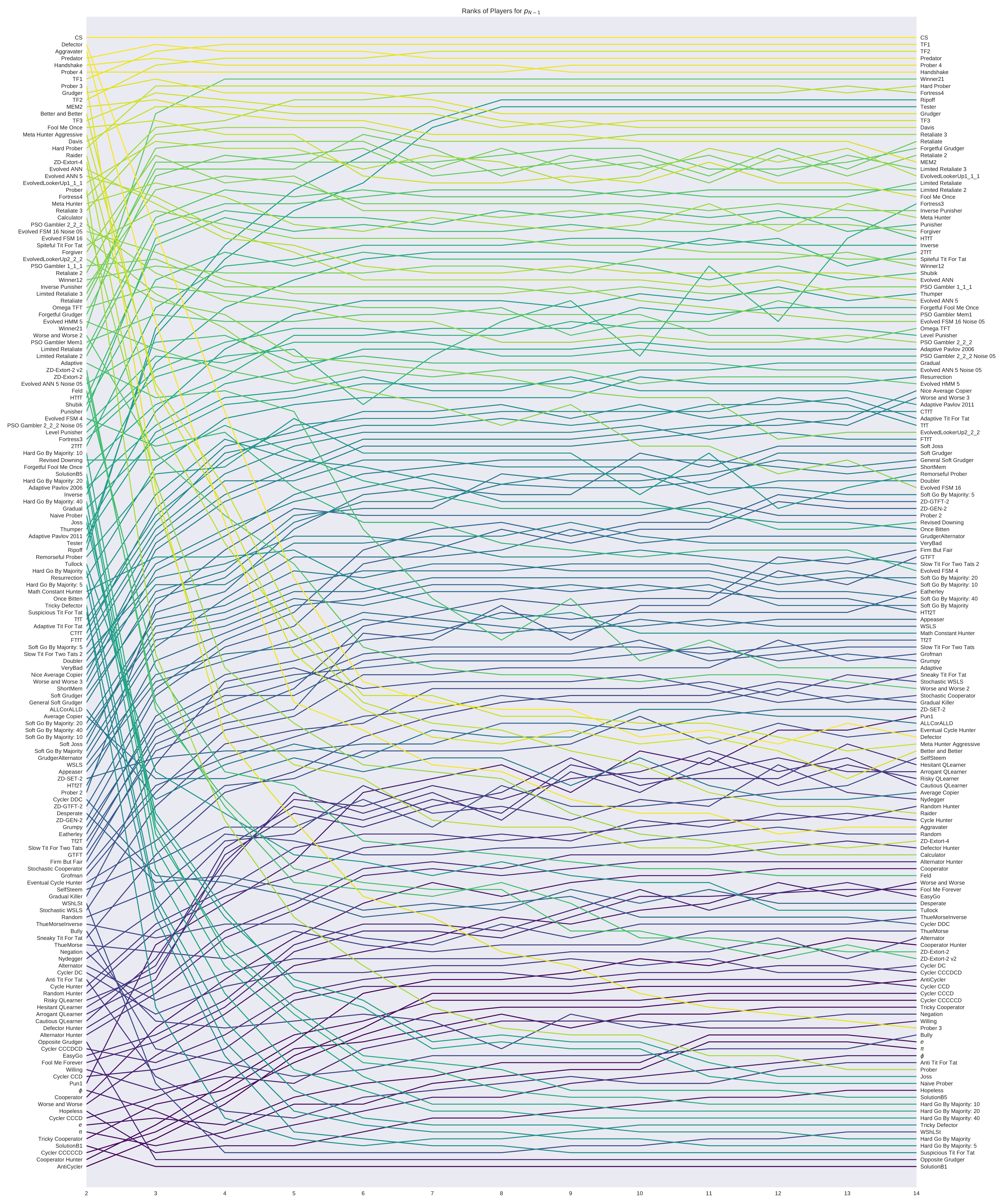}
    \caption{\textbf{Resistance}: Ranks of all strategies according to \(x_{N-1}\) for different
    population sizes.}
    \label{fig:ranks_v_size_resist}
\end{figure}

In addition to identifying good invaders, strategies resistant to invasion by
other strategies are identified by examining the distribution of $x_{N-1}$ for
each strategy. The ranks of each strategy for all considered values of \(N\)
according to mean \(x_{N-1}\) are shown in
Figures~\ref{fig:ranks_v_size_resist}.

\begin{figure}[!hbtp]
    \centering
    \begin{subfigure}{.3\textwidth}
        \centering
        \includegraphics[height=.9\textheight]{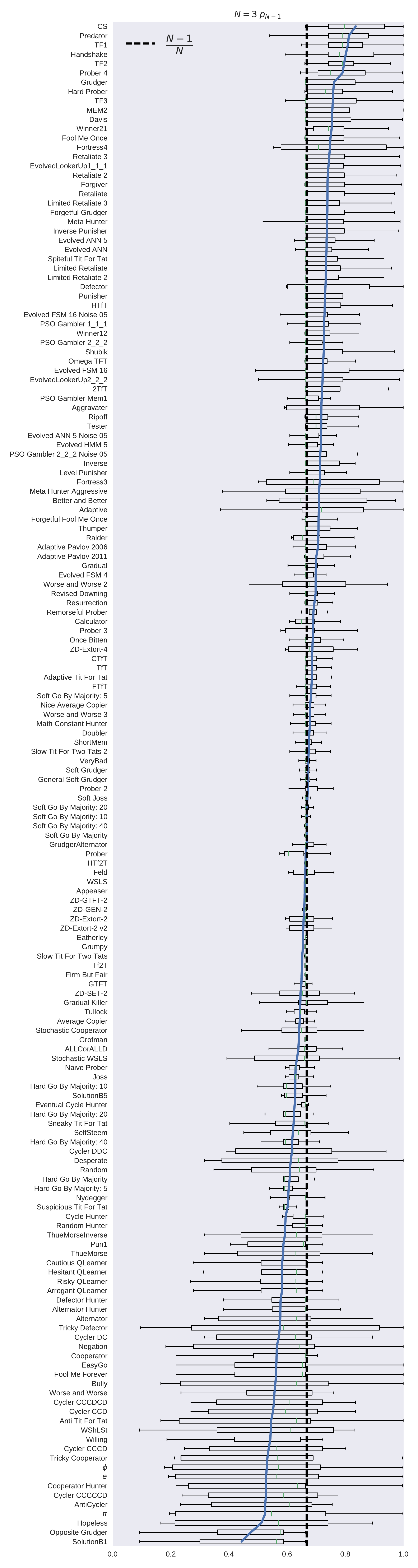}
        \caption{\(N=3\)}
        \label{fig:boxplot_3_resist}
    \end{subfigure}%
    ~
    \begin{subfigure}{.3\textwidth}
        \centering
        \includegraphics[height=.9\textheight]{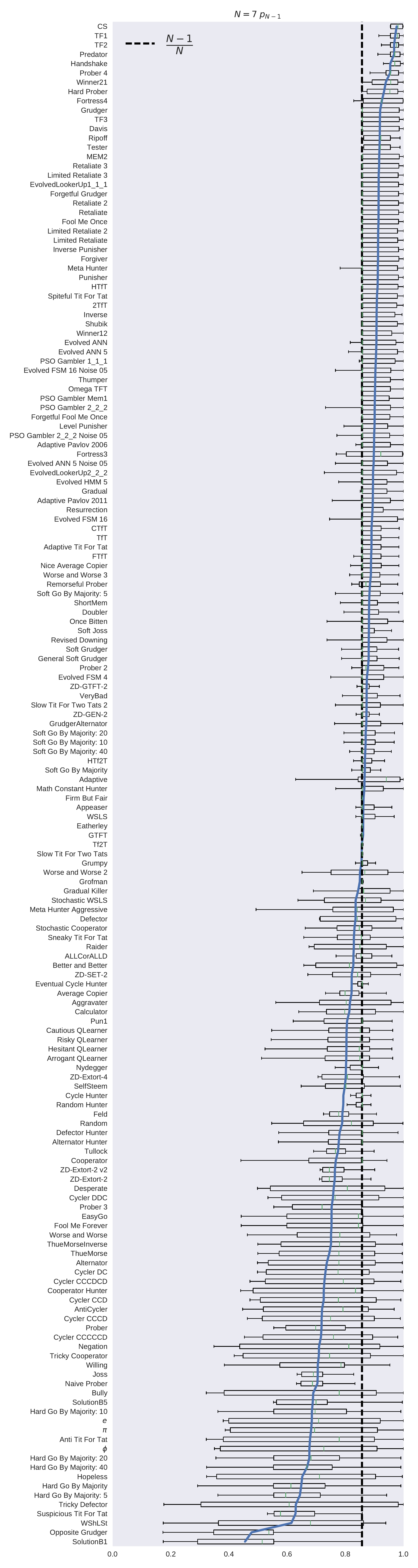}
        \caption{\(N=7\)}
        \label{fig:boxplot_7_resist}
    \end{subfigure}%
    ~
    \begin{subfigure}{.3\textwidth}
        \centering
        \includegraphics[height=.9\textheight]{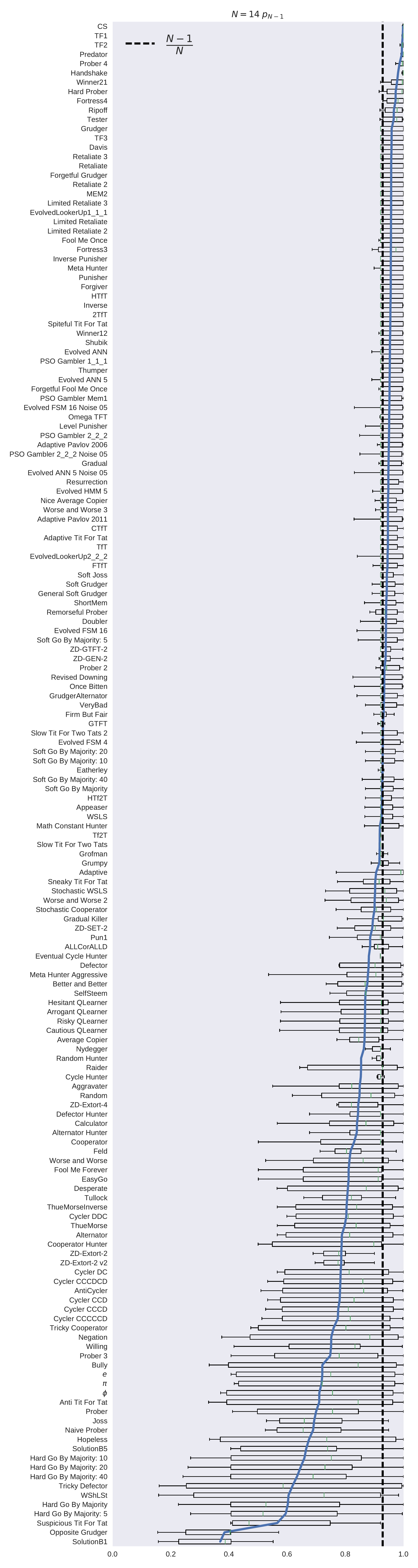}
        \caption{\(N=14\)}
        \label{fig:boxplot_14_resist}
    \end{subfigure}
    \caption{The fixation probabilities \(x_{N-1}\).}
\end{figure}

The fixation probabilities are shown in
Figures~\ref{fig:boxplot_3_resist},~\ref{fig:boxplot_7_invade}
and~\ref{fig:boxplot_14_resist} for \(N\in\{3, 7, 14\}\) showing the mean
fixation as well as the neutral fixation for each given scenario.

Table~\ref{tbl:top_resist} shows the top strategies when ranked
according to \(x_{N-1}\) for \(N\in\{3, 7, 14\}\).
Once again none of the short memory strategies from
Section~\ref{sec:two_individuals} perform well for high \(N\).

\begin{table}[!hbtp]
    \scriptsize
    \centering
    \begin{subfigure}[t]{.5\textwidth}
        \centering
        \begin{tabular}{llr}
\toprule
{} &                Player &  Mean $p_{N-1}$ \\
\midrule
1  &                    CS &          0.8359 \\
2  &              Predator &          0.8121 \\
3  &                   \textbf{TF1} &          0.8087 \\
4  &             Handshake &          0.8014 \\
5  &                   \textbf{TF2} &          0.7957 \\
6  &              Prober 4 &          0.7905 \\
7  &               Grudger &          0.7612 \\
8  &           Hard Prober &          0.7582 \\
9  &                   \textbf{TF3} &          0.7570 \\
10 &                  MEM2 &          0.7554 \\
11 &                 Davis &          0.7536 \\
12 &              Winner21 &          0.7529 \\
13 &          Fool Me Once &          0.7489 \\
14 &             Fortress4 &          0.7467 \\
15 &           Retaliate 3 &          0.7448 \\
16 &  EvolvedLookerUp1\_1\_1 &          0.7422 \\
\bottomrule
\end{tabular}

        \caption{\(N=3\)}
    \end{subfigure}%
    ~
    \begin{subfigure}[t]{.5\textwidth}
        \centering
        \begin{tabular}{llr}
\toprule
{} &       Player &  Mean $p_{N-1}$ \\
\midrule
1  &           CS &          0.9765 \\
2  &          \textbf{TF1} &          0.9714 \\
3  &          \textbf{TF2} &          0.9677 \\
4  &     Predator &          0.9677 \\
5  &    Handshake &          0.9547 \\
6  &     Prober 4 &          0.9540 \\
7  &     Winner21 &          0.9392 \\
8  &  Hard Prober &          0.9331 \\
9  &    Fortress4 &          0.9255 \\
10 &      Grudger &          0.9198 \\
11 &          \textbf{TF3} &          0.9189 \\
12 &        Davis &          0.9186 \\
13 &       Ripoff &          0.9183 \\
14 &       Tester &          0.9176 \\
15 &         MEM2 &          0.9165 \\
16 &  Retaliate 3 &          0.9161 \\
\bottomrule
\end{tabular}

        \caption{\(N=7\)}
    \end{subfigure}

    \begin{subfigure}[t]{.3\textwidth}
        \centering
        \begin{tabular}{llr}
\toprule
{} &       Player &  Mean $p_{N-1}$ \\
\midrule
1  &           CS &          0.9984 \\
2  &          \textbf{TF1} &          0.9973 \\
3  &          \textbf{TF2} &          0.9949 \\
4  &     Predator &          0.9941 \\
5  &     Prober 4 &          0.9863 \\
6  &    Handshake &          0.9812 \\
7  &     Winner21 &          0.9778 \\
8  &  Hard Prober &          0.9731 \\
9  &    Fortress4 &          0.9726 \\
10 &       Ripoff &          0.9669 \\
11 &       Tester &          0.9662 \\
12 &      Grudger &          0.9592 \\
13 &          \textbf{TF3} &          0.9589 \\
14 &        Davis &          0.9588 \\
15 &  Retaliate 3 &          0.9580 \\
16 &    Retaliate &          0.9576 \\
\bottomrule
\end{tabular}

        \caption{\(N=14\)}
    \end{subfigure}
    \caption{Top resistors for \(N\in\{3, 7, 14\}\)}
    \label{tbl:top_resist}
\end{table}

Interestingly none of these strategies is stochastic: this is explained by
the need of strategies to have a steady hand when interacting with their own
kind. Acting stochastically increases the chance of friendly fire.
However it is possible to design a strategy with a stochastic or error-correcting
handshake that is an excellent resistor even in noisy environments \cite{Lee2015}.

There are are only two new strategies that appear in the top ranks for
\(x_{N-1}\): TF1 and TF2. These two strategies are with CS the strongest
resistors. They all have handshakes, and whilst the handshakes of CS and
Handshake (which ranks highly for the smaller values of \(N\)) were
programmed, the handshakes of TF1 and TF2 evolved through an evolutionary
process without any priming.

As described in Section~\ref{sec:strong_invaders} the strategies trained with
the payoff maximizing objective are among the best invaders in the library
however they are not as resistant to invasion as the strategies trained using a
Moran objective function. These strategies include trained finite state machine
strategies, but they do not appear to have handshaking mechanisms. Therefore it
is reasonable to conclude that the objective function is the cause of the
emergence of handshaking mechanisms. More specifically, TF1 and TF2 evolved
handshakes for high invasion resistance. TF3 is a better total payoff maximizer
which makes it a better invader along with the strategies
trained to maximize total payoff since successful fitness proportionate selection
is necessary for invasion. Training with an objective with initial population
mix other than $(N/2, N/2)$ may favor invasion or resistance.

The payoff maximizing strategies typically will not defect before the opponent's
first defection, possibly because the training strategy collection contains some
strategies such as Grudger and Fool Me Once that retaliate harshly by defecting
for the remainder of the match if the opponent has more than a small number of
cumulative defections. Paradoxically it is advantageous to defect (as a signal)
in order to achieve mutual cooperation with opponents using the same strategy
but not with other opponents. Nevertheless an evolutionary process is able to
tunnel through the costs and risks associated to early defections to find more
optimal solutions, so it is not surprising in hindsight that handshaking
strategies emerge from the evolutionary training process.

A handshake requires at least one defection and there is
selective pressure to defect as few times as possible to achieve the
self-recognition mechanism. It is also unwise to defect on the first move as
some strategies additionally retaliate first round defections. So the
handshakes used by TF1, TF2, and CS are in some sense optimal.

It is evident through
Sections~\ref{sec:two_individuals},~\ref{sec:strong_invaders}
and~\ref{sec:strong_resistors} that performance of strategies not only depends
on the initial population distribution but also that there seems to be a
difference depending on whether or not \(N>2\). This will be explored further in
the next section, looking not only at \(x_1\) and \(x_{N-1}\) but also consider
\(x_{N/2}\).

\subsection{The effect of population size}\label{sec:population_size}

\begin{figure}[!hbtp]
    \centering
    \includegraphics[height=.9\textheight]{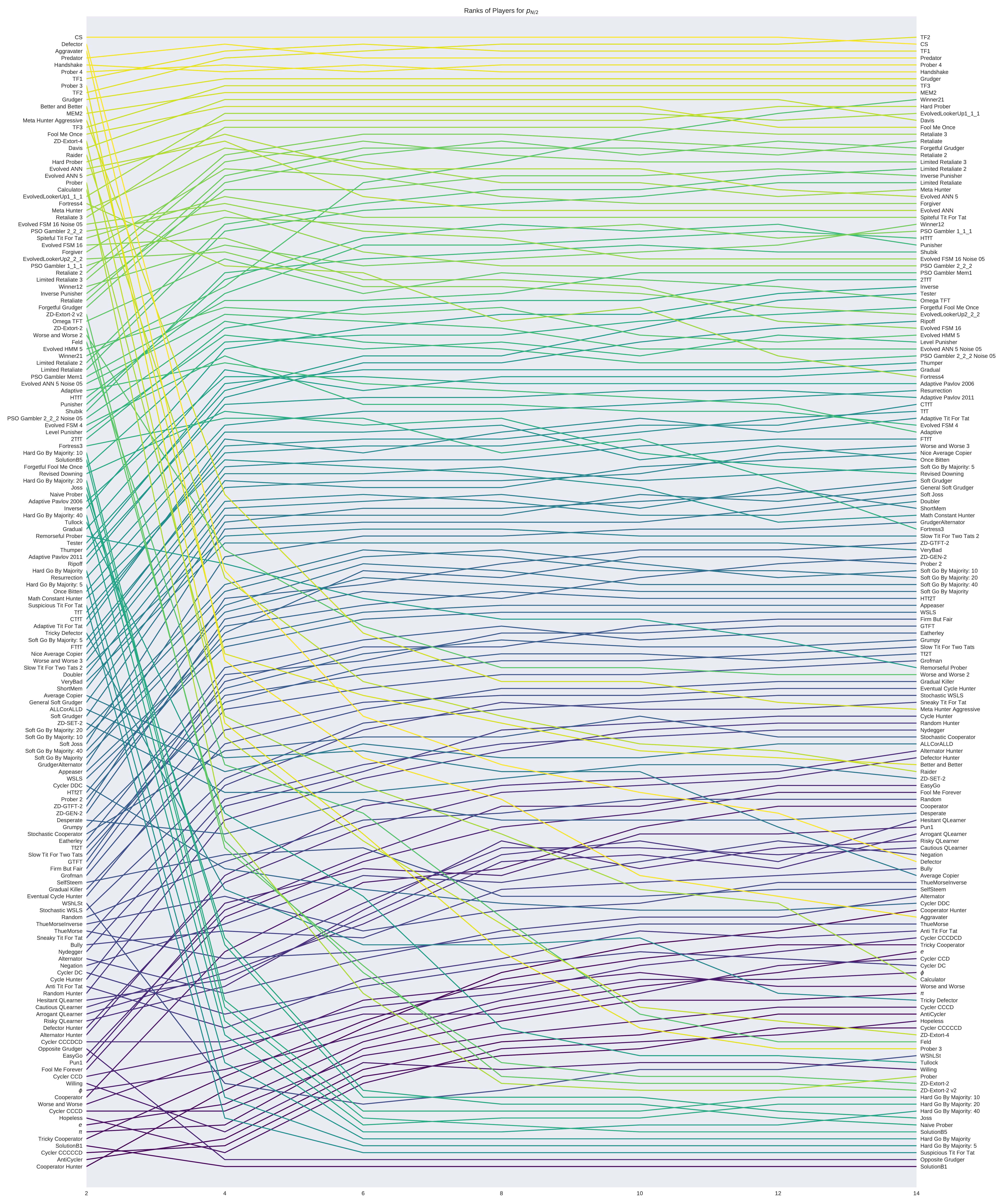}
    \caption{Fixation ranks of all strategies according to \(x_{N/2}\) for different
    population sizes.}
    \label{fig:ranks_v_size_coexist}
\end{figure}

To complement Figures~\ref{fig:ranks_v_size_invade}
and~\ref{fig:ranks_v_size_resist}, Figure~\ref{fig:ranks_v_size_coexist} shows
the rank of each strategy based on \(x_{N/2}\).
Tables~\ref{tbl:ranks_v_size_invade},~\ref{tbl:ranks_v_size_resist}
and~\ref{tbl:ranks_v_size_coexist} show the same information for a selection of
strategies:

\begin{itemize}
    \item The strategies that ranked highly for \(N=2\);
    \item The strategies that ranked highly for \(N=14\);
    \item The zero determinant strategies.
\end{itemize}

The results for \(x_{N/2}\) show similarities to the results for \(x_{N-1}\) and
in particular TF1, TF2 and TF3 ranked one, three and eight. This is to be
expected since, as described in Section~\ref{sec:strategies} these strategies
were trained in an initial population of \((N/2, N/2)\) individuals.

For all starting populations
\(i\in\{1, N/2, N-1\}\) the ranks of strategies are relatively stable across the
different values of \(N>2\) however for \(N=2\) there is a distinct difference.
This highlights that there is little that can be inferred about the evolutionary
performance of a strategy in a large population from its performance in a small
population. This is confirmed by the performance of the zero determinant strategies: while
some do rank relatively highly for \(N=2\) (ZD-extort-4 has rank 16) this rank
does not translate to larger populations.

\begin{table}[!hbtp]
    \centering
    \scriptsize
    \begin{tabular}{lrrrrrrrrrrrrr}
\toprule
                 Size &      2 &      3 &      4 &      5 &      6 &      7 &      8 &      9 &     10 &     11 &     12 &     13 &     14 \\
\midrule
                   CS &    1 &    1 &    2 &   11 &    9 &   11 &   13 &   21 &   16 &   22 &   17 &   25 &   23 \\
             Defector &    2 &   43 &   80 &   91 &   89 &   87 &   87 &  103 &   97 &  105 &   94 &  103 &  101 \\
           Aggravater &    3 &   50 &   89 &   99 &  102 &  103 &  108 &  113 &  114 &  115 &  115 &  116 &  117 \\
             Predator &    4 &    8 &   24 &   35 &   28 &   33 &   31 &   43 &   36 &   43 &   34 &   45 &   35 \\
            Handshake &    5 &   17 &   40 &   46 &   43 &   46 &   46 &   49 &   48 &   49 &   47 &   50 &   49 \\
\midrule
       Evolved FSM 16 &   31 &   11 &    6 &    2 &    1 &    1 &    1 &    1 &    1 &    1 &    1 &    1 &    1 \\
    PSO Gambler 2\_2\_2 &   29 &   14 &   10 &    6 &    4 &    2 &    2 &    2 &    2 &    2 &    2 &    2 &    2 \\
 EvolvedLookerUp2\_2\_2 &   33 &   18 &   11 &    9 &   10 &    6 &    6 &    5 &    3 &    5 &    3 &    3 &    3 \\
          Evolved ANN &   20 &   10 &    8 &    7 &    8 &    5 &    3 &    3 &    4 &    3 &    4 &    4 &    4 \\
        Evolved ANN 5 &   21 &    9 &    7 &    8 &    7 &    4 &    5 &    4 &    5 &    4 &    5 &    5 &    5 \\
\midrule
                  \textbf{TF1} &    7 &   13 &   33 &   38 &   30 &   39 &   42 &   46 &   42 &   46 &   41 &   46 &   46 \\
                  \textbf{TF2} &    9 &   19 &   29 &   33 &   19 &   28 &   29 &   38 &   27 &   34 &   26 &   32 &   30 \\
                  \textbf{TF3} &   14 &    4 &    5 &    5 &    6 &    9 &   11 &   11 &   12 &   14 &   13 &   13 &   16 \\
\midrule
          ZD-Extort-4 &   16 &   81 &  107 &  120 &  135 &  136 &  142 &  140 &  142 &  142 &  144 &  144 &  145 \\
       ZD-Extort-2 v2 &   41 &  105 &  126 &  140 &  152 &  152 &  153 &  152 &  153 &  153 &  153 &  152 &  153 \\
          ZD-Extort-2 &   43 &  107 &  125 &  139 &  151 &  151 &  152 &  153 &  152 &  152 &  152 &  153 &  152 \\
             ZD-SET-2 &  100 &  111 &  117 &  117 &  122 &  127 &  131 &  128 &  131 &  131 &  130 &  132 &  131 \\
            ZD-GTFT-2 &  112 &   92 &   82 &   80 &   81 &   82 &   84 &   72 &   81 &   71 &   78 &   72 &   70 \\
             ZD-GEN-2 &  113 &   96 &   87 &   83 &   85 &   88 &   90 &   82 &   87 &   82 &   86 &   83 &   91 \\
\bottomrule
\end{tabular}
    \caption{Invasion: Fixation ranks of some strategies according to \(x_1\) for different
    population sizes}
    \label{tbl:ranks_v_size_invade}
\end{table}

\begin{table}[!hbtp]
    \centering
    \scriptsize
    \begin{tabular}{lrrrrrrrrrrrrr}
\toprule
           Size &      2 &      3 &      4 &      5 &      6 &      7 &      8 &      9 &     10 &     11 &     12 &     13 &     14 \\
\midrule
             CS &    1 &    1 &    1 &    1 &    1 &    1 &    1 &    1 &    1 &    1 &    1 &    1 &    1 \\
       Defector &    2 &   29 &   55 &   79 &   94 &   97 &   98 &   98 &  102 &  101 &  103 &  100 &  102 \\
     Aggravater &    3 &   42 &   71 &   97 &  101 &  106 &  107 &  111 &  113 &  113 &  116 &  115 &  115 \\
       Predator &    4 &    2 &    3 &    3 &    3 &    4 &    4 &    4 &    4 &    4 &    4 &    4 &    4 \\
      Handshake &    5 &    4 &    5 &    5 &    5 &    5 &    5 &    6 &    6 &    6 &    6 &    6 &    6 \\
\midrule
            \textbf{TF1} &    7 &    3 &    2 &    2 &    2 &    2 &    2 &    2 &    2 &    2 &    2 &    2 &    2 \\
            \textbf{TF2} &   10 &    5 &    4 &    4 &    4 &    3 &    3 &    3 &    3 &    3 &    3 &    3 &    3 \\
       Prober 4 &    6 &    6 &    6 &    6 &    6 &    6 &    6 &    5 &    5 &    5 &    5 &    5 &    5 \\
\midrule
            \textbf{TF3} &   13 &    9 &   10 &   11 &   11 &   11 &   13 &   14 &   13 &   13 &   13 &   13 &   13 \\
\midrule
    ZD-Extort-4 &   19 &   68 &   98 &  106 &  108 &  114 &  115 &  115 &  118 &  118 &  117 &  118 &  117 \\
 ZD-Extort-2 v2 &   49 &   98 &  111 &  121 &  123 &  124 &  124 &  130 &  130 &  132 &  134 &  132 &  134 \\
    ZD-Extort-2 &   50 &   97 &  112 &  123 &  124 &  125 &  123 &  126 &  131 &  131 &  132 &  133 &  133 \\
       ZD-SET-2 &  108 &  105 &  104 &  104 &  103 &  103 &  100 &  100 &  101 &   99 &   98 &   98 &   98 \\
      ZD-GTFT-2 &  112 &   95 &   88 &   84 &   75 &   72 &   71 &   73 &   71 &   71 &   67 &   68 &   68 \\
       ZD-GEN-2 &  114 &   96 &   89 &   86 &   77 &   75 &   72 &   74 &   72 &   72 &   68 &   69 &   69 \\
\bottomrule
\end{tabular}
    \caption{Resistance: Fixation ranks of some strategies according to \(x_{N-1}\) for different
    population sizes}
    \label{tbl:ranks_v_size_resist}
\end{table}

\begin{table}[!hbtp]
    \centering
    \scriptsize
    \begin{tabular}{lrrrrrrr}
\toprule
           Size &      2 &      4 &      6 &      8 &     10 &     12 &     14 \\
\midrule
             CS &    1 &    1 &    1 &    1 &    1 &    1 &    2 \\
       Defector &    2 &   78 &   99 &  106 &  110 &  113 &  120 \\
     Aggravater &    3 &   91 &  105 &  111 &  122 &  125 &  128 \\
       Predator &    4 &    2 &    4 &    4 &    4 &    4 &    4 \\
      Handshake &    5 &    6 &    5 &    6 &    6 &    6 &    6 \\
\midrule
            \textbf{TF2} &    9 &    4 &    3 &    2 &    2 &    2 &    1 \\
            \textbf{TF1} &    7 &    3 &    2 &    3 &    3 &    3 &    3 \\
       Prober 4 &    6 &    5 &    6 &    5 &    5 &    5 &    5 \\
\midrule
            \textbf{TF3} &   14 &    8 &    8 &    8 &    8 &    8 &    8 \\
\midrule
    ZD-Extort-4 &   16 &  102 &  117 &  129 &  141 &  143 &  145 \\
 ZD-Extort-2 v2 &   41 &  118 &  135 &  151 &  152 &  152 &  153 \\
    ZD-Extort-2 &   43 &  117 &  136 &  149 &  151 &  151 &  152 \\
       ZD-SET-2 &  100 &  110 &  110 &  108 &  106 &  106 &  108 \\
      ZD-GTFT-2 &  112 &   82 &   80 &   77 &   75 &   75 &   74 \\
       ZD-GEN-2 &  113 &   85 &   81 &   82 &   79 &   77 &   76 \\
\bottomrule
\end{tabular}
    \caption{Ranks of some strategies according to \(x_{N/2}\) for different
    population sizes}
    \label{tbl:ranks_v_size_coexist}
\end{table}

\begin{figure}[!htbp]
    \centering
    \begin{subfigure}[t]{.3\textwidth}
        \centering
        \includegraphics[width=.9\textwidth]{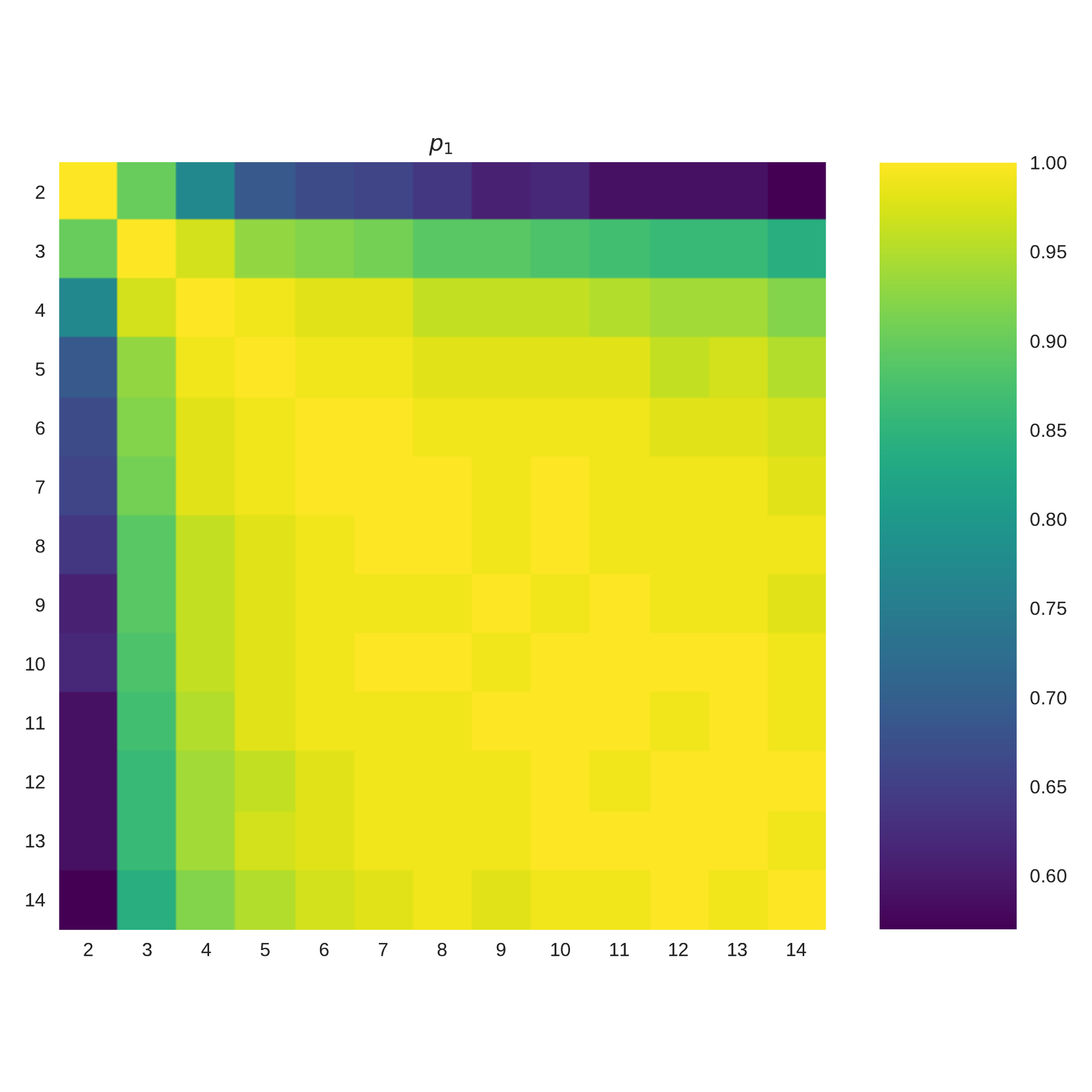}
        \caption{Rank based on \(x_1\)}
    \end{subfigure}
    ~
    \begin{subfigure}[t]{.3\textwidth}
        \centering
        \includegraphics[width=.9\textwidth]{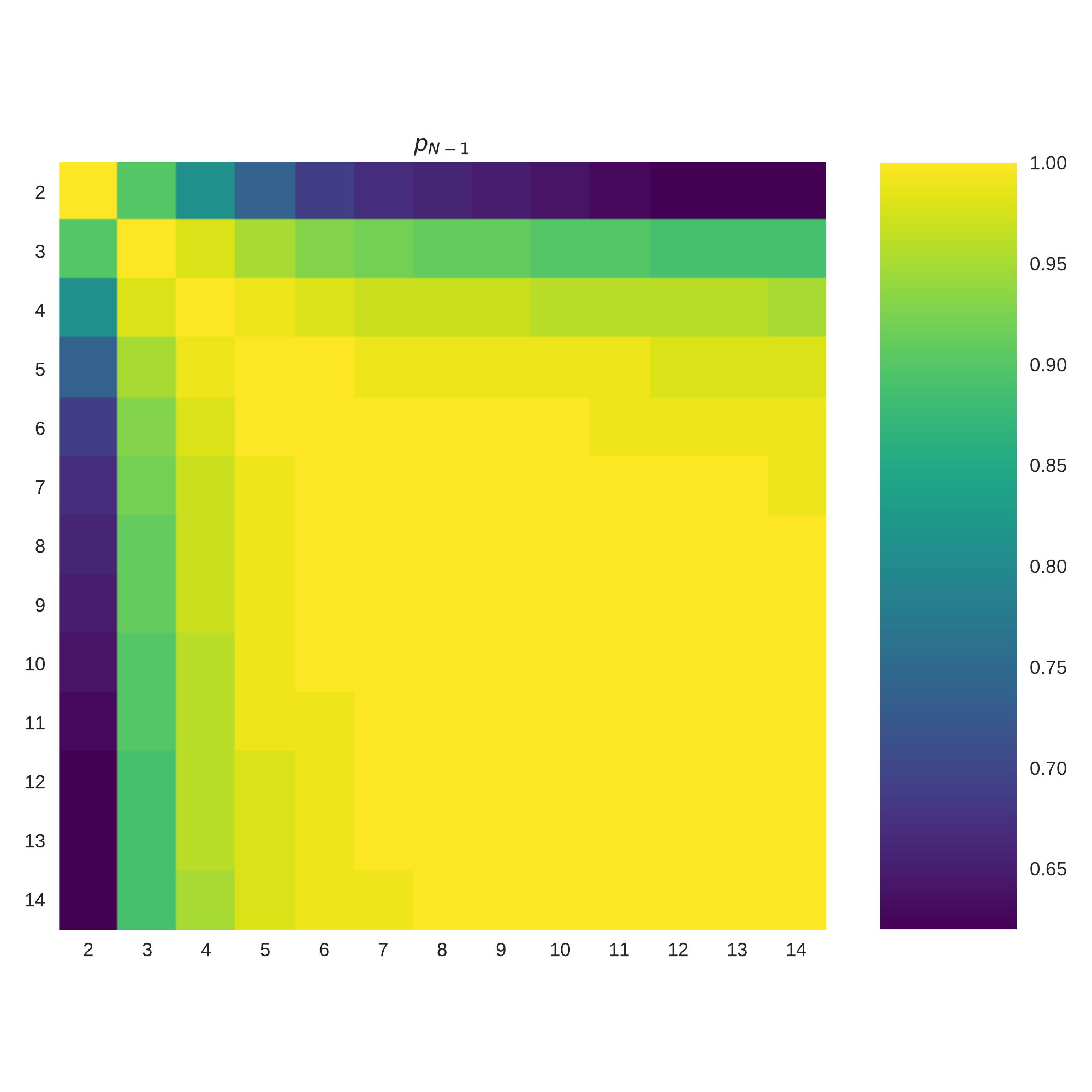}
        \caption{Rank based on \(x_{N - 1}\)}
    \end{subfigure}
    ~
    \begin{subfigure}[t]{.3\textwidth}
        \centering
        \includegraphics[width=.9\textwidth]{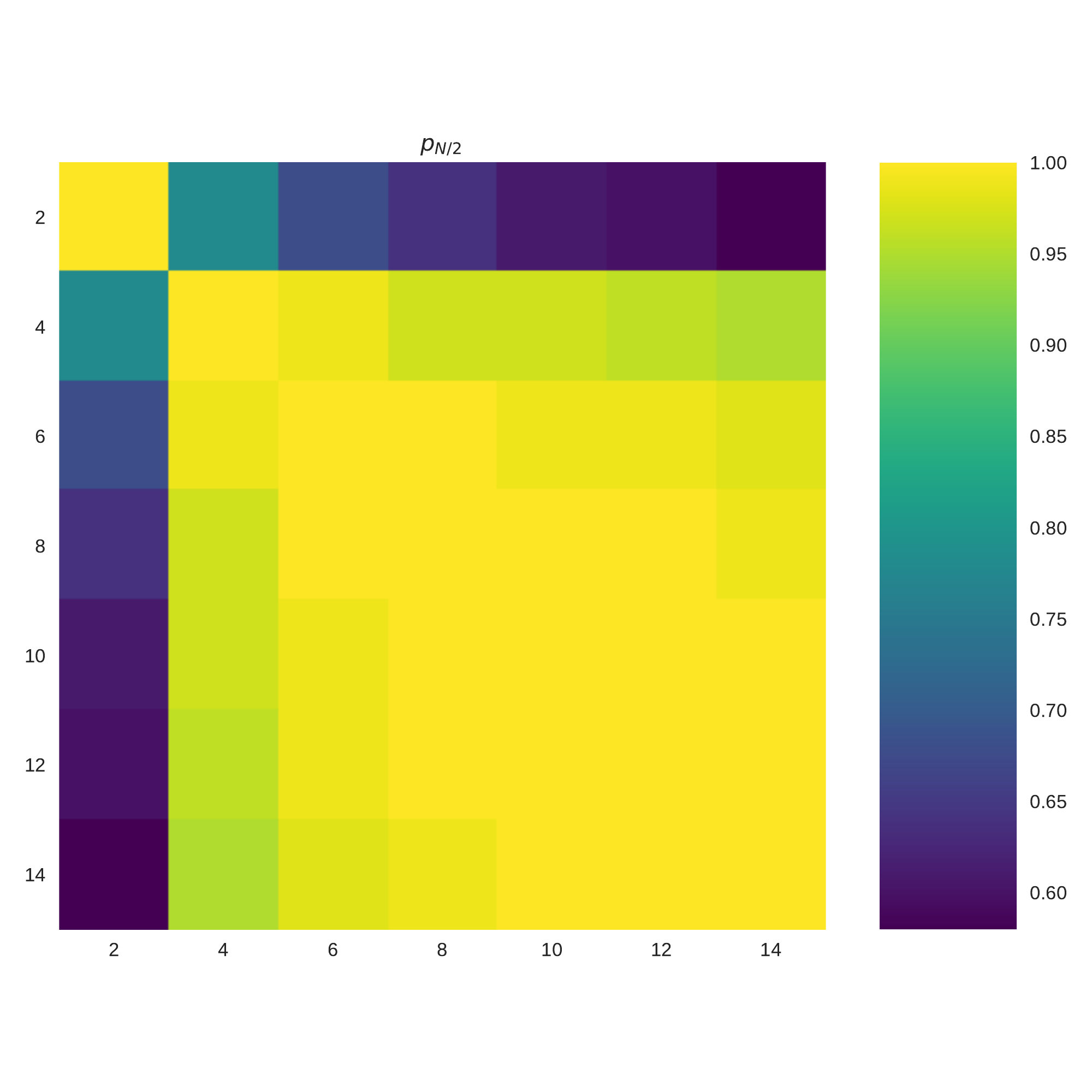}
        \caption{Rank based on \(x_{N/2}\)}
    \end{subfigure}
    \caption{Heatmap of correlation coefficients of rankings by population size.}
    \label{fig:correlation_coefficients}
\end{figure}

Figure~\ref{fig:correlation_coefficients} show the correlation coefficients
of the ranks of strategies in differing population size. How well a strategy
performs in any Moran process for \(N>2\) has
little to do with the performance for \(N=2\). This illustrates why the strong
performance of zero determinant strategies predicted in \cite{Press2012} does
not extend to larger populations. This was discussed theoretically in
\cite{Adami2013} and observed empirically in these simulations.

\section{Discussion}

Training strategies to excel
at the Moran process leads to the evolution of cooperation, but only with like
individuals in the case of TF1 and TF2. This may have significant implications
for human social interactions such as the evolution of ingroup/outgroup mechanisms
and other sometimes costly rituals that reinforce group behavior.

While TF1 and TF2 are competent invaders, the best invaders
in the study do not appear to employ strict handshakes, and are generally
cooperative strategies. TF3, which does not use a handshake, is a better invader
than TF1 and TF2 but not as good a resistor. Nevertheless it was the result
of the same kind of training procceses and is a better combined invader-resistor
than the invaders that were trained previously to maximize payout.

The strategies trained to maximize payoff in head-to-head matches are generally
cooperative and are effective invaders.
Combined with the fact that handshaking strategies are stronger resisters,
this suggests that while maximizing individual payoff can lead to the evolution
of cooperation, these strategies are not the most evolutionarily stable
in the long run. A strategy with a handshaking mechanism is still capable of
invading and is more resistant to subsequent invasions. Moreover, the
best resistor of the payoff maximally trained strategies (Evolved Looker Up
1\_1\_1),
which always defects if the opponent defects in the first round, is effectively
employing a one-shot handshake of C. Similarly, Grudger (also known as Grim),
which emerged from training memory one strategies for the Moran process,
also effectively employs a handshake of always cooperating, as it defects
for the remainder of the match if the opponent ever defects.

The insights that payoff maximizers are better invaders and that handshakers
are better resistors suggests that a strategy
aware of the population distribution could choose to become a handshaker at
a critical threshold and use a strategy better for invasion when in the
minority. Information about the population distribution was not available
to our strategies. Previous work has showed that strategies able to retain
memory across matches can infer the population distribution and act in such
a manner, resulting in a strategy effective at invasion and resistance
\cite{Lee2015}.

We did not attempt other objective functions that may serve to select for both
invasion and resistance better than training at a starting population of
$(N/2, N/2)$. Nevertheless our results suggest that there is not much room for
improvement. Any handshake more sophisticated than always cooperate necessarily involves
a defection. (A strategy with a handshake consisting of a long sequence of cooperations is
effectively a grudger.) For TF3 or EvolvedLookerUp1\_1\_1 to become better resistors
they need a longer or more strict handshake. But if this handshake involves
a defection then likely the invasion ability is diminished for $N > 2$: the top
invaders for larger $N$ are nice strategies that do not defect before their
opponents. This is because good invaders need to maximize match payoff to benefit
from fitness proportionate selection,
and so in the absence of a handshake mechanism, knowledge of the population
distribution, or some identifying label on the opponent,
a strategy must be generally cooperative. Aggressive strategies
are only effective invaders for the smallest $N$, dropping dramatically in rank
as the population size increases.

We did, however, attempt to evolve CS using finite state machines and lookup table
based players,
which resulted in some very similar strategies. In particular we evolved a
lookup strategy that had a handshake of DC and played TFT with other players
after a correct handshake while defecting otherwise, which is quite close in
function to CS (full grudging is not possible with a lookup table of limited
depth).

Finally we note that it may be possible to achieve similar results with smaller
capacity finite state machine players.

\section{Conclusion}\label{sec:conclusion}

A detailed empirical analysis of 164 strategies of the IPD within a pairwise
Moran process has been carried out. All \(\binom{164}{2}=13,366\) possible
ordered pairs of strategies have been placed in a Moran process with different
starting values allowing the each strategy to attempt to invade the other.
This is the largest such experiment carried out and has led to many insights.

When studying evolutionary processes it is vital to consider \(N>2\) since
results for \(N=2\) cannot be used to extrapolate performance in larger
populations. This was shown both observationally in
Sections~\ref{sec:strong_invaders} and~\ref{sec:strong_resistors} but also by
considering the correlation of the ranks in different population sizes in
Section~\ref{sec:population_size}.

Memory one strategies do not perform as well as longer memory strategies in general
in this study. Several longer memory strategies were high performers for invasion,
particularly the strategies which have been trained using a number of reinforcement
learning algorithms. Interestingly they have been trained to perform well in
tournaments and not Moran processes specifically. In some cases these strategies
utilize all the history of play (the neural network strategies and the lookup
table strategies, the latter using the first round and some number of trailing rounds).

There are no memory one strategies in the top 5 performing strategies
for \(N>3\). Training memory-one strategies specifically for the Moran process
typically led to Grudger / Grim, a memory-one strategy with
four-vector (1, 0, 0, 0). It appears to be the best resistor of the memory-one strategies.
The highest performing memory-one strategy for invasion is PSO Gambler Mem 1,
training to maximize total payout, which has four-vector $(1, 0.52173487, 0, 0.12050939)$.
For comparison, training for maximum score difference between the player and the
opponent resulted in a strategy nearly the same as Grudger, with four-vector
$(0.9459, 0, 0, 0)$ (not included in the study).

\begin{figure}[!hbtp]
    \centering
    \begin{subfigure}[t]{.5\textwidth}
        \centering
        \includegraphics[width=\textwidth]{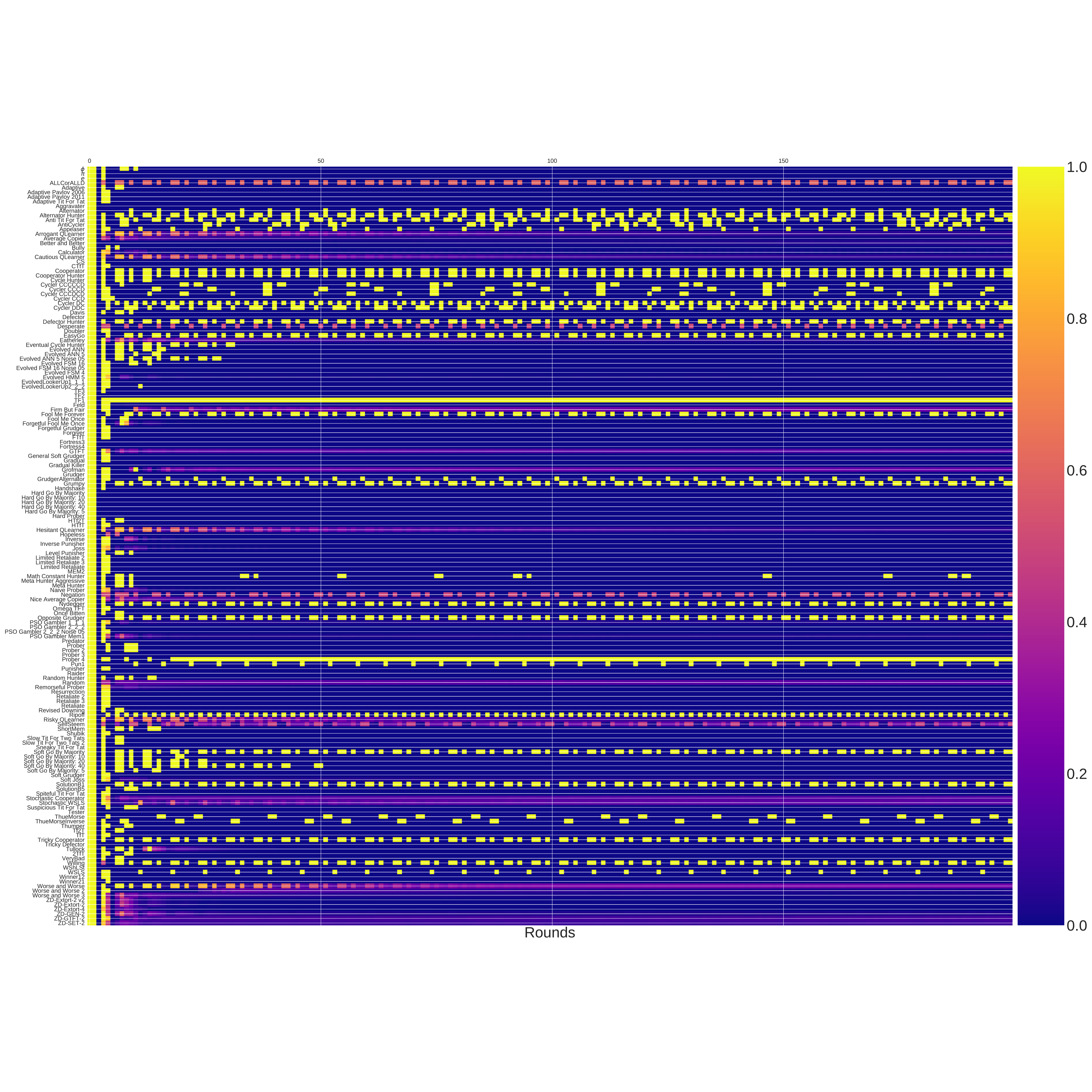}
        \caption{TF1}
    \end{subfigure}%
    ~
    \begin{subfigure}[t]{.5\textwidth}
        \centering
        \includegraphics[width=\textwidth]{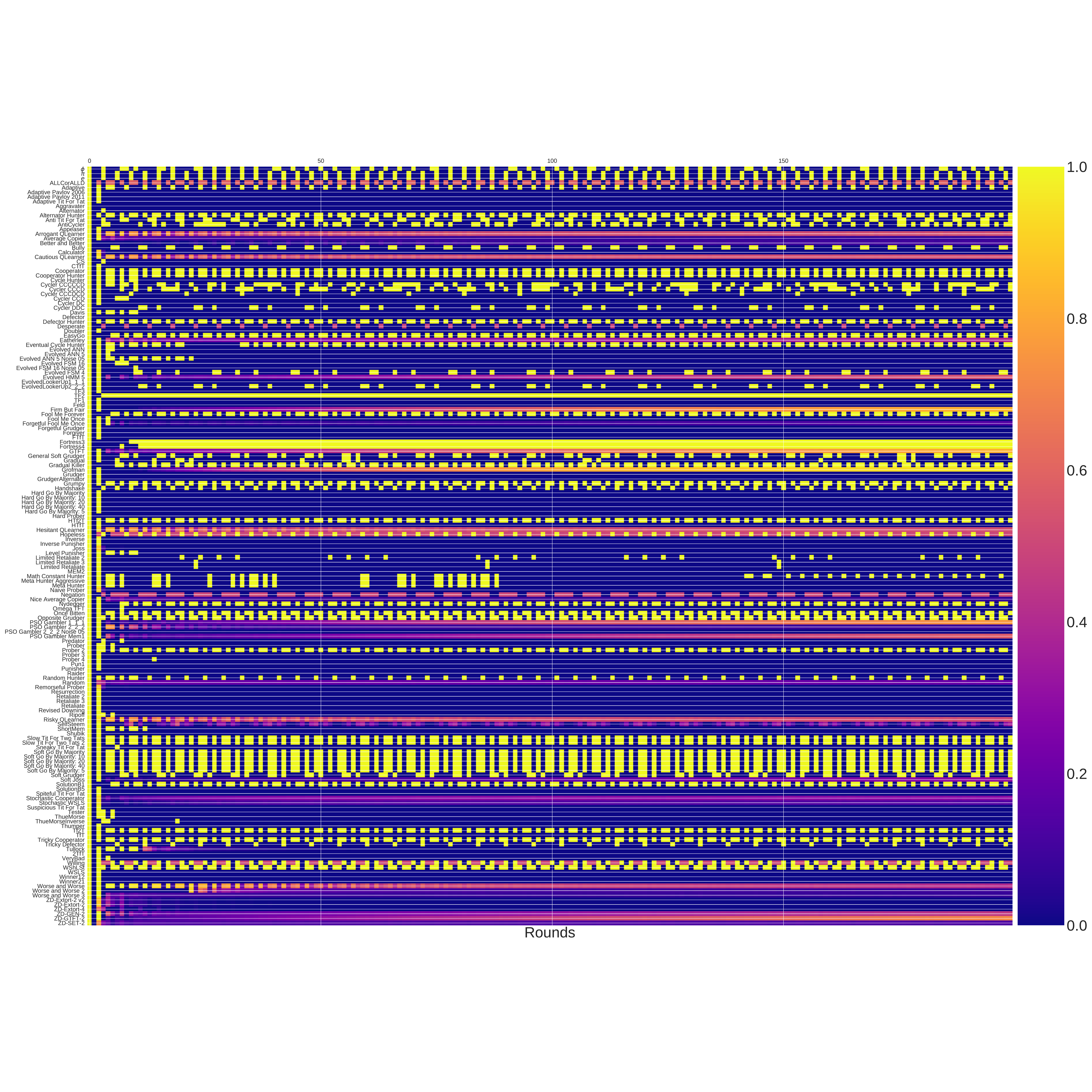}
        \caption{TF2}
    \end{subfigure}

    \begin{subfigure}[t]{.5\textwidth}
        \centering
        \includegraphics[width=\textwidth]{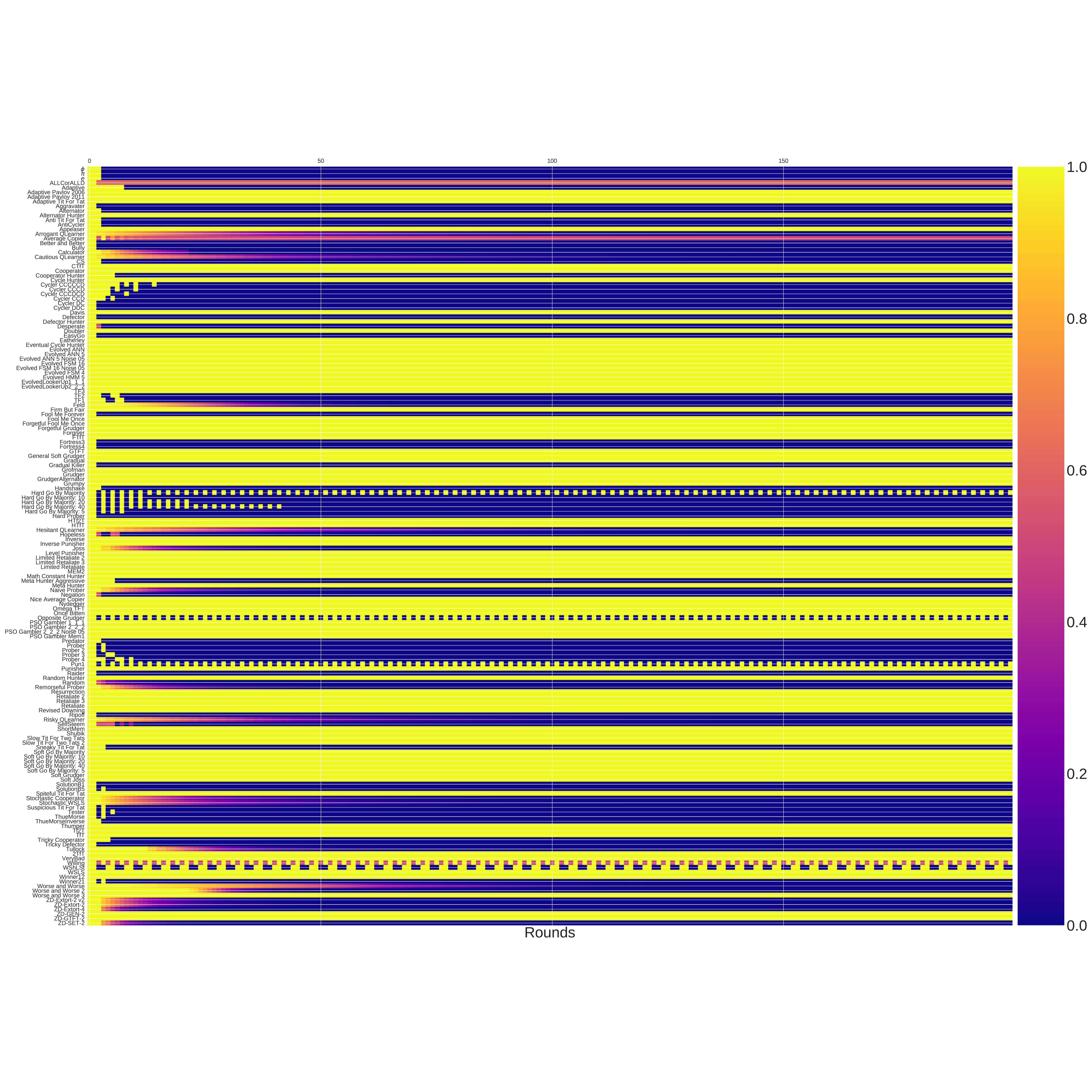}
        \caption{TF3}
    \end{subfigure}%
    ~
    \begin{subfigure}[t]{.5\textwidth}
        \centering
        \includegraphics[width=\textwidth]{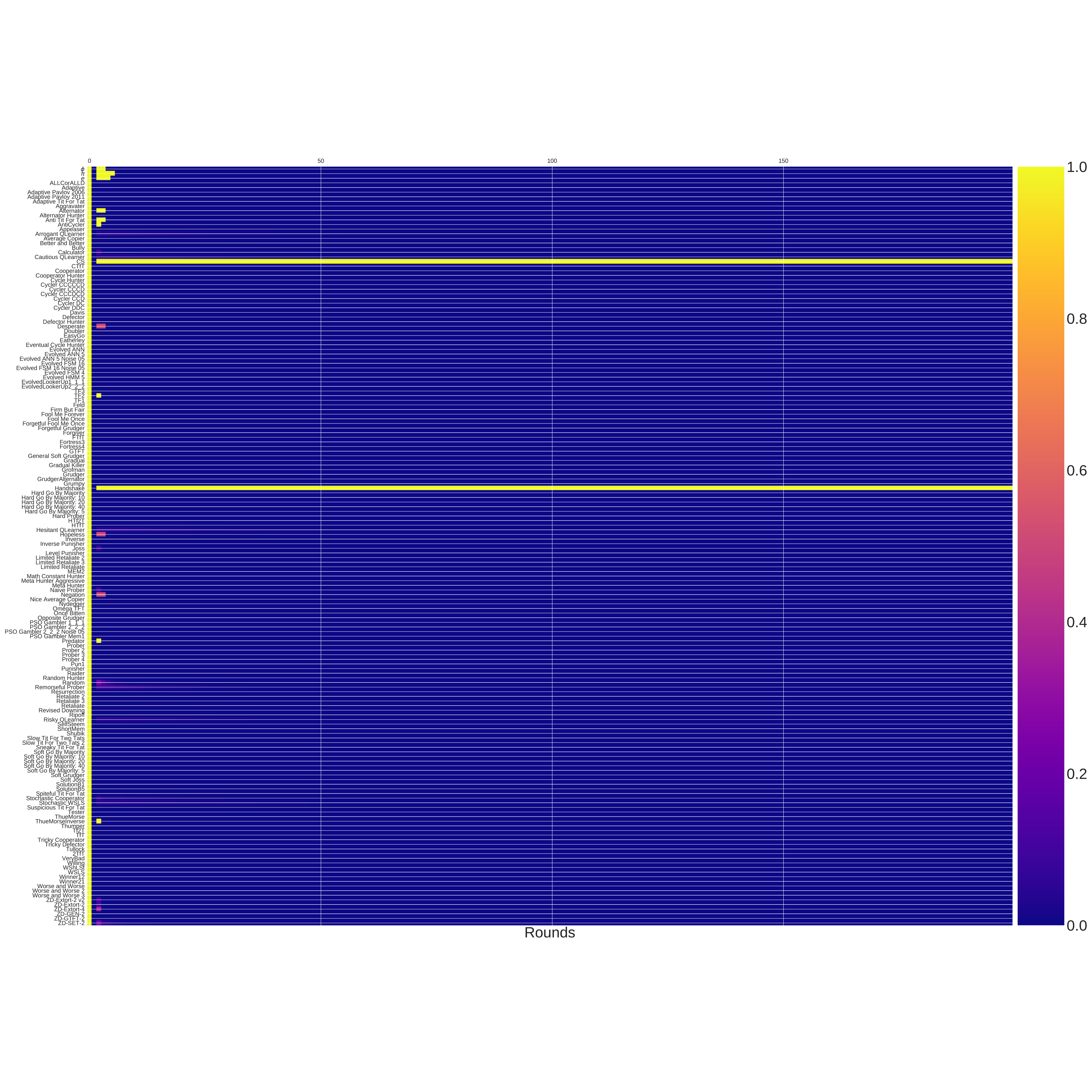}
        \caption{CS}
    \end{subfigure}
    \caption{Cooperation rate per round (over 10000 repetitions).}
    \label{fig:cooperation_rates}
\end{figure}

One of the major findings discussed in Section~\ref{sec:strong_resistors}, is
the ability of strategies with a handshake mechanism to resist invasion. This
was not only revealed for CS (a human designed strategy) but also for
two FSM strategies (TF1 and TF2) specifically trained through an evolutionary
process. In these two cases, the handshake mechanism was a product of the
evolutionary process. Figure~\ref{fig:cooperation_rates} shows the cooperation
rate of TF1, TF2, TF3 and CS
for each round of a match against all the opponents in this study. While TF3
does not have a strict handshake mechanism it is clear that all these strategies
start a match by cooperating. It is then evident that TF3 cooperates more than
the other strategies thus explaining the difference in performance. It is also
clear that CS only cooperates with itself and Handshake: it is a very aggressive
strategy.

These findings are important for the ongoing understanding of
population dynamics and offer evidence for some of the shortcomings of low
memory which has started to be recognised by the community \cite{Hilbe2017}.

All source code for this work has been written in a sustainable manner: it is
open source, under version control and tested which ensures that all results can
be reproduced \cite{Prlic2012, Sandve2013, Wilson2014}. The raw data as well as
the processed data has also been properly archived and can be found
at \cite{data}.

There are many opportunities to build on this work. In particular, an analysis
of the effect of noise should offer insights regarding the stability of the findings,
particularly for the handshaking strategies. They may be less dominant for
larger amounts of noise since the handshaking mechanisms may become brittle.
There are many other variations to explore including populations with more
than one type, spatial structure, and mutation.

\section*{Acknowledgements}

This work was performed using the computational facilities of the Advanced
Research Computing @ Cardiff (ARCCA) Division, Cardiff University.

A variety of software libraries have been used in this work:

\begin{itemize}
    \item The Axelrod library (IPD strategies and Moran processes)
        \cite{axelrodproject}.
    \item The matplotlib library (visualisation) \cite{hunter2007matplotlib}.
    \item The pandas and numpy libraries (data manipulation)
        \cite{mckinney2010data, walt2011numpy}.
\end{itemize}

\newpage
\bibliographystyle{plain}
\bibliography{./tex/bibliography.bib}

\appendix

\section{List of players}\label{app:list_of_players}

\begin{multicols}{2}
	\begin{enumerate}
		\item $\phi$ - \textit{Deterministic} - \textit{Memory depth}: \(\infty\). \cite{axelrodproject}
\item $\pi$ - \textit{Deterministic} - \textit{Memory depth}: \(\infty\). \cite{axelrodproject}
\item $e$ - \textit{Deterministic} - \textit{Memory depth}: \(\infty\). \cite{axelrodproject}
\item ALLCorALLD - \textit{Stochastic} - \textit{Memory depth}: 1. \cite{axelrodproject}
\item Adaptive - \textit{Deterministic} - \textit{Memory depth}: \(\infty\). \cite{Li2011}
\item Adaptive Pavlov 2006 - \textit{Deterministic} - \textit{Memory depth}: \(\infty\). \cite{kendall2007iterated}
\item Adaptive Pavlov 2011 - \textit{Deterministic} - \textit{Memory depth}: \(\infty\). \cite{Li2011}
\item Adaptive Tit For Tat: 0.5 - \textit{Deterministic} - \textit{Memory depth}: \(\infty\). \cite{Tzafestas2000}
\item Aggravater - \textit{Deterministic} - \textit{Memory depth}: \(\infty\). \cite{axelrodproject}
\item Alternator - \textit{Deterministic} - \textit{Memory depth}: 1. \cite{Axelrod1984, Mittal2009}
\item Alternator Hunter - \textit{Deterministic} - \textit{Memory depth}: \(\infty\). \cite{axelrodproject}
\item Anti Tit For Tat - \textit{Deterministic} - \textit{Memory depth}: 1. \cite{Hilbe2013}
\item AntiCycler - \textit{Deterministic} - \textit{Memory depth}: \(\infty\). \cite{axelrodproject}
\item Appeaser - \textit{Deterministic} - \textit{Memory depth}: \(\infty\). \cite{axelrodproject}
\item Arrogant QLearner - \textit{Stochastic} - \textit{Memory depth}: \(\infty\). \cite{axelrodproject}
\item Average Copier - \textit{Stochastic} - \textit{Memory depth}: \(\infty\). \cite{axelrodproject}
\item Better and Better - \textit{Stochastic} - \textit{Memory depth}: \(\infty\). \cite{Prison1998}
\item Bully - \textit{Deterministic} - \textit{Memory depth}: 1. \cite{Nachbar1992}
\item Calculator - \textit{Stochastic} - \textit{Memory depth}: \(\infty\). \cite{Prison1998}
\item Cautious QLearner - \textit{Stochastic} - \textit{Memory depth}: \(\infty\). \cite{axelrodproject}
\item CollectiveStrategy(\textbf{CS}) - \textit{Deterministic} - \textit{Memory depth}: \(\infty\). \cite{Li2009}
\item Contrite Tit For Tat(\textbf{CTfT}) - \textit{Deterministic} - \textit{Memory depth}: 3. \cite{Axelrod1995}
\item Cooperator - \textit{Deterministic} - \textit{Memory depth}: 0. \cite{Axelrod1984, Mittal2009, Press2012}
\item Cooperator Hunter - \textit{Deterministic} - \textit{Memory depth}: \(\infty\). \cite{axelrodproject}
\item Cycle Hunter - \textit{Deterministic} - \textit{Memory depth}: \(\infty\). \cite{axelrodproject}
\item Cycler CCCCCD - \textit{Deterministic} - \textit{Memory depth}: 5. \cite{axelrodproject}
\item Cycler CCCD - \textit{Deterministic} - \textit{Memory depth}: 3. \cite{axelrodproject}
\item Cycler CCCDCD - \textit{Deterministic} - \textit{Memory depth}: 5. \cite{axelrodproject}
\item Cycler CCD - \textit{Deterministic} - \textit{Memory depth}: 2. \cite{Mittal2009}
\item Cycler DC - \textit{Deterministic} - \textit{Memory depth}: 1. \cite{axelrodproject}
\item Cycler DDC - \textit{Deterministic} - \textit{Memory depth}: 2. \cite{Mittal2009}
\item Davis: 10 - \textit{Deterministic} - \textit{Memory depth}: \(\infty\). \cite{Axelrod1980}
\item Defector - \textit{Deterministic} - \textit{Memory depth}: 0. \cite{Axelrod1984, Mittal2009, Press2012}
\item Defector Hunter - \textit{Deterministic} - \textit{Memory depth}: \(\infty\). \cite{axelrodproject}
\item Desperate - \textit{Stochastic} - \textit{Memory depth}: 1. \cite{Berg2015}
\item Doubler - \textit{Deterministic} - \textit{Memory depth}: \(\infty\). \cite{Prison1998}
\item EasyGo - \textit{Deterministic} - \textit{Memory depth}: \(\infty\). \cite{Li2011, Prison1998}
\item Eatherley - \textit{Stochastic} - \textit{Memory depth}: \(\infty\). \cite{Axelrod1980b}
\item Eventual Cycle Hunter - \textit{Deterministic} - \textit{Memory depth}: \(\infty\). \cite{axelrodproject}
\item Evolved ANN - \textit{Deterministic} - \textit{Memory depth}: \(\infty\). \cite{axelrodproject}
\item Evolved ANN 5 - \textit{Deterministic} - \textit{Memory depth}: \(\infty\). \cite{axelrodproject}
\item Evolved ANN 5 Noise 05 - \textit{Deterministic} - \textit{Memory depth}: \(\infty\). \cite{axelrodproject}
\item Evolved FSM 16 - \textit{Deterministic} - \textit{Memory depth}: 16. \cite{axelrodproject}
\item Evolved FSM 16 Noise 05 - \textit{Deterministic} - \textit{Memory depth}: 16. \cite{axelrodproject}
\item Evolved FSM 4 - \textit{Deterministic} - \textit{Memory depth}: 4. \cite{axelrodproject}
\item Evolved HMM 5 - \textit{Stochastic} - \textit{Memory depth}: 5. \cite{axelrodproject}
\item EvolvedLookerUp1\_1\_1 - \textit{Deterministic} - \textit{Memory depth}: \(\infty\). \cite{axelrodproject}
\item EvolvedLookerUp2\_2\_2 - \textit{Deterministic} - \textit{Memory depth}: \(\infty\). \cite{axelrodproject}
\item FSM Player: [(0, 'C', 0, 'C'), (0, 'D', 3, 'C'), (1, 'C', 5, 'D'), (1, 'D', 0, 'C'), (2, 'C', 3, 'C'), (2, 'D', 2, 'D'), (3, 'C', 4, 'D'), (3, 'D', 6, 'D'), (4, 'C', 3, 'C'), (4, 'D', 1, 'D'), (5, 'C', 6, 'C'), (5, 'D', 3, 'D'), (6, 'C', 6, 'D'), (6, 'D', 6, 'D'), (7, 'C', 7, 'D'), (7, 'D', 5, 'C')], 0, C(\textbf{TF3}) - \textit{Deterministic} - \textit{Memory depth}: \(\infty\).
\item FSM Player: [(0, 'C', 13, 'D'), (0, 'D', 12, 'D'), (1, 'C', 3, 'D'), (1, 'D', 4, 'D'), (2, 'C', 14, 'D'), (2, 'D', 9, 'D'), (3, 'C', 0, 'C'), (3, 'D', 1, 'D'), (4, 'C', 1, 'D'), (4, 'D', 2, 'D'), (5, 'C', 12, 'C'), (5, 'D', 6, 'C'), (6, 'C', 1, 'C'), (6, 'D', 14, 'D'), (7, 'C', 12, 'D'), (7, 'D', 2, 'D'), (8, 'C', 7, 'D'), (8, 'D', 9, 'D'), (9, 'C', 8, 'D'), (9, 'D', 0, 'D'), (10, 'C', 2, 'C'), (10, 'D', 15, 'C'), (11, 'C', 7, 'D'), (11, 'D', 13, 'D'), (12, 'C', 3, 'C'), (12, 'D', 8, 'D'), (13, 'C', 7, 'C'), (13, 'D', 10, 'D'), (14, 'C', 10, 'D'), (14, 'D', 7, 'D'), (15, 'C', 15, 'C'), (15, 'D', 11, 'D')], 0, C(\textbf{TF2}) - \textit{Deterministic} - \textit{Memory depth}: \(\infty\).
\item FSM Player: [(0, 'C', 7, 'C'), (0, 'D', 1, 'C'), (1, 'C', 11, 'D'), (1, 'D', 11, 'D'), (2, 'C', 8, 'D'), (2, 'D', 8, 'C'), (3, 'C', 3, 'C'), (3, 'D', 12, 'D'), (4, 'C', 6, 'C'), (4, 'D', 3, 'C'), (5, 'C', 11, 'C'), (5, 'D', 8, 'D'), (6, 'C', 13, 'D'), (6, 'D', 14, 'C'), (7, 'C', 4, 'D'), (7, 'D', 2, 'D'), (8, 'C', 14, 'D'), (8, 'D', 8, 'D'), (9, 'C', 0, 'C'), (9, 'D', 10, 'D'), (10, 'C', 8, 'C'), (10, 'D', 15, 'C'), (11, 'C', 6, 'D'), (11, 'D', 5, 'D'), (12, 'C', 6, 'D'), (12, 'D', 9, 'D'), (13, 'C', 9, 'D'), (13, 'D', 8, 'D'), (14, 'C', 8, 'D'), (14, 'D', 13, 'D'), (15, 'C', 4, 'C'), (15, 'D', 5, 'C')], 0, C(\textbf{TF1}) - \textit{Deterministic} - \textit{Memory depth}: \(\infty\).
\item Feld: 1.0, 0.5, 200 - \textit{Stochastic} - \textit{Memory depth}: 200. \cite{Axelrod1980}
\item Firm But Fair - \textit{Stochastic} - \textit{Memory depth}: 1. \cite{Frean1994}
\item Fool Me Forever - \textit{Deterministic} - \textit{Memory depth}: \(\infty\). \cite{axelrodproject}
\item Fool Me Once - \textit{Deterministic} - \textit{Memory depth}: \(\infty\). \cite{axelrodproject}
\item Forgetful Fool Me Once: 0.05 - \textit{Stochastic} - \textit{Memory depth}: \(\infty\). \cite{axelrodproject}
\item Forgetful Grudger - \textit{Deterministic} - \textit{Memory depth}: 10. \cite{axelrodproject}
\item Forgiver - \textit{Deterministic} - \textit{Memory depth}: \(\infty\). \cite{axelrodproject}
\item Forgiving Tit For Tat(\textbf{FTfT}) - \textit{Deterministic} - \textit{Memory depth}: \(\infty\). \cite{axelrodproject}
\item Fortress3 - \textit{Deterministic} - \textit{Memory depth}: 3. \cite{Ashlock2006b}
\item Fortress4 - \textit{Deterministic} - \textit{Memory depth}: 4. \cite{Ashlock2006b}
\item GTFT: 0.33 - \textit{Stochastic} - \textit{Memory depth}: 1. \cite{Gaudesi2016, Nowak1993}
\item General Soft Grudger: n=1,d=4,c=2 - \textit{Deterministic} - \textit{Memory depth}: \(\infty\). \cite{axelrodproject}
\item Gradual - \textit{Deterministic} - \textit{Memory depth}: \(\infty\). \cite{Beaufils1997}
\item Gradual Killer: ('D', 'D', 'D', 'D', 'D', 'C', 'C') - \textit{Deterministic} - \textit{Memory depth}: \(\infty\). \cite{Prison1998}
\item Grofman - \textit{Stochastic} - \textit{Memory depth}: \(\infty\). \cite{Axelrod1980}
\item Grudger - \textit{Deterministic} - \textit{Memory depth}: 1. \cite{Axelrod1980, Banks1990, Beaufils1997, Berg2015, Li2011}
\item GrudgerAlternator - \textit{Deterministic} - \textit{Memory depth}: \(\infty\). \cite{Prison1998}
\item Grumpy: Nice, 10, -10 - \textit{Deterministic} - \textit{Memory depth}: \(\infty\). \cite{axelrodproject}
\item Handshake - \textit{Deterministic} - \textit{Memory depth}: \(\infty\). \cite{Robson1990}
\item Hard Go By Majority - \textit{Deterministic} - \textit{Memory depth}: \(\infty\). \cite{Mittal2009}
\item Hard Go By Majority: 10 - \textit{Deterministic} - \textit{Memory depth}: 10. \cite{axelrodproject}
\item Hard Go By Majority: 20 - \textit{Deterministic} - \textit{Memory depth}: 20. \cite{axelrodproject}
\item Hard Go By Majority: 40 - \textit{Deterministic} - \textit{Memory depth}: 40. \cite{axelrodproject}
\item Hard Go By Majority: 5 - \textit{Deterministic} - \textit{Memory depth}: 5. \cite{axelrodproject}
\item Hard Prober - \textit{Deterministic} - \textit{Memory depth}: \(\infty\). \cite{Prison1998}
\item Hard Tit For 2 Tats(\textbf{HTf2T}) - \textit{Deterministic} - \textit{Memory depth}: 3. \cite{Stewart2012}
\item Hard Tit For Tat(\textbf{HTfT}) - \textit{Deterministic} - \textit{Memory depth}: 3. \cite{PD2017}
\item Hesitant QLearner - \textit{Stochastic} - \textit{Memory depth}: \(\infty\). \cite{axelrodproject}
\item Hopeless - \textit{Stochastic} - \textit{Memory depth}: 1. \cite{Berg2015}
\item Inverse - \textit{Stochastic} - \textit{Memory depth}: \(\infty\). \cite{axelrodproject}
\item Inverse Punisher - \textit{Deterministic} - \textit{Memory depth}: \(\infty\). \cite{axelrodproject}
\item Joss: 0.9 - \textit{Stochastic} - \textit{Memory depth}: 1. \cite{Axelrod1980, Stewart2012}
\item Level Punisher - \textit{Deterministic} - \textit{Memory depth}: \(\infty\). \cite{Eckhart2015}
\item Limited Retaliate 2: 0.08, 15 - \textit{Deterministic} - \textit{Memory depth}: \(\infty\). \cite{axelrodproject}
\item Limited Retaliate 3: 0.05, 20 - \textit{Deterministic} - \textit{Memory depth}: \(\infty\). \cite{axelrodproject}
\item Limited Retaliate: 0.1, 20 - \textit{Deterministic} - \textit{Memory depth}: \(\infty\). \cite{axelrodproject}
\item MEM2 - \textit{Deterministic} - \textit{Memory depth}: \(\infty\). \cite{Li2014}
\item Math Constant Hunter - \textit{Deterministic} - \textit{Memory depth}: \(\infty\). \cite{axelrodproject}
\item Meta Hunter Aggressive: 7 players - \textit{Deterministic} - \textit{Memory depth}: \(\infty\). \cite{axelrodproject}
\item Meta Hunter: 6 players - \textit{Deterministic} - \textit{Memory depth}: \(\infty\). \cite{axelrodproject}
\item Naive Prober: 0.1 - \textit{Stochastic} - \textit{Memory depth}: 1. \cite{Li2011}
\item Negation - \textit{Stochastic} - \textit{Memory depth}: 1. \cite{PD2017}
\item Nice Average Copier - \textit{Stochastic} - \textit{Memory depth}: \(\infty\). \cite{axelrodproject}
\item Nydegger - \textit{Deterministic} - \textit{Memory depth}: 3. \cite{Axelrod1980}
\item Omega TFT: 3, 8 - \textit{Deterministic} - \textit{Memory depth}: \(\infty\). \cite{kendall2007iterated}
\item Once Bitten - \textit{Deterministic} - \textit{Memory depth}: 12. \cite{axelrodproject}
\item Opposite Grudger - \textit{Deterministic} - \textit{Memory depth}: \(\infty\). \cite{axelrodproject}
\item PSO Gambler 1\_1\_1 - \textit{Stochastic} - \textit{Memory depth}: \(\infty\). \cite{axelrodproject}
\item PSO Gambler 2\_2\_2 - \textit{Stochastic} - \textit{Memory depth}: \(\infty\). \cite{axelrodproject}
\item PSO Gambler 2\_2\_2 Noise 05 - \textit{Stochastic} - \textit{Memory depth}: \(\infty\). \cite{axelrodproject}
\item PSO Gambler Mem1 - \textit{Stochastic} - \textit{Memory depth}: 1. \cite{axelrodproject}
\item Predator - \textit{Deterministic} - \textit{Memory depth}: 9. \cite{Ashlock2006b}
\item Prober - \textit{Deterministic} - \textit{Memory depth}: \(\infty\). \cite{Li2011}
\item Prober 2 - \textit{Deterministic} - \textit{Memory depth}: \(\infty\). \cite{Prison1998}
\item Prober 3 - \textit{Deterministic} - \textit{Memory depth}: \(\infty\). \cite{Prison1998}
\item Prober 4 - \textit{Deterministic} - \textit{Memory depth}: \(\infty\). \cite{Prison1998}
\item Pun1 - \textit{Deterministic} - \textit{Memory depth}: 2. \cite{Ashlock2006}
\item Punisher - \textit{Deterministic} - \textit{Memory depth}: \(\infty\). \cite{axelrodproject}
\item Raider - \textit{Deterministic} - \textit{Memory depth}: 3. \cite{Ashlock2014}
\item Random Hunter - \textit{Deterministic} - \textit{Memory depth}: \(\infty\). \cite{axelrodproject}
\item Random: 0.5 - \textit{Stochastic} - \textit{Memory depth}: 0. \cite{Axelrod1980, Tzafestas2000}
\item Remorseful Prober: 0.1 - \textit{Stochastic} - \textit{Memory depth}: 2. \cite{Li2011}
\item Resurrection - \textit{Deterministic} - \textit{Memory depth}: 1. \cite{Eckhart2015}
\item Retaliate 2: 0.08 - \textit{Deterministic} - \textit{Memory depth}: \(\infty\). \cite{axelrodproject}
\item Retaliate 3: 0.05 - \textit{Deterministic} - \textit{Memory depth}: \(\infty\). \cite{axelrodproject}
\item Retaliate: 0.1 - \textit{Deterministic} - \textit{Memory depth}: \(\infty\). \cite{axelrodproject}
\item Revised Downing: True - \textit{Deterministic} - \textit{Memory depth}: \(\infty\). \cite{Axelrod1980}
\item Ripoff - \textit{Deterministic} - \textit{Memory depth}: 2. \cite{Ashlock2008}
\item Risky QLearner - \textit{Stochastic} - \textit{Memory depth}: \(\infty\). \cite{axelrodproject}
\item SelfSteem - \textit{Stochastic} - \textit{Memory depth}: \(\infty\). \cite{Andre2013}
\item ShortMem - \textit{Deterministic} - \textit{Memory depth}: 10. \cite{Andre2013}
\item Shubik - \textit{Deterministic} - \textit{Memory depth}: \(\infty\). \cite{Axelrod1980}
\item Slow Tit For Two Tats - \textit{Deterministic} - \textit{Memory depth}: 2. \cite{axelrodproject}
\item Slow Tit For Two Tats 2 - \textit{Deterministic} - \textit{Memory depth}: 2. \cite{Prison1998}
\item Sneaky Tit For Tat - \textit{Deterministic} - \textit{Memory depth}: \(\infty\). \cite{axelrodproject}
\item Soft Go By Majority - \textit{Deterministic} - \textit{Memory depth}: \(\infty\). \cite{Axelrod1984, Mittal2009}
\item Soft Go By Majority: 10 - \textit{Deterministic} - \textit{Memory depth}: 10. \cite{axelrodproject}
\item Soft Go By Majority: 20 - \textit{Deterministic} - \textit{Memory depth}: 20. \cite{axelrodproject}
\item Soft Go By Majority: 40 - \textit{Deterministic} - \textit{Memory depth}: 40. \cite{axelrodproject}
\item Soft Go By Majority: 5 - \textit{Deterministic} - \textit{Memory depth}: 5. \cite{axelrodproject}
\item Soft Grudger - \textit{Deterministic} - \textit{Memory depth}: 6. \cite{Li2011}
\item Soft Joss: 0.9 - \textit{Stochastic} - \textit{Memory depth}: 1. \cite{Prison1998}
\item SolutionB1 - \textit{Deterministic} - \textit{Memory depth}: 3. \cite{Ashlock2015}
\item SolutionB5 - \textit{Deterministic} - \textit{Memory depth}: 5. \cite{Ashlock2015}
\item Spiteful Tit For Tat - \textit{Deterministic} - \textit{Memory depth}: \(\infty\). \cite{Prison1998}
\item Stochastic Cooperator - \textit{Stochastic} - \textit{Memory depth}: 1. \cite{Adami2013}
\item Stochastic WSLS: 0.05 - \textit{Stochastic} - \textit{Memory depth}: 1. \cite{axelrodproject}
\item Suspicious Tit For Tat - \textit{Deterministic} - \textit{Memory depth}: 1. \cite{Beaufils1997, Hilbe2013}
\item Tester - \textit{Deterministic} - \textit{Memory depth}: \(\infty\). \cite{Axelrod1980b}
\item ThueMorse - \textit{Deterministic} - \textit{Memory depth}: \(\infty\). \cite{axelrodproject}
\item ThueMorseInverse - \textit{Deterministic} - \textit{Memory depth}: \(\infty\). \cite{axelrodproject}
\item Thumper - \textit{Deterministic} - \textit{Memory depth}: 2. \cite{Ashlock2008}
\item Tit For 2 Tats(\textbf{Tf2T}) - \textit{Deterministic} - \textit{Memory depth}: 2. \cite{Axelrod1984}
\item Tit For Tat(\textbf{TfT}) - \textit{Deterministic} - \textit{Memory depth}: 1. \cite{Axelrod1980}
\item Tricky Cooperator - \textit{Deterministic} - \textit{Memory depth}: 10. \cite{axelrodproject}
\item Tricky Defector - \textit{Deterministic} - \textit{Memory depth}: \(\infty\). \cite{axelrodproject}
\item Tullock: 11 - \textit{Stochastic} - \textit{Memory depth}: 11. \cite{Axelrod1980}
\item Two Tits For Tat(\textbf{2TfT}) - \textit{Deterministic} - \textit{Memory depth}: 2. \cite{Axelrod1984}
\item VeryBad - \textit{Deterministic} - \textit{Memory depth}: \(\infty\). \cite{Andre2013}
\item Willing - \textit{Stochastic} - \textit{Memory depth}: 1. \cite{Berg2015}
\item Win-Shift Lose-Stay: D(\textbf{WShLSt}) - \textit{Deterministic} - \textit{Memory depth}: 1. \cite{Li2011}
\item Win-Stay Lose-Shift: C(\textbf{WSLS}) - \textit{Deterministic} - \textit{Memory depth}: 1. \cite{Kraines1989, Nowak1993, Stewart2012}
\item Winner12 - \textit{Deterministic} - \textit{Memory depth}: 2. \cite{Mathieu2015}
\item Winner21 - \textit{Deterministic} - \textit{Memory depth}: 2. \cite{Mathieu2015}
\item Worse and Worse - \textit{Stochastic} - \textit{Memory depth}: \(\infty\). \cite{Prison1998}
\item Worse and Worse 2 - \textit{Stochastic} - \textit{Memory depth}: \(\infty\). \cite{Prison1998}
\item Worse and Worse 3 - \textit{Stochastic} - \textit{Memory depth}: \(\infty\). \cite{Prison1998}
\item ZD-Extort-2 v2: 0.125, 0.5, 1 - \textit{Stochastic} - \textit{Memory depth}: 1. \cite{Kuhn2017}
\item ZD-Extort-2: 0.1111111111111111, 0.5 - \textit{Stochastic} - \textit{Memory depth}: 1. \cite{Stewart2012}
\item ZD-Extort-4: 0.23529411764705882, 0.25, 1 - \textit{Stochastic} - \textit{Memory depth}: 1. \cite{axelrodproject}
\item ZD-GEN-2: 0.125, 0.5, 3 - \textit{Stochastic} - \textit{Memory depth}: 1. \cite{Kuhn2017}
\item ZD-GTFT-2: 0.25, 0.5 - \textit{Stochastic} - \textit{Memory depth}: 1. \cite{Stewart2012}
\item ZD-SET-2: 0.25, 0.0, 2 - \textit{Stochastic} - \textit{Memory depth}: 1. \cite{Kuhn2017}

	\end{enumerate}
\end{multicols}

\end{document}